\documentclass[11pt]{article}

\usepackage{epsfig,latexsym,amsfonts,amsmath,amsthm,amssymb,amsbsy,multirow,slashed,wasysym,textcomp,subfigure,wrapfig,datetime,comment,mathtools,cancel,cite,mathrsfs,tikz}
\usetikzlibrary{calc,math,arrows.meta}
\usepackage{datetime2}
\usepackage[font={footnotesize,sl},bf]{caption}

\usepackage[normalem]{ulem}

\usepackage{setspace}
\usepackage{hyperref}

\def\bea{\begin{eqnarray}}
\def\eea{\end{eqnarray}}
\newcommand{\beq}{\begin{eqnarray}}
\newcommand{\eqq}{\end{eqnarray}}
 \newcommand{\badat}{\begin{alignedat}}
 \newcommand{\eadat}{\end{alignedat}}

\newcommand{\eal}[1]{\be \begin{aligned} #1 \end{aligned}\end{equation}} 
\newcommand{\eqn}[1]{\be #1 \end{equation}} 
\newcommand{\eqa}[1]{\bea  #1\end{eqnarray}}

\newcommand{\bz}{\bar{z}}

\newcommand{\scri}{\mathscr{I}}

\newcommand{\R}{\mathbb{R}}
\newcommand{\RP}{\mathbb{RP}}

\newcommand{\rd}{\mathrm{d}}
\newcommand{\re}{\mathrm{e}}

\newcommand{\sq}[1]{[#1]}
\newcommand{\an}[1]{\langle#1\rangle}
\newcommand{\mo}{\mathcal{O}}
\newcommand{\mc}{\mathcal{C}}
\newcommand{\mA}{\mathcal{A}}

\newcommand{\my}{\bar{\mathcal{Y}}}
\newcommand{\zb}{\bar{z}}
\newcommand{\hb}{\bar{h}}

\newcommand{\hypfo}{{}_1F_{1}}
\newcommand{\D}{\Delta}
\newcommand{\e}{\epsilon}
\newcommand{\om}{\omega}
\newcommand{\mU}{\mathcal{U}}

\newcommand{\tu}{\tilde{u}}

\newtheorem{lemma}{Lemma}[section]

\long\def\new#1\endnew{{\bf #1}}		
\long\def\del#1\enddel{}

\def\del{\partial}

\usepackage{xcolor}
\definecolor{oldmauve}{rgb}{0.4, 0.19, 0.28}
\definecolor{pansypurple}{rgb}{0.47, 0.09, 0.29}
\definecolor{burgundy}{rgb}{0.5, 0.0, 0.13}
\definecolor{carminepink}{rgb}{0.92, 0.3, 0.26}
\definecolor{blue(pigment)}{rgb}{0.2, 0.2, 0.6}
\definecolor{darkseagreen}{rgb}{0.56, 0.74, 0.56}
\definecolor{darkspringgreen}{rgb}{0.09, 0.45, 0.27}
\definecolor{ceruleanblue}{rgb}{0.16, 0.32, 0.75}

\newcommand{\p}{\partial}

\newcommand{\be}{\begin{eqnarray}}
\newcommand{\en}{\end{eqnarray}}

\def\bz{{\bar z}}

\author{}
\numberwithin{equation}{section} 

\begin{document}

\begin{titlepage}

  \thispagestyle{empty}

 \begin{flushright}
 \end{flushright}


  \begin{center}  
{\LARGE\textbf{Carrollian Amplitudes and Celestial Symmetries}}

\vskip1cm
Lionel Mason\footnote{\fontsize{8pt}{10pt}\selectfont\ \href{mailto:Lionel.Mason@maths.ox.ac.uk}{lionel.mason@maths.ox.ac.uk}}, Romain Ruzziconi\footnote{\fontsize{8pt}{10pt}\selectfont\ \href{mailto:Romain.Ruzziconi@maths.ox.ac.uk}{romain.ruzziconi@maths.ox.ac.uk}}, 
 Akshay Yelleshpur Srikant\footnote{\fontsize{8pt}{10pt}\selectfont \ \href{mailto:Akshay.YelleshpurSrikant@maths.ox.ac.uk}{akshay.yelleshpur@maths.ox.ac.uk}}
\vskip0.5cm

\normalsize
\medskip

\textit{Mathematical Institute, University of Oxford, \\ Andrew Wiles Building, Radcliffe Observatory Quarter, \\
Woodstock Road, Oxford, OX2 6GG, UK}

\end{center}

\vskip0.5cm

\begin{abstract}
Carrollian holography aims to express gravity in four-dimensional asymptotically flat spacetime in terms of a dual three-dimensional Carrollian CFT living at null infinity. Carrollian amplitudes are massless scattering amplitudes written in terms of asymptotic or null data at $\scri$. These position space amplitudes at $\mathscr I$ are to be re-interpreted as correlation functions in the putative dual Carrollian CFT. We derive basic results concerning tree-level Carrollian amplitudes yielding dynamical constraints on the holographic dual. We obtain surprisingly compact expressions for $n$-point MHV gluon and graviton amplitudes in position space at $\scri$. We discuss the UV/IR behaviours of Carrollian amplitudes and investigate their collinear limit, which allows us to define a notion of Carrollian OPE. By smearing the OPE along the generators of null infinity, we obtain the action of the celestial symmetries -- namely, the $S$ algebra for Yang-Mills theory and $Lw_{1+\infty}$ for gravity -- on the Carrollian operators. As a consistency check, we systematically relate our results with celestial amplitudes using the link between the two approaches. Finally, we initiate a direct connection between twistor space and Carrollian amplitudes. 

\end{abstract}

\end{titlepage}
\setcounter{page}{2}

\setcounter{tocdepth}{1}
\tableofcontents

\section{Introduction}

Formulating the holographic principle in asymptotically flat spacetime is an intriguing but difficult problem that has recently attracted much attention. Using bottom-up approaches and properties of the bulk theory, one can deduce kinematical and dynamical constraints on putative dual theories. Ultimately, the hope is that these constraints will be sufficiently restrictive to identify explicit candidates for dual theories. In this endeavor, two main directions of investigation have been explored. The first approach, coined as Carrollian holography \cite{Arcioni:2003xx,Arcioni:2003td,Dappiaggi:2005ci,Barnich:2006av,Barnich:2010eb,Bagchi:2010zz,Barnich:2012xq,Barnich:2012rz,Bagchi:2012xr,Bagchi:2014iea,Bagchi:2015wna,Bagchi:2016bcd,Ciambelli:2018wre,Donnay:2022aba,Bagchi:2022emh,Donnay:2022wvx}, suggests that gravity in four-dimensional (4D) asymptotically flat spacetime is dual to a 3D Carrollian CFT (also called conformal Carrollian field theory, or just BMS field theory) living at null infinity $\mathscr{I}$. Carrollian CFTs are field theories exhibiting BMS symmetries as spacetime symmetries, and can typically be obtained from standard relativistic CFTs by taking the speed of light to zero, i.e. $c \to 0$ \cite{Leblond}. In the second approach, called celestial holography \cite{deBoer:2003vf,He:2015zea,Pasterski:2016qvg,Cheung:2016iub,Pasterski:2017kqt,Strominger:2017zoo,Pasterski:2017ylz} (see also \cite{Pasterski:2021rjz,Raclariu:2021zjz,McLoughlin:2022ljp,Donnay:2023mrd} for reviews), the putative dual theory is a 2D CFT living on the celestial sphere, which is referred to as the celestial CFT (or CCFT for short). Although these two roads seem a priori disconnected, it has been shown in \cite{Donnay:2022aba,Bagchi:2022emh,Donnay:2022wvx} that they are actually related, offering a beautiful and non-trivial interplay between Carrollian physics and celestial amplitudes to tackle the problem of flat space holography. Interestingly, twistor theory is intertwined with both these approaches: twistor spaces are naturally constructed at null infinity \cite{Ko:1977gw,Hansen:1978jz} in such a way that Carrollian data can be re-expressed as twistor data \cite{Eastwood:1982,Mason:1986} providing top-down approaches to both Carrollian holography \cite{Adamo:2014yya,Geyer:2014lca, Adamo:2020yzi,Adamo:2021bej,Adamo:2021lrv,Adamo:2022mev} and celestial holography \cite{Adamo:2019ipt, Costello:2022jpg,Costello:2022upu,Mason:2022hly}.

Carrollian and celestial holography have been successfully applied in different frameworks. The celestial approach has been intimately tied to the $\mathcal{S}$-matrix since its inception: it involves rewriting scattering amplitudes in the boost eigenstate basis rather than in the usual energy eigenstate basis, hence highlighting the conformal properties of the amplitudes. For massless particles, this change of basis is implemented via a Mellin transform and the resulting amplitudes are referred to as celestial amplitudes. They are then naturally interpreted as correlation functions of a 2D CFT on the celestial sphere. The Lorentz symmetries and the subleading soft graviton theorem of scattering amplitudes provide the global and local 2D conformal symmetries of this theory \cite{Cachazo:2014fwa,Adamo:2014yya,Kapec:2014opa,Kapec:2016jld}. This dictionary has been successfully applied to extract CCFT correlation functions from scattering amplitudes. At tree-level, the functional form of the four point celestial amplitude involving massless particles of arbitrary helicity has been identified in \cite{Arkani-Hamed:2020gyp, Puhm:2019zbl, Pasterski:2017ylz}. The MHV and NMHV celestial gluon amplitudes have been derived in \cite{Schreiber:2017jsr} and the resulting functions are generalized hypergeometric functions. It is worth pointing out that even at tree-level, these functions are as complex as those obtained in the evaluation of multi-loop Feynman integrals. Consequently, loop-level computations of celestial amplitudes have largely been restricted to four points \cite{Banerjee:2017jeg, Gonzalez:2020tpi}, except for the rational one-loop amplitudes that have special configurations of helicities \cite{Albayrak:2020saa}. 

The celestial dictionary provides new insights into  structural statements about amplitudes using the  CFT framework. For example, the soft theorems for scattering amplitudes are rewritten as conformally soft theorems for celestial amplitudes \cite{Pasterski:2017kqt,Donnay:2018neh,Adamo:2019ipt,Puhm:2019zbl,Guevara:2019ypd}, which then have a natural interpretation as Ward identities constraining the 2D CCFT correlators. Moreover, the collinear limits of amplitudes yield the celestial OPEs \cite{Fan:2019emx, Pate:2019lpp}, from which symmetry algebras of soft operators can be deduced \cite{Guevara:2021abz,Strominger:2021mtt, Mago:2021wje, Ren:2022sws, Bhardwaj:2022anh, Jiang:2021ovh, Drozdov:2023qoy, Ahn:2021erj, Ahn:2022oor}. These celestial symmetry algebras can be alternatively derived from the classical phase space of gravity and Yang-Mills theory \cite{Freidel:2021ytz,Freidel:2023gue}, and they find a natural geometric interpretation in twistor space \cite{Adamo:2021lrv,Mason:2022hly,Bu:2022iak,Bittleston:2023bzp} where they are realized as local symmetries. 

Analogously, one could attempt to reformulate the $\mathcal{S}$-matrix in a Carrollian language. In fact, as advocated in \cite{Donnay:2022aba,Bagchi:2022emh,Donnay:2022wvx}, using a Fourier transform, scattering amplitudes can be written in position space and then naturally interpreted as correlation functions of Carrollian operators at null infinity. For this reason, we will refer to these position space amplitudes at $\mathscr I$ as Carrollian amplitudes. However, by contrast with the celestial case, few results exist in that regard: the two-point Carrollian amplitude has been computed in \cite{Donnay:2022wvx,Liu:2022mne}, while the three-point amplitude has been derived in split signature in \cite{Salzer:2023jqv} using an embedding space formalism, and in Lorentzian signature in \cite{Nguyen:2023miw} by pushing bulk correlators of massless fields to infinity. These low-point amplitudes are completely determined by the symmetries and do not impose any dynamical constraints on the putative dual theory. Four-point position space amplitudes have been discussed in \cite{Banerjee:2019prz} using a modified Mellin transform.

In this paper, pursuing the analysis initiated in \cite{Donnay:2022aba,Bagchi:2022emh,Donnay:2022wvx}, we derive some fundamental results concerning Carrollian amplitudes, which are subsequently interpreted in the language of Carrollian CFT. More specifically, we focus on tree-level amplitudes in Yang-Mills and Einstein gravity and compute their Carrollian counterparts, offering strong dynamical constraints on the putative Carrollian CFT at null infinity. We provide an expression for $n$-point MHV amplitudes of arbitrary multiplicities for generic kinematics. Surprisingly, we find that these Carrollian amplitudes involve only simple rational functions and exponentials of the kinematics in contrast to the transcendental behaviour of their celestial counterparts. Moreover, we show that the collinear limit of amplitudes yields a notion of Carrollian OPE in the putative dual theory at null infinity. We check that this definition of OPE is meaningful, which requires much care due to the ultra-local nature of Carrollian CFTs. Smearing the Carrollian OPEs on the generators of null infinity allows to obtain the action of the celestial symmetry algebras ($Lw_{1+\infty}$ and $S$-algebra for gravity and Yang-Mills, respectively) in the Carrollian CFT. At each step of the presentation, we relate our Carrollian results with their celestial counterparts using the dictionary established in \cite{Donnay:2022aba,Donnay:2022wvx}, and comment on the similarities and differences between the two approaches. Finally, we provide a novel derivation of the Fourier correspondence between asymptotic data and momentum space data that manifests the full Lorentz symmetries using homogeneous coordinates on $\scri$ and on-shell momentum space. In this framework, conformal invariance can be checked simply by ensuring that weights balance.  We go on to use this framework to relate twistor wave functions with the asymptotic data at null infinity, offering a direct connection between twistor space and Carrollian CFT in split signature.

The paper is organized as follows. In Section \ref{sec:Elements of Carrollian CFT}, we present some elements of Carrollian CFT that are necessary for the Carrollian interpretation of position space amplitudes. In section \ref{sec:Carrollian and celestial amplitudes}, we explain how to compute Carrollian amplitudes from the usual scattering amplitudes in momentum space. We also review the link with the celestial amplitudes, and relate the integral transforms considered in \cite{Donnay:2022aba,Donnay:2022wvx} with the modified Mellin transform discussed in \cite{Bagchi:2022emh}. In Sections \ref{sec:Two-point amplitude}, \ref{sec:Three-point amplitude} and \ref{sec:Four-point amplitude}, we respectively derive the two-, three- and four-point Carrollian amplitudes and connect with previous literature. In Section \ref{sec:MHV Carrollian amplitudes}, we extend the previous results to $n$-point tree-level Carrollian amplitudes in the MHV sector. We comment on the simplicity of the Carrollian expressions compared to their celestial counterparts. In Section \ref{sec:Soft limits and memories}, we comment on the UV/IR behaviours of Carrollian amplitudes in a similar spirit than the discussion presented in \cite{Arkani-Hamed:2020gyp} for celestial amplitudes. In Section \ref{sec:Collinear limits and Carrollian OPEs}, we express the collinear limits of amplitudes in a Carrollian language and find that it leads to the definition of Carrollian OPE. In Section \ref{sec:Celestial symmetries}, using the Carrollian OPEs, we explain how celestial symmetry algebras act on the Carrollian operators at $\mathscr{I}$. Section \ref{sec:twistors} gives a new proof of the relationship between momentum space and Carrollian wavefunctions and connects Carrollian CFT with twistor space wave functions in split signature. Finally, in section \ref{sec:Discussion}, we discuss some implication of our work for future endeavours.

\section{Elements of Carrollian CFT}
\label{sec:Elements of Carrollian CFT}

In Carrollian holography, the putative dual theory lives at null infinity\footnote{As argued in \cite{Donnay:2022aba,Donnay:2022wvx}, the dual Carrollian CFT lives on both $\mathscr{I}^+$ and $\mathscr{I}^-$ in Lorentzian signature. For simplicity, we will only consider the component $\mathscr{I}^+$ in the presentation, but we refer to these references for the detailed construction.} $\mathscr{I} \simeq \mathbb{R} \times \mathscr{S}$, where $\mathscr{S}$ is a two-dimensional surface, which will be the celestial Riemann sphere in Lorentzian signature ($\mathscr{S} \simeq S^2$) or the celestial Lorentzian torus in split signature ($\mathscr{S} \simeq \mathcal{LT}_2 = S^1 \times S^1 /\mathbb{Z}_2$). We denote $x^a = (u,z,\bar{z})$ the coordinates at $\mathscr{I}$, where $u$ is a null or retarded time coordinate, and $(z,\bar{z})$ are coordinates on $\mathscr{S}$ that are complex conjugates in Lorentzian signature, and real in split signature. From a geometric perspective, the BMS symmetries are the symmetries of $\scri$ endowed with its null angle structure or strong conformal structure \cite{Penrose:1962ij,Penrose:1964ge,Newman:1966ub, Penrose:1986uia}. This has more recently been recast in terms of a Carrollian structure at $\mathscr{I}$ \cite{Duval:2014uva,Duval:2014lpa}.  Both formulations include  degenerate metrics $ds^2=q_{ab}dx^a dx^b = 0 du^2 + 2 dz d\bar{z}$ defined up to conformal rescalings $q_{ab} \sim \omega^2 q_{ab}$ by  some function $\omega(z, \bar z)>0$.  The null angles are a specification of the ratio $du:ds$ whereas the Carrollian structure introduces the vector field $n^a \partial_a = \partial_u$ in the kernel of the metric, $q_{ab} n^b = 0$ subject to the conformal rescaling law $n^a \sim \omega^{-1} n^a$. This freedom comes from the ambiguity to choose the finite part of the conformal factor in the conformal compactification of the bulk spacetime \cite{Penrose:1986uia}.\footnote{In this paper, we work with the flat metric on $\mathscr{S}$, which can be obtained in Lorentzian signature at the expanse of sending a point to infinity (often the round metric is used but this leads to more complicated formulae).} One often talks about conformal Carrollian structure, or universal structure \cite{1977asst.conf....1G,Ashtekar:2014zsa}. Hence, the BMS symmetries, or equivalently, the conformal Carrollian symmetries, are generated by vector fields $\xi = \xi^a \partial_a$ at $\mathscr{I}$ preserving the Carrollian structure, up to some scaling $\alpha (z, \bar z)$,
\begin{equation}
    \mathcal{L}_{\xi} q_{ab} = 2\alpha q_{ab}, \qquad  \mathcal{L}_{\xi} n^a = -\alpha n^a  .
    \label{conformal Carroll symmetries}
\end{equation} The solution $\xi$ of \eqref{conformal Carroll symmetries} is given explicitly by
\begin{equation}
    \xi = (\mathcal{T} + {u} \alpha ) \partial_u  + \mathcal{Y}\partial_z + \bar{\mathcal{Y}}\partial_{\bar{z}}, \qquad \alpha = \frac{1}{2}(\partial_z \mathcal{Y} + \partial_{\bar{z}} \bar{\mathcal{Y}} )  
    \label{BMS vectors}
\end{equation} where $\mathcal{T}(z, \bar z)$ is the supertranslation parameter and $(\mathcal{Y}(z), \bar{\mathcal{Y}}(\bar{z}))$ are the superrotation parameters satisfying the conformal Killing equation in two dimensions, i.e. $\partial_z \bar{\mathcal{Y}}= 0 = \partial_{\bar z} \mathcal{Y}$. The vector fields \eqref{BMS vectors} form an algebra called the (extended) BMS \cite{Bondi:1962px,Sachs:1962wk,Sachs:1962zza,Barnich:2009se} or conformal Carrollian algebra \cite{Duval:2014uva,Duval:2014lpa} (we will now use these two terminologies interchangeably). 

The conformal symmetries in two dimensions form an infinite-dimensional algebra, out of which one can define the global conformal subalgebra $SL(2,\mathbb C)$, which is finite dimensional. Similarly, the conformal Carrollian symmetries in three dimensions form an infinite-dimensional algebra, out of which one can define the global conformal Carrollian algebra by taking
\begin{equation}
    \mathcal{T}(z, \bar z) = 1, z, \bar{z}, z\bar{z}, \qquad \mathcal{Y} (z) = 1, z, z^2, \qquad \bar{\mathcal{Y}}(\bar{z}) = 1, \bar{z}, \bar{z}^2
    \label{Poincaré}
\end{equation} for the parameters in \eqref{BMS vectors}. The conformal Carrollian algebra in three dimensions is isomorphic to the Poincaré algebra in four dimensions, the latter being symmetry of the vacuum in the bulk theory (we refer e.g. to Appendix B of \cite{Donnay:2022wvx} for the explicit isomorphism). 

In Carrollian holography, the putative dual theory is a \textit{Carrollian CFT}, which is defined as a field theory exhibiting conformal Carrollian symmetries as spacetime symmetries. As we will see, while the symmetries \eqref{BMS vectors} might look reminiscent of those of a 2D CFT, the additional null time coordinate $u$ has some profound implications. Examples of Carrollian CFTs have been broadly discussed in the literature and are often obtained by taking the Carrollian limit of CFTs, which corresponds to sending the speed of light to zero, i.e. $c \to 0$. The simplest example is probably the 2D conformal Carrollian scalar field \cite{Barnich:2012rz,Bagchi:2022eav}, or its conformally coupled analogue in higher dimension \cite{Gupta:2020dtl,Rivera-Betancour:2022lkc,Baiguera:2022lsw}. Other examples of Carrollian field theories include e.g. Carrollian electrodynamics and Yang-Mills \cite{Duval:2014uoa,Bagchi:2019xfx,Henneaux:2021yzg,Chen:2021xkw,deBoer:2021jej,Chen:2023pqf,Nguyen:2023vfz}, Carrollian gravity \cite{Grumiller:2020elf,Gomis:2020wxp,Henneaux:2021yzg,Hansen:2021fxi,Perez:2021abf,Fuentealba:2022gdx,Campoleoni:2022ebj,deBoer:2023fnj}, Carrollian fermions \cite{Hao:2022xhq,Bagchi:2022eui,Banerjee:2022ocj,Yu:2022bcp}, SUSY extensions \cite{Barducci:2018thr,Bagchi:2022owq}, and fractonic realizations \cite{Bidussi:2021nmp,Figueroa-OFarrill:2023vbj,Perez:2023uwt}. There exist two families of Carrollian CFTs, one called electric and the other magnetic, which essentially depend on the choice of scalings of the fields with respect to the speed of light in the limit. A systematic way to build Carrollian CFT actions without resorting to a limit process is through the BMS geometric action construction presented in \cite{Barnich:2017jgw,Barnich:2022bni}. This construction has already been shown to be useful in the holographic description of gravity in 3D \cite{Merbis:2019wgk} and 4D non-radiative spacetimes \cite{Barnich:2021dta,Barnich:2022bni} in terms of a magnetic-type of boundary Carrollian CFT. In this paper, we study the complementary sector, namely the one associated with the radiation at null infinity which is relevant for massless scattering processes. This sector will be instead described by an electric-type of Carrollian CFT.

In 2D CFT, one defines a notion of (quasi-)conformal primary field of conformal weights $(h,\bar{h})$ as a field $\mathcal{O}_{(h,\bar{h})}(z,\bar z)$ which transforms as
\begin{equation}
    \delta_{\mathcal{Y}, \bar{\mathcal{Y}}} \mathcal{O}_{(h,\bar{h})} =  [ \mathcal{Y} \partial_z + \bar{\mathcal{Y}} \partial_{\bar z} + h \partial_z \mathcal{Y} + \bar{h}  \partial_{\bar z} \bar{\mathcal{Y}}]\mathcal{O}_{(h,\bar{h})}
    \label{conformal primaries}
\end{equation} under (global) conformal transformations. This encodes the statement that the operator $\mathcal{O}_{(h,\bar h)}$ takes values in $(\Omega_{\mathscr{S}}^{1,0})^h\otimes (\Omega_{\mathscr{S}}^{0,1})^{\bar h} $ so that it should be thought of as coming with a factor of $dz^h \otimes d\bar z^{\bar h}$, as a tensor on the Riemann surface $\mathscr{S}$.
Correlators in a 2D CFT obey the global conformal Ward identities
\begin{equation}
    \sum_{i=0}^n [ \mathcal{Y}(z_i) \partial_{z_i} + \bar{\mathcal{Y}}(\bar{z}_i) {\partial}_{\bar{z}_i} + h_i \partial_{z_i} \mathcal{Y}(z_i) + \bar{h}_i  \partial_{\bar{z}_i} \bar{\mathcal{Y}} (\bar z_i)] \langle \mathcal{O}_{(h_1,\bar{h}_1)}(z_1, \bar{z}_1) \ldots \mathcal{O}_{(h_n,\bar{h}_n)}(z_n, \bar{z}_n) \rangle = 0\, ,
    \label{CFT WI}
\end{equation} with $(\mathcal{Y}(z_i), \bar{\mathcal{Y}}(\bar z_i))$ taken as in \eqref{Poincaré}. 

In 3D Carrollian CFT, the variable $u$ takes values in $(\Omega_{\mathscr{S}}^{1,0})^{-\frac12} \otimes (\Omega_{\mathscr{S}}^{0,1})^{-\frac12} $ on $\mathscr{S}$ so that  one defines a notion of (quasi-)conformal Carrollian primary of weights $(k,\bar k)$ as a field $\Phi_{(k,\bar k)} (u, z, \bar z)$ which transforms as follows under (global) conformal Carrollian symmetries \cite{Ciambelli:2018xat,Ciambelli:2018wre,Freidel:2021qpz,Donnay:2022aba,Donnay:2022wvx,Nguyen:2023vfz}:
\begin{equation}    \delta_{(\mathcal{T},\mathcal{Y}, \bar{\mathcal{Y}})} \Phi_{(k,\bar{k})} = \Big[\Big(\mathcal{T} + \frac{u}{2}(\partial_z \mathcal{Y} + \partial_{\bar{z}}\bar{\mathcal{Y}}) \Big)  \partial_u + \mathcal{Y} \partial_z + \bar{\mathcal{Y}} \partial_{\bar z} + k \partial_z \mathcal{Y} + \bar{k}  \partial_{\bar z} \bar{\mathcal{Y}} \Big] \Phi_{(k,\bar{k})} \, .
\label{carrollian primary}
\end{equation} There are two types of descendants in a Carrollian CFT: the descendants with respect to $\partial_z$ and $\partial_{\bar z}$, which will very much look like descendants in a 2D CFT, and the descendants with respect to $\partial_u$. The latter have the nice property that, if $\Phi_{(k,\bar k)}$ is a conformal Carrollian primary, then $\partial_u^m \Phi_{(k,\bar k)}$ is also a conformal Carrollian primary of weights $(k+ \tfrac{m}{2}, \bar{k} + \tfrac{m}{2})$, for any positive integer $m$. The correlators in a Carrollian CFT, 
\begin{equation}
    \langle \Phi_{(k_1,\bar k_1)}(u_1, z_1, \bar{z}_1) \ldots \Phi_{(k_n,\bar k_n)}(u_n, z_n, \bar{z}_n) \rangle , 
\end{equation} obey the (global) conformal Carrollian Ward identities \cite{Chen:2021xkw,Donnay:2022wvx}
\begin{multline}
\label{Carrollian WI}
    \sum_{i=0}^n \Big[ \Big( \mathcal{T}(z_i, \bar{z}_i) + \frac{u_i}{2} (\partial_{z_i}\mathcal{Y}(z_i) + {\partial}_{\bar{z}_i} \bar{\mathcal{Y}}(\bar{z}_i ) )\Big)\partial_{u_i} + \mathcal{Y}(z_i) \partial_{z_i} + \bar{\mathcal{Y}}(\bar{z}_i) {\partial}_{\bar{z}_i}
    \\
     + k_i \partial_{z_i} \mathcal{Y}(z_i) + \bar{k}_i  \partial_{\bar{z}_i} \bar{\mathcal{Y}} (\bar z_i) \Big] \langle \Phi_{(k_1,\bar k_1)}(u_1, z_1, \bar{z}_1) \ldots \Phi_{(k_n,\bar k_n)}(u_n, z_n, \bar{z}_n) \rangle = 0 
\end{multline} where the parameters $(\mathcal{T}(z_i,\bar z_i), \mathcal{Y}(z_i), \bar{\mathcal{Y}}(\bar z_i))$ are taken as in \eqref{Poincaré}. As we will explain in the next section, the Poincaré invariance of bulk scattering amplitudes can be holographically recast as the global conformal Carrollian invariance \eqref{Carrollian WI}, provided one introduces the notion of Carrollian amplitudes.

\section{Carrollian and celestial amplitudes}
\label{sec:Carrollian and celestial amplitudes}

In this paper, we focus on scattering of massless particles (scalars, gluons or gravitons\footnote{For half-integral spin, some of the conventions below require some minor adjustments.}). We start with amplitudes in momentum space and use Mellin and Fourier transforms to obtain celestial and Carrollian amplitudes, respectively. 

We parametrize particle's momenta by 
\begin{equation}
    \label{eq:mompar31}
    p^\mu =  \om q^{\mu} = \e \om \left(1+z \zb, z + \zb, -i \left(z - \zb \right), 1 - z \zb \right)
\end{equation}
where $\e = \pm 1$ tells us if the particle is outgoing ($+1$) or incoming ($-1$), $\omega >0$ is the energy and $(z,\bar z)$ are stereographic  coordinates on the celestial sphere parametrizing the direction of the null momentum. We will refer to $\mathcal{S}$-matrix elements in momentum space as scattering amplitudes and denote them by
\begin{equation}
    \mA_n \left(\left\lbrace \om_1, z_1, \zb_1\right\rbrace_{J_1}^{\epsilon_1}, \dots ,\left\lbrace \om_n, z_n, \zb_n\right\rbrace_{J_n}^{\epsilon_n}\right)
\label{scattering amplitude}    
\end{equation} where $n$ denotes the total number of particles, and $J_i$ the particle helicities. \textit{Celestial amplitudes} are obtained from momentum space scattering amplitudes \eqref{scattering amplitude} by performing Mellin transforms \cite{deBoer:2003vf,He:2015zea,Pasterski:2016qvg,Cheung:2016iub,Pasterski:2017kqt,Strominger:2017zoo,Pasterski:2017ylz} , 
\begin{multline}
\label{celestial amplitudes}
    \mathcal{M}_n\left(\left\lbrace \Delta_1, z_1, \zb_1\right\rbrace_{J_1}^{\epsilon_1}, \dots , \left\lbrace \Delta_n, z_n, \zb_n\right\rbrace_{J_n}^{\epsilon_n}\right) \\
    = \prod^n_{i=1} \left( \int_0^{+\infty}  d\omega_i \, \omega_i^{\Delta_i-1} \right) \mA_n \left(\left\lbrace \om_1, z_1, \zb_1\right\rbrace_{J_1}^{\epsilon_1} , \dots , \left\lbrace \om_n, z_n, \zb_n\right\rbrace_{J_n}^{\epsilon_n}\right)
\end{multline} where $\Delta_i$ denotes the eigenvalue with respect to the Lorentz boost generator along the direction fixed by the null momentum $p^\mu_i$. These amplitudes being expressed in a boost eigenstates basis, they exhibit nice transformation properties under conformal transformations on the celestial sphere. In particular, they satisfy the 2D CFT Ward identities \eqref{CFT WI} for $h_i=\frac{\Delta_i + J_i}{2}$ and $\bar{h}_i = \frac{\Delta_i - J_i}{2}$, which allows us to interpret \eqref{celestial amplitudes} as a 2D CFT correlation function of operators \eqref{conformal primaries} inserted on the celestial sphere
\begin{equation}
\label{celestial dictionary}
   \mathcal{M}_n\left(\left\lbrace \Delta_1, z_1, \zb_1\right\rbrace_{J_1}^{\epsilon_1}, \dots , \left\lbrace \Delta_n, z_n, \zb_n\right\rbrace_{J_n}^{\epsilon_n}\right) \equiv  \langle \mathcal{O}^{\epsilon_1}_{\Delta_1,J_1} (z_1, \bz_1)  \ldots \mathcal{O}^{\epsilon_n}_{\Delta_n, J_n} (z_n, \bz_n)  \rangle \, .
\end{equation} This identification constitutes the key ingredient of the celestial holography dictionary.

Analogously, as discussed in \cite{Ashtekar:1981hw, Ashtekar:1981sf,Donnay:2022wvx}, see also the proof  in Section \ref{Fourier-scri}, momentum space scattering amplitudes \eqref{scattering amplitude} are transformed to position space amplitudes at $\scri$ by Fourier transforms
\begin{multline}
\label{eq:AtoC}
    \mc_n\left(\left\lbrace u_1, z_1, \zb_1\right\rbrace_{J_1}^{\epsilon_1}, \dots , \left\lbrace u_n, z_n, \zb_n\right\rbrace_{J_n}^{\epsilon_n}\right) \\
    = \prod_{i=1}^n  \left(\int_0^{+\infty}  \frac{d\om_i}{2\pi} \, e^{i\e_i \om_i u_i} \right) \mA_n \left(\left\lbrace \om_1, z_1, \zb_1\right\rbrace_{J_1}^{\epsilon_1}, \dots , \left\lbrace \om_n, z_n, \zb_n\right\rbrace_{J_n}^{\epsilon_n}\right).
\end{multline} 
For standard momentum space wave-functions, i.e., as defined by \eqref{Fourier}, then by \eqref{third-} and \eqref{third+}, the corresponding wave functions at $\scri$ will be a potential for the leading radiation field at $\scri$ \cite{Newman:1961qr,Penrose:1986uia}: it follows from Lemma \ref{Fourier-scri-lemma} that to obtain the radiation field from the Carrollian wave function we must apply $\p_u^{|\epsilon J|}$. Thus  for gravity this transform gives the asymptotic shear $\bar\sigma=C_{zz }$ which is related to the leading radiation field by $\Psi^0_4=\ddot{\bar\sigma}^0$ and for Maxwell it gives the asymptotic potential $a_{z}$ whose $\p_u$ derivative gives the Maxwell radiation field. We will refer to position space amplitudes at $\mathscr I$ \eqref{eq:AtoC} as \textit{Carrollian amplitudes}, as by construction, they satisfy\footnote{As discussed in Section \ref{sec:Elements of Carrollian CFT}, the global conformal Carrollian algebra is isomorphic to the Poincaré algebra and the statement \eqref{Carrollian WI amplitudes} is just a holographic reformulation of the Poincaré invariance of the bulk amplitudes. We refer to \cite{Donnay:2022wvx} for a detailed derivation of \eqref{Carrollian WI amplitudes} for amplitudes in position space at $\scri$.} 
\begin{multline}
\label{Carrollian WI amplitudes}
    \sum_{i=0}^n \Big[ \Big( \mathcal{T}(z_i, \bar{z}_i) + \frac{u}{2} (\partial_{z_i}\mathcal{Y}(z_i) + {\partial}_{\bar{z}_i} \bar{\mathcal{Y}}(\bar{z}_i )) \Big)\partial_{u_i} 
    + \mathcal{Y}(z_i) \partial_{z_i} + \bar{\mathcal{Y}}(\bar{z}_i) {\partial}_{\bar{z}_i} \\ + \frac{1+\epsilon_i J_i}{2} \partial_{z_i} \mathcal{Y}(z_i) + \frac{1- \epsilon_i J_i}{2}  \partial_{\bar z_i} \bar{\mathcal{Y}}(\bar z_i) \Big]  \mc_n\left(\left\lbrace u_1, z_1, \zb_1\right\rbrace_{J_1}^{\epsilon_1}, \dots , \left\lbrace u_n, z_n, \zb_n\right\rbrace_{J_n}^{\epsilon_n}\right) = 0\, ,
\end{multline} which corresponds to the global conformal Carrollian Ward identities at null infinity \eqref{Carrollian WI} after setting
\begin{equation}
    k_i = \frac{1+\epsilon_i J_i}{2}, \qquad \bar k_i = \frac{1- \epsilon_i J_i}{2} \, .
    \label{fiexed Carrollian weights}
\end{equation}
These symmetry properties follow directly from the manifestly conformally invariant formulation of the Fourier transform \eqref{eq:AtoC} discussed in Section \ref{Fourier-scri}.

Thus, by analogy with the celestial case, Carrollian amplitudes \eqref{eq:AtoC} will be interpreted as Carrollian CFT correlators of operators \eqref{carrollian primary} inserted at null infinity \cite{Donnay:2022aba,Donnay:2022wvx}
\begin{equation}
\label{carrollian identification}
    \mc_n\left(\left\lbrace u_1, z_1, \zb_1\right\rbrace_{J_1}^{\epsilon_1}, \dots , \left\lbrace u_n, z_n, \zb_n\right\rbrace_{J_n}^{\epsilon_n}\right) \equiv \langle \Phi^{\epsilon_1}_{(k_1,\bar{k}_1)} (u_1, z_1, \bz_1) \ldots \Phi^{\epsilon_n}_{(k_n,\bar{k}_n)} (u_n, z_n, \bz_n)  \rangle 
\end{equation} 
where the Carrollian weights $(k_i, \bar k_i)$ are fixed in terms of the particle helicites through \eqref{fiexed Carrollian weights}. In the relation \eqref{carrollian identification}, the Carrollian CFT operators are identified with the boundary values of the bulk operators obtained by using the stationary phase approximation in the limit $r\to \infty$ \cite{Donnay:2022wvx}. Moreover, as shown in \cite{Nguyen:2023vfz}, the transformations \eqref{carrollian primary} with weights \eqref{fiexed Carrollian weights} precisely match with the unitary representations of the global conformal Carrollian algebra induced at null infinity. Hence, inspired by the AdS/CFT terminology, the identification \eqref{carrollian identification} can be seen as the \textit{extrapolate dictionary for Carrollian holography}.

As explained below \eqref{carrollian primary}, $\partial_u$-descendants of conformal Carrollian primaries are also primaries. Therefore, the following expression
\begin{multline}
\label{descendants amplitudes}
 \mathcal{C}_n^{m_1\ldots m_n} \left(\left\lbrace u_1, z_1, \zb_1\right\rbrace_{J_1}^{\epsilon_1}, \dots , \left\lbrace u_n, z_n, \zb_n\right\rbrace_{J_n}^{\epsilon_n}\right) 
   = \partial_{u_1}^{m_1} \ldots \partial_{u_n}^{m_n} \mc_n\left(\left\lbrace u_1, z_1, \zb_1\right\rbrace_{J_1}^{\epsilon_1}, \dots , \left\lbrace u_n, z_n, \zb_n\right\rbrace_{J_n}^{\epsilon_n}\right) \\
   =  \prod_{i=1}^n  \left(\int_0^{+\infty}  \frac{d\om_i}{2\pi} \, (i \epsilon \omega_i)^{m_i} e^{i\e_i \om_i u_i} \right) \mA_n \left(\left\lbrace \om_1, z_1, \zb_1\right\rbrace_{J_1}^{\epsilon_1}, \dots , \left\lbrace \om_n, z_n, \zb_n\right\rbrace_{J_n}^{\epsilon_n}\right) \\
   = \langle \partial_{u_1}^{m_1}\Phi^{\epsilon_1}_{(k_1,\bar{k}_1)} (u_1, z_1, \bz_1)  \ldots \partial_{u_n}^{m_n}\Phi^{\epsilon_n}_{(k_n,\bar{k}_n)} (u_n, z_n, \bz_n)  \rangle 
\end{multline} also satisfies the global conformal Carrollian Ward identities \eqref{Carrollian WI}, but with shifted wights $k_i = \tfrac{1+ m_i + \epsilon_i J_i }{2}$ and $\bar{k}_i = \tfrac{1+ m_i - \epsilon_i J_i }{2}$. We will show in Section \ref{sec:Collinear limits and Carrollian OPEs} that the Carrollian OPEs presented in Equations \eqref{eq:carrollOPEformula1}, \eqref{eq:carrollOPEformula2}, \eqref{eq:carrollOPEformula3} and \eqref{eq:OPEnews} require the inclusion of all the $\partial_u$-descendants in the theory. Besides the particular case $\mathcal{C}_n \equiv \mathcal{C}_n^{0 \ldots 0}$ defined in \eqref{eq:AtoC}, it will also be interesting for holographic purposes to consider the case $\tilde{\mathcal{C}}_n \equiv \mathcal{C}_n^{1 \ldots 1}$ given explicitly by
 \begin{multline}
 \label{eq:AtoCnews}
\tilde{\mc}_n\left(\left\lbrace u_1, z_1, \zb_1\right\rbrace_{J_1}^{\epsilon_1}, \dots , \left\lbrace u_n, z_n, \zb_n\right\rbrace_{J_n}^{\epsilon_n}\right) 
 := \partial_{u_1} \ldots \partial_{u_n} \mc_n\left(\left\lbrace u_1, z_1, \zb_1\right\rbrace_{J_1}^{\epsilon_1}, \dots , \left\lbrace u_n, z_n, \zb_n\right\rbrace_{J_n}^{\epsilon_n}\right) \\
 = \prod_{i=1}^n \left( \int \frac{d\om_i}{2\pi} \, i \epsilon_i \om_i\, e^{i\e_i \om_i u_i} \right) \mA_n \left(\left\lbrace \om_1, z_1, \zb_1\right\rbrace_{J_1}^{\epsilon_1}, \dots , \left\lbrace \om_n, z_n, \zb_n\right\rbrace_{J_n}^{\epsilon_n}\right)
 \end{multline} which, in practice, exhibits nicer properties than \eqref{eq:AtoC}: it is free from divergences and logarithms \cite{Donnay:2022wvx}, see e.g. Section \ref{sec:Two-point amplitude}. 
 Another reason why the correlator of first descendants $\partial_{u}\Phi^{\epsilon}_{(k,\bar{k})}$ \eqref{eq:AtoCnews} is better behaved than \eqref{eq:AtoC} can be seen in the case of both Maxwell theory and gravity. For Maxwell $\Phi^{\epsilon}_{(k,\bar{k})}$ is identified with the asymptotic potential which is subject to residual gauge freedom, memory and so on. Upon taking $\p_u$ we obtain the radiation field which is gauge invariant and not subject to memory. Analogously, for gravity, $\Phi^{\epsilon}_{(k,\bar{k})}$ is identified with the asymptotic shear $C_{zz}$, which transforms inhomogeneously under supertranslations (i.e. as a quasi-conformal Carrollian primary in the terminology of \eqref{carrollian primary}), while the Bondi news $N_{zz} = \partial_u C_{zz}$ transforms homogeneously under supertranslations and is free from displacement memory. Furthermore, both the Maxwell radiation field and the Bondi news, are sufficient to classically reconstruct the bulk radiative field through the Kirchhoff-d'Adh\'emar formula given in \eqref{dAdhemar} following \cite{Penrose1980GoldenON,Penrose:1985bww} (see also \cite{Donnay:2022wvx} for a review of this formula in the current notation); in Section \ref{Fourier-scri} we derive its connection with the Fourier transform above in a Lorentz invariant framework.

 Notice that, in \eqref{descendants amplitudes}, we can analytically continue $m_i = \delta_i - 1$ to the complex plane ($\delta_i \in \mathbb{C}$), to obtain the integral transform
 \begin{equation}
    \prod_{i=1}^n  \left(\int_0^{+\infty}  \frac{d\om_i}{2\pi} \, (i \epsilon \omega_i)^{\delta_i - 1} e^{i\e_i \om_i u_i} \right) \mA_n \left(\left\lbrace \om_1, z_1, \zb_1\right\rbrace_{J_1}^{\epsilon_1}, \dots , \left\lbrace \om_n, z_n, \zb_n\right\rbrace_{J_n}^{\epsilon_n}\right) \, ,
    \label{modMellin}
 \end{equation} 
 which precisely corresponds to the modified Mellin transform originally introduced in \cite{Banerjee:2018gce, Banerjee:2018fgd} to regulate the Mellin transform of graviton amplitudes. It was later used in \cite{Bagchi:2022emh} to relate scattering amplitudes with Carrollian correlators. However, the value of $\delta_i$ was unspecified much like $\Delta_i$ in the original Mellin transform \eqref{celestial amplitudes}. This clarifies the role of this integral transform in the present framework of Carrollian amplitudes. In particular, from the above considerations, the expression \eqref{modMellin} transforms as a correlator of $n$ conformal Carrollian primary fields with weights $k_i= \tfrac{\delta_i + \epsilon_i J_i }{2}$ and $\bar{k}_i = \tfrac{\delta_i - \epsilon_i J_i }{2}$, in agreement with the above references.

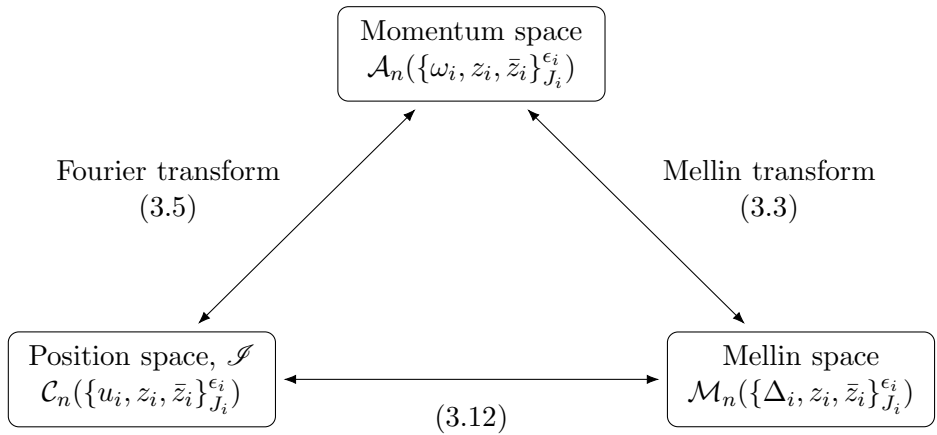
\begin{figure}[ht!]
    \centering
    \begin{tikzpicture}
        \tikzmath{\a = 2.5; \b = sqrt(3)*\a;} 
        \tikzstyle{nb} = [rectangle,draw,fill=white,rounded corners,outer sep=3pt,text width=3.2cm,align=center,inner sep=5pt];
        \coordinate (A) at (-\a, 0);
        \coordinate (B) at ( \a, 0);
        \coordinate (C) at ( 0, \b);
        \draw[white,opacity=0] ($(A)+(-4,-1.5)$) -- ($(B)+(+4,-1.5)$) -- ($(B)+(0,\b)+(+4,1)$) -- ($(A)+(0,\b)+(-4,1)$) -- cycle;
        \node[nb,left] (An) at (A) {Position space,  $\scri$  $\mathcal C_n (\lbrace u_i, z_i, \zb_i\rbrace_{J_i}^{\epsilon_i})$};
        \node[nb,right] (Bn) at (B) {Mellin space $\mathcal M_n (\lbrace \Delta_i, z_i, \zb_i\rbrace_{J_i}^{\epsilon_i})$};
        \node[nb] (Cn) at (C) {Momentum space $\mA_n (\lbrace \om_i, z_i, \zb_i\rbrace_{J_i}^{\epsilon_i})$};
        \draw[Latex-Latex] (An) -- (Bn);
        \draw[Latex-Latex] (Bn) -- (Cn);
        \draw[Latex-Latex] (An) -- (Cn);
        \node[below,align=center,text width=2cm] at ($(A)!0.5!(B)-(0,0.2)$) {\eqref{Btransform1}};
        \node[align=center,anchor=south east] at ($(An)!0.45!(Cn)$) {Fourier transform \\ \eqref{eq:AtoC}};
        \node[align=center,anchor=south west] at ($(Bn)!0.45!(Cn)$) {Mellin transform \\ \eqref{celestial amplitudes}};
    \end{tikzpicture}
    \caption{Interplay between the three bases of scattering in flat spacetime.}
    \label{fig:FBM}
\end{figure}

Finally, one can obtain Carrollian amplitudes \eqref{eq:AtoC} and \eqref{eq:AtoCnews} directly from the celestial amplitudes \eqref{celestial amplitudes}, which is one of the key ingredients in the Carrollian/celestial correspondence discussed in \cite{Donnay:2022aba,Donnay:2022wvx}. This is done by combining an inverse Mellin transform with a Fourier transform in \eqref{celestial amplitudes}, which yields explicitly \cite{Donnay:2022aba,Donnay:2022wvx,Fiorucci:2023lpb}:
\begin{multline}
 \label{Btransform1}
    \mathcal{M}_n\left(\left\lbrace \Delta_1, z_1, \zb_1\right\rbrace_{J_1}^{\epsilon_1}, \dots , \left\lbrace \Delta_n, z_n, \zb_n\right\rbrace_{J_n}^{\epsilon_n}\right) \\ 
    =     \prod_{i=1}^n \left(  ( -i \epsilon_i)^{\Delta_i}\Gamma[\Delta_i] \int_{-\infty}^{+\infty}  \frac{d u_i}{(u_i - i \epsilon_i \varepsilon)^{\Delta_i}} \right) \, \mc_n\left(\left\lbrace u_1, z_1, \zb_1\right\rbrace_{J_1}^{\epsilon_1}, \dots , \left\lbrace u_n, z_n, \zb_n\right\rbrace_{J_n}^{\epsilon_n}\right)
\end{multline} or
\begin{multline}
 \label{Btransfom2}  
   \mathcal{M}_n\left(\left\lbrace \Delta_1, z_1, \zb_1\right\rbrace_{J_1}^{\epsilon_1}, \dots , \left\lbrace \Delta_n, z_n, \zb_n\right\rbrace_{J_n}^{\epsilon_n}\right) \\ 
    =     \prod_{i=1}^n \left( \, (- i \epsilon_i)^{\Delta_i}\Gamma[\Delta_i-1] \int_{-\infty}^{+\infty} \frac{d u_i}{(u_i - i \epsilon_i \varepsilon)^{\Delta_i-1}} \right) \, \tilde{\mc}_n\left(\left\lbrace u_1, z_1, \zb_1\right\rbrace_{J_1}^{\epsilon_1}, \dots , \left\lbrace u_n, z_n, \zb_n\right\rbrace_{J_n}^{\epsilon_n}\right)
\end{multline} where we take the regulator $\varepsilon \to 0^+$. Using the dictionaries \eqref{celestial dictionary} and \eqref{carrollian identification}, the above integral transforms simply trade the time dependence of the Carrollian operators at $\mathscr I$ for the conformal dimension of the celestial CFT correlators. In particular, the Carrollian extrapolate dictionary discussed below \eqref{carrollian identification} is in perfect agreement with the celestial extrapolate dictionary introduced in \cite{Pasterski:2021dqe}. Notice that this recasting of the information from the 3D Carrollian CFT to the 2D celestial CFT is not anodyne: it has dramatic consequences on the 2D CFT structure, such as the appearance of distributional branches for the low-point correlation functions. In the next sections, we will illustrate these features and systematically check our results through the above integral formulae to close the triangle in Figure \ref{fig:FBM}.   

\section{Two-point amplitudes}
\label{sec:Two-point amplitude}
As a warm-up, we review the computation of the two-point Carrollian amplitude discussed in \cite{Donnay:2022wvx,Liu:2022mne}. We also comment on the advantage of working with Carrollian amplitudes of the form \eqref{eq:AtoCnews} instead of \eqref{eq:AtoC}. We will then derive the corresponding celestial amplitude by implementing the transform \eqref{Btransform1}. The 2-point tree-level scattering amplitude reads as
\begin{equation}
    \mathcal A_2(\{\omega_1,z_1,\bar z_1\}^{-}_{J_1},\{\omega_2,z_2,\bar z_2\}^{+}_{J_2}) = \kappa^2_{J_1,J_2}\,\pi\,\frac{\delta(\omega_1-\omega_2)}{\omega_1}\,\delta^{(2)}(z_1-z_2)\,\delta_{J_1,J_2},
\label{two point amplitude in fourier}
\end{equation} 
where we took the first and the second particle as incoming and outgoing, respectively. In this expression, $\kappa_{J_1,J_2}$ is the normalization that will depend on the particles involved.
\subsection{Carrollian amplitude}
Applying the successive integral transforms as in \eqref{eq:AtoC}, we find 
\begin{align}
        \mathcal C_2(\{u_1,z_1,\bar z_1\}_{J_1}^-,\{u_2,z_2,\bar z_2\}_{J_2}^+) &= \frac{1}{4\pi^2} \int_0^{+\infty}d\omega_1\int_0^{+\infty}d\omega_2\,e^{-i\omega_1 u_1}e^{i\omega_2 u_2}  \mathcal A_2(\{\omega_1,z_1,\bar z_1\}^{-}_{J_1},\{\omega_2,z_2,\bar z_2\}^{+}_{J_2}) \nonumber \\
        &= \frac{\kappa^2_{J_1,J_2}}{4\pi} \int_0^{+\infty}\frac{d\omega}{\omega}\,e^{-i\omega( u_1- u_2)}\,\delta^{(2)}(z_1-z_2)\,\delta_{J_1,J_2} \, .
     \label{W12 def}
\end{align} As discussed in \cite{Donnay:2022wvx,Liu:2022mne}, the integral in the last line
\begin{equation}
\label{divergent integral}
    \mathcal{I}_0 (u_1 - u_2)   =\int_0^{+\infty}\frac{d\omega}{\omega}\,e^{-i\omega(u_1-u_2)}
\end{equation} is divergent but can nevertheless be regulated as
\begin{equation}
    \begin{split}
        \mathcal I_\beta(x) &= \lim_{\varepsilon\to 0^+}\int_0^{+\infty} d\omega\,\omega^{\beta-1}\,e^{-i\omega x-\omega\varepsilon} = \lim_{\varepsilon\to 0^+}\frac{\Gamma[\beta](-i)^\beta}{(x-i\varepsilon)^\beta}.
    \end{split} \label{I beta}
\end{equation}
In the limit $\beta\to 0^+$, we obtain 
\begin{equation}
        \mathcal I_\beta(x) = \frac{1}{\beta}-\left[\gamma+\ln|x| + \frac{i\pi}{2}\text{sign}(x) \right] + \mathcal O(\beta), \label{curly I beta}
\end{equation}
where $\gamma$ is the Euler-Mascheroni constant. So \eqref{W12 def} yields the regularized 2-point Carrollian amplitude 
\begin{equation}
   \mathcal C_2(\{u_1,z_1,\bar z_1\}_{J_1}^-,\{u_2,z_2,\bar z_2\}_{J_2}^+) = \lim_{\beta \to 0}  \frac{\kappa^2_{J_1,J_2}}{4\pi}  \left[\frac{1}{\beta} -\left(\gamma + \ln|u_{12}| + \frac{i\pi}{2}\text{sign}(u_{12}) \right) \right]\delta^{(2)}(z_{12})\,\delta_{J_1,J_2}  
    \label{Wuup}
\end{equation} where $u_{ij} = u_i - u_j$ and $z_{ij} = z_i - z_j$. The $u$-independent regulated divergence $\sim \beta^{-1}$ and the logarithmic behaviour might be seen as undesirable features for a 2-point correlation function. However, they are essential to satisfy the Carrollian Ward identities, see Section 5.4 of \cite{Donnay:2022wvx}. A similar situation occurs for 2D CFT. For instance, consider a free scalar field in 2D: the correlation function of the scalar field with itself is logarithmic, and one would need to add a regulator to satisfy the conformal Ward identities. Usually, to circumvent these subtleties, one considers correlation functions of descendants of the scalar field, which exhibit standard 2D CFT correlation functions. Similarly, in a 3D Carrollian CFT, one can consider correlation functions of $\partial_u$-Carrollian descendants as in \eqref{descendants amplitudes}. Focusing on the first descendants \eqref{eq:AtoCnews}, we get the very simple expression 
\begin{equation}
   \tilde{\mathcal C}_2(\{u_1,z_1,\bar z_1\}_{J_1}^-,\{u_2,z_2,\bar z_2\}_{J_2}^+) = \lim_{\varepsilon\to 0^+}    \frac{\kappa^2_{J_1,J_2}}{4\pi}   \frac{1}{(u_{12}-i \varepsilon)^2}\delta^{(2)}(z_{12})\,\delta_{J_1,J_2} 
\label{2point News}
\end{equation} which matches with the usual electric branch of solutions of the Carrollian Ward identities. This already suggests that the putative dual theory governing the Carrollian amplitudes is an electric-type of Carrollian CFT.  

\subsection{Celestial from Carrollian}

Let us now check the consistency of the triangle in Figure \ref{fig:FBM}. The 2-point celestial amplitude has been computed in \cite{Pasterski:2017ylz} and is obtained by applying \eqref{celestial amplitudes} on \eqref{two point amplitude in fourier} \cite{Donnay:2022wvx}:
\begin{equation}
    \begin{split}
        &\mathcal M_2 (\{\Delta_1,z_1,\bar z_1\}_{J_1}^-,\{\Delta_2,z_2,\bar z_2\}_{J_2}^+)  \\
        &= \int_0^{+\infty}d\omega_1\,\omega_1^{\Delta_1-1}\int_0^{+\infty}d\omega_2\,\omega_2^{\Delta_2-1}\, \mathcal \mathcal A_2(\{\omega_1,z_1,\bar z_1\}^{-}_{J_1},\{\omega_2,z_2,\bar z_2\}^{+}_{J_2}) \\
        &= \kappa^2_{J_1,J_2}\,\pi\int_0^{+\infty}d\omega\,\omega^{\Delta_1+\Delta_2-3}\,\delta^{(2)}(z_1-z_2)\,\delta_{\alpha_1,\alpha_2} = 2\pi^2\,\kappa^2_{J_1,J_2}\,\delta(\nu_1+\nu_2)\,\delta^{(2)}(z_1-z_2)\,\delta_{J_1,J_2},
    \end{split}
\label{2pt celestial M2}
\end{equation}
where we assumed that the conformal dimension is on the principal series, $\Delta_i = 1+i\nu_i$ ($\nu_i \in \mathbb R$), and used
\begin{equation}
\int_0^{+\infty}d\omega\, \omega^{i\nu-1} = 2\pi\,\delta(\nu) \, . \label{delta represented in mellin}
\end{equation} Starting from \eqref{W12 def} (or alternatively from \eqref{2point News}), we can apply the integral transforms \eqref{Btransform1} (respectively \eqref{Btransfom2}). The first integral over $u_1$ gives
\begin{equation}
    \begin{split}
         &4\pi \, i^{\Delta_1+1}\Gamma[\Delta_1]\lim_{\varepsilon\to 0^+}\int_{-\infty}^{+\infty} \frac{d u_1}{(u_1+i\varepsilon)^{\Delta_1}}\mathcal I_0(u_1-u_2) \\
        &= 4\pi i \lim_{\varepsilon\to 0^+}\int_{-\infty}^{+\infty}d u_1\int_0^{+\infty}\omega^{\Delta_1-1}\,e^{i\omega u_1-\omega\varepsilon}\int_0^{+\infty}\frac{d\omega'}{\omega'}e^{-i\omega'(u_1-u_2)} \\
        &=8\pi^2 i \lim_{\varepsilon\to 0^+}\int_0^{+\infty} d\omega\,\omega^{\Delta_1-2}\,e^{i\omega u_2-\omega \varepsilon} = 8\pi^2 i \lim_{\varepsilon\to 0^+} \frac{i^{\Delta_1-1}\Gamma[\Delta_1-1]}{(u_2+i\varepsilon)^{\Delta_1-1}}.
    \end{split}
\end{equation} Now, the second integral yields
\begin{equation}
    \begin{split}
        &4\pi \, (-i)^{\Delta_1+1}\Gamma[\Delta_1]\lim_{\varepsilon\to 0^+}\int_{-\infty}^{+\infty} \frac{d u_1} {(u_1-i\varepsilon)^{\Delta_1}} \left(  8\pi^2 i \frac{i^{\Delta_1-1}\Gamma[\Delta_1-1]}{(u_2+i\varepsilon)^{\Delta_1-1}} \right) \\
        &= 32\pi^3 \lim_{\varepsilon\to 0^+} \int_{-\infty}^{+\infty}d u_2\, \frac{i^{\Delta_1-1}\Gamma[\Delta_1-1]}{(u_2+i\varepsilon)^{\Delta_1-1}} \frac{(-i)^{\Delta_2}\Gamma[\Delta_2]}{(u_2-i\varepsilon)^{\Delta_2}} = 128\pi^5 \delta(\nu_1+\nu_2),
    \end{split}
\end{equation} where in the last equality, we assumed $\Delta_i=1+i\nu_i$ and identified the integral representation of the delta function. Putting all together, we find the 2-point celestial amplitude
\begin{equation}
        \mathcal M_2 (\{\Delta_1,z_1,\bar z_1\}_{J_1}^-,\{\Delta_2,z_2,\bar z_2\}_{J_2}^+) = 2\pi^2\, \kappa^2_{J_1,J_2} \,\delta(\nu_1+\nu_2)\,\delta^{(2)}(z_{12})\,\delta_{J_1,J_2},
\end{equation}
which reproduces \eqref{2pt celestial M2} and closes the triangle.

\section{Three-point amplitudes}
\label{sec:Three-point amplitude}

Three point amplitudes involving massless particles vanish in a $(3,1)$ signature spacetime. Thus, in this section, we work in Klein space which is a split signature spacetime with metric $\eta = \text{diag} \left(-,+,-,+\right)$ and whose conformal boundary is $\scri=\R\times \mathscr{S}$ where now $\mathscr{S}$ is the Lorentzian torus $\mathcal{LT}_2 = S^1 \times S^1 /\mathbb{Z}_2$ \cite{Atanasov:2021oyu}. Null momenta in Klein space can be parameterized in a manner similar to \eqref{eq:mompar31} as\footnote{This parameterization is reached by Wick rotating the third component of \eqref{eq:mompar31}.}
\begin{equation}
    \label{eq:mompar22}
    p_i^\mu = \e_i q_i^{\mu} = \e_i \om_i \left(1+z_i \zb_i, z_i + \zb_i, z_i - \zb_i, 1 - z_i \zb_i \right).
\end{equation}
Here $(z_i, \zb_i)$ are coordinates on a Poincar{\'e} patch of $\mathcal{LT}_2$, $\om_i$ is the energy and $\e_i = \pm 1$ labels the Poincar{\'e} patches which are now connected. For more details on how they relate to global coordinates on $\mathcal{LT}_2$, we refer the reader to \cite{Atanasov:2021oyu,Mason:2022hly}. Null momenta admit a decomposition into real spinor helicity variables 
\begin{align}
    p_{\alpha \dot{\alpha}} \equiv \sigma^{\mu}_{\alpha \dot{\alpha}}p_{\mu} = \kappa_{\alpha} \tilde{\kappa}_{\dot{\alpha}}. 
\end{align}
These are defined up to a little group scaling $\kappa \to t \kappa, \tilde{\kappa} \to \frac{1}{t} \tilde{\kappa}$ ($t\in \mathbb R/\lbrace 0 \rbrace$) and for the parametrization in \eqref{eq:mompar22}, we can set
\begin{align}
    \label{eq:shvars}
    \kappa_i = \sqrt{2 \omega_i} \e_i\begin{pmatrix}
        1\\ z_i
    \end{pmatrix}, \qquad  \tilde{\kappa}_i = \sqrt{2 \omega_i}\begin{pmatrix}
        1\\ \zb_i
    \end{pmatrix}.
\end{align} 
Using the standard notation $\mathcal A_3(\{\omega_1,z_1,\bar z_1\}^{\epsilon_1}_{J_1},\{\omega_2,z_2,\bar z_2\}^{\epsilon_2}_{J_2}, \{\omega_3,z_3,\bar z_3\}^{\epsilon_3}_{J_3}) \equiv \mathcal{A}_3(1^{J_1},2^{J_2},3^{J_3})$, the tree-level three-point scattering amplitudes are completely fixed by their helicities to be
\begin{equation}
\label{eq:threepoints}
    \mathcal{A}_3(1^{J_1},2^{J_2},3^{J_3}) = \left\{
    \begin{aligned}
        & \kappa_{J_1,J_2,J_3} \lbrack 12 \rbrack^{J_1+J_2-J_3} \lbrack 23 \rbrack^{J_2+J_3-J_1} \lbrack 31 \rbrack^{J_3+J_1-J_2} , && \text{if } J_1+J_2+J_3 > 0\,, \\
        & \kappa_{J_1,J_2,J_3} \langle 12 \rangle^{J_3-J_1-J_2} \langle 23 \rangle^{J_1-J_2-J_3} \langle 31 \rangle^{J_2-J_1-J_3} , && \text{if } J_1+J_2+J_3 < 0 
    \end{aligned}
    \right.
\end{equation} 
where we used the spinor-helicity notations $[ij] = \tilde{\kappa}_{i\dot{\alpha}}  \tilde{\kappa}_j^{\dot{\alpha}}$ and $ \langle i j \rangle = \kappa_i^\alpha  \kappa_{j\alpha}$. We have left the momentum conserving $\delta$ function implicit in the above formulae. We disregard amplitudes with $J_1+J_2+J_3 = 0$ as they do not lead to consistent four-particle interactions\footnote{The only exception is the $\phi^3$ interaction.} \cite{McGady:2013sga}. We will refer to the amplitudes on the first line of \eqref{eq:threepoints} as $\overline{\text{MHV}}$ and the second as MHV. Finally, for some combinations of integer spins, it will be necessary to introduce an extra colour index for these amplitudes to be compatible with spin-statistics. The case that will be relevant to this paper is Yang-Mills theory with gauge group $SU(N)$, corresponding to $J_1=J_2=-J_3 = 1$, where each particle transforms in the adjoint and the amplitude must be multiplied by the structure constant $f^{abc}$. In the rest of this paper, we will tacitly suppress this, restoring it only when it is crucial.

\subsection{Carrollian amplitude}

We will first focus on the $\overline{\text{MHV}}$ amplitudes. The corresponding Carrollian amplitude is obtained by applying the definition \eqref{eq:AtoC} on (the first line of) \eqref{eq:threepoints}:
\begin{multline}
\label{step1}
    \mc^{\overline{\text{MHV}}}_3 (\{u_1,z_1,\bar z_1\}^{\epsilon_1}_{J_1},\{u_2,z_2,\bar z_2\}^{\epsilon_2}_{J_2}, \{u_3,z_3,\bar z_3\}^{\epsilon_3}_{J_3}) =\kappa_{J_1,J_2,J_3}  \frac{1}{\left(2\pi\right)^3}  \zb_{12}^{J_1+J_2-J_3}\zb_{23}^{J_2+J_3-J_1}\zb_{13}^{J_1+J_3-J_2} \\ \times\int d\om_1\, d\om_2 \, d\om_3 \, e^{i \e_1 \om_1 u_1+i \e_2 \om_2 u_2+i \e_3 \om_3 u_3} \om_1^{J_1} \om_2^{J_2} \om_3^{J_3} \delta^{(4)}\left(p_1+p_2+p_3\right) \, .
\end{multline}
We can rewrite the delta function by solving $p_1+ p_2+ p_3 = 0$ for $\om_2, \om_3, z_1, z_2$:
\begin{equation}
    \delta^{(4)}\left(p_1+p_2+p_3\right) = \frac{1}{4\left|\zb_{12}\zb_{13}\right|\omega_1^2} \delta\left(z_{12}\right)\delta\left(z_{23}\right) \delta\left(\om_{2} + \frac{\zb_{13}}{\zb_{23}}\e_1 \e_2 \om_1\right)\delta\left(\om_{3} - \frac{\zb_{12}}{\zb_{23}}\e_1 \e_3 \om_1\right).
\end{equation}
Re-injecting this into \eqref{step1} yields the three-point Carrollian amplitude
\begin{align}
\label{eq:3pt} 
    \nonumber\mc^{\overline{\text{MHV}}}_3 &= \kappa_{J_1,J_2,J_3}\frac{\delta\left(z_{12}\right)\delta\left(z_{23}\right)}{4\left(2\pi\right)^3\left|\zb_{12}\zb_{13}\right|}\left|\zb_{12}\right|^{J_1+J_2}\left|\zb_{23}\right|^{-J_1} \left|\zb_{31}\right|^{J_3+J_1} \Theta\left(-\frac{\zb_{13}}{\zb_{23}}\e_1\e_2\right)\Theta\left(\frac{\zb_{12}}{\zb_{23}}\e_1\e_3\right) \left(\text{sign }\zb_{12}\right)^{J_1+J_2-J_3} \\
    \nonumber &\quad\times \left(\text{sign }\zb_{23}\right)^{J_2+J_3-J_1}\left(\text{sign }\zb_{13}\right)^{J_1+J_3-J_2}\int_0^{+\infty} d\om_1 \, e^{i \e_1 \left(u_1 - \frac{\zb_{13}}{\zb_{23}}u_2+ \frac{\zb_{12}}{\zb_{23}}u_3\right)\om_1} \, \om_1^{J_1+J_2+J_3-2} \\
     \nonumber &= \kappa_{J_1,J_2,J_3} \frac{ \delta\left(z_{12}\right)\delta\left(z_{23}\right)}{4 \left(2\pi\right)^3} 
     \Theta\left(-\frac{\zb_{13}}{\zb_{23}}\e_1\e_2\right)\Theta\left(\frac{\zb_{12}}{\zb_{23}}\e_1\e_3\right)X\left(\zb_{ij}, J_i\right) \mathcal{S}\left(\zb_{ij}, J_i\right)\\
     &\hspace{3cm} \times \frac{ \left(i \, \e_1 \, \text{sign}\left(\zb_{23}\right)\right)^{J_1+J_2+J_3-1}\Gamma\left(J_1+J_2+J_3-1\right)}{\left(\zb_{23}u_1 - \zb_{13}u_2+\zb_{12}u_3 + i \e_1\, \text{sign} (\zb_{23}) \varepsilon\right)^{J_1+J_2+J_3-1}} 
\end{align}
where\footnote{The quantity $\mathcal{S}\left(\zb_{ij}, J_i \right)$ defined in \eqref{defXS} is invariant under Lorentz transformations for integer $J_i$. For half integers, it picks up a phase.}
\begin{equation}
    \begin{split}
& X \left(\zb_{ij}, J_i \right) = \left|\zb_{12}\right|^{J_1+J_2-1}\left|\zb_{23}\right|^{J_2+J_3-1} \left|\zb_{31}\right|^{J_3+J_1-1} ,\\
& \mathcal{S} \left(\zb_{ij}, J_i \right) = \left(\text{sign }\zb_{12}\right)^{J_1+J_2-J_3} \left(\text{sign }\zb_{23}\right)^{J_2+J_3-J_1}\left(\text{sign }\zb_{13}\right)^{J_1+J_3-J_2} \, .
\end{split} \label{defXS}
\end{equation}
Equation \eqref{eq:3pt} should be regarded as a formal expression for the amplitude since the $\Gamma$ function is divergent when $J_1 + J_2 + J_3 = 1$. In these cases, it should be regulated in a manner similar to \eqref{divergent integral}. We can obtain the three-point Carrollian amplitude of descendants \eqref{eq:AtoCnews} by shifting $J_i \to J_i + 1$ and multiplying by the appropriate factors:
\begin{align}
\label{eq:3ptnews}
    \nonumber\tilde{\mc}^{\overline{\text{MHV}}}_3 &=\kappa_{J_1,J_2,J_3} \frac{-i \e_1 \e_2 \e_3 \,\delta\left(z_{12}\right)\delta\left(z_{23}\right)}{4 \left(2\pi\right)^3} 
     \Theta\left(-\frac{\zb_{13}}{\zb_{23}}\e_1\e_2\right)\Theta\left(\frac{\zb_{12}}{\zb_{23}}\e_1\e_3\right)X\left(\zb_{ij}, J_i+1\right) \mathcal{S}\left(\zb_{ij}, J_i+1\right)\\
     &\hspace{3cm} \times \frac{ \left(i \, \e_1 \, \text{sign}\left(\zb_{23}\right)\right)^{J_1+J_2+J_3+2}\Gamma\left(J_1+J_2+J_3+2\right)}{\left(\zb_{23}u_1 - \zb_{13}u_2+\zb_{12}u_3 + i \e_1\, \text{sign} (\zb_{23}) \varepsilon\right)^{J_1+J_2+J_3+2}}. 
\end{align} 
This expression is always finite since $J_1+J_2+J_3+2 >0, \, \forall J_1+J_2+J_3>0$. For completeness, we also display the three-point MHV Carrollian amplitudes of descendants which are simply obtained from the corresponding $\overline{\text{MHV}}$ ones by the replacements $z_{ij} \leftrightarrow \zb_{ij}$ and $J_i \to -J_i$: 
\begin{align}
\label{eq:3ptnewsneg}
     \nonumber\tilde{\mc}^{\text{MHV}}_3 &= \kappa_{J_1,J_2,J_3}\frac{-i \e_1 \e_2 \e_3 \,\delta\left(\zb_{12}\right)\delta\left(\zb_{23}\right)}{4 \left(2\pi\right)^3} 
     \Theta\left(-\frac{z_{13}}{z_{23}}\e_1\e_2\right)\Theta\left(\frac{z_{12}}{z_{23}}\e_1\e_3\right)X\left(z_{ij}, -J_i+1\right) \mathcal{S}\left(z_{ij}, -J_i+1\right)\\
     &\hspace{3cm} \times \frac{ \left(i \, \e_1 \, \text{sign}\left(z_{23}\right)\right)^{J_1+J_2+J_3+2}\Gamma\left(J_1+J_2+J_3+2\right)}{\left(z_{23}u_1 - z_{13}u_2+z_{12}u_3 + i \e_1\, \text{sign} (z_{23}) \varepsilon\right)^{J_1+J_2+J_3+2}}. 
\end{align}
Notice that the functional dependence of the holographic three-point Carrollian amplitudes found here is compatible with the result derived in \cite{Salzer:2023jqv} using embedding space formalism. Moreover, the expression obtained for the 3 graviton $\overline{\text{MHV}}$ Carrollian amplitude from \eqref{eq:3pt} is in agreement with the one computed in \cite{Banerjee:2019prz} via the modified Mellin transform \footnote{In order to compare the two, we must set $\D_i = 1$ in that reference.}. 

\subsection{Celestial from Carrollian}

\label{sec:crosscheck3pt}
Finally, as a cross-check of our diagram \ref{fig:FBM}, we can perform the integral transform \eqref{Btransform1} on \eqref{eq:3pt} or \eqref{Btransfom2} on \eqref{eq:3ptnews} to obtain the three-point celestial amplitude $\mathcal{M}_3$. Defining
\begin{equation}
    \label{eq:3pturescaling}
    \tilde{u}_1 = u_1 \zb_{23}, \qquad  \tilde{u}_2 = u_2 \zb_{13}, \qquad \tilde{u}_3 = u_3 \zb_{12}
\end{equation}
in \eqref{eq:3pt} and implementing the integral transform, we obtain
\begin{align}
    \mathcal{M}_3  &= \kappa_{J_1,J_2,J_3} \frac{ \delta\left(z_{12}\right)\delta\left(z_{23}\right)}{4 \left(2\pi\right)^3} 
     \Theta\left(-\frac{\zb_{13}}{\zb_{23}}\e_1\e_2\right)\Theta\left(\frac{\zb_{12}}{\zb_{23}}\e_1\e_3\right)X\left(\zb_{ij}, J_i\right) \mathcal{S}\left(\zb_{ij}, J_i\right)\\
     &\nonumber \frac{\mathcal{N}\left(\D_i, \e_i \right)}{\left|\zb_{12}\zb_{23}\zb_{13}\right|}\int_{-\infty}^{\infty} d\tu_1 \, d\tu_2 \, d\tu_3   \left(\frac{\tu_1}{\zb_{23} }+ i \e_1 \varepsilon \right)^{-\D_1}  \left(\frac{\tu_2}{\zb_{31}}+ i \e_2 \varepsilon \right)^{-\D_2}  \left(\frac{\tu_3}{\zb_{12}}+ i \e_3 \varepsilon \right)^{-\D_3}\\
     &\hspace{3cm} \times \frac{ \left(i \, \e_1 \, \text{sign}\left(\zb_{23}\right)\right)^{J_1+J_2+J_3-1}\Gamma\left(J_1+J_2+J_3-1\right)}{\left(\tu_1 - \tu_2+\tu_3 + i \e_1\, \text{sign} (\zb_{23}) \varepsilon\right)^{J_1+J_2+J_3-1}}. 
\end{align}
Here $\mathcal{N}\left(\D_i, \e_i \right) = \prod_{k=1}^3\left(-i\e_k\right)^{\D_k}\Gamma\left(\D_k\right)$ contains all the factors arising from the integral transform \eqref{Btransform1}. The contours for $\tilde{u}_1, \tilde{u}_2, \tu_3$ can be deformed such that they pick up the discontinuities across the branch cuts along the negative $\tu_i$ axes, leading to 
\begin{align}
    \mathcal{M}_3  &= \frac{2i\kappa_{J_1,J_2,J_3} \delta\left(z_{12}\right)\delta\left(z_{23}\right)}{\left(2\pi\right)^3} 
     \Theta\left(-\frac{\zb_{13}}{\zb_{23}}\e_1\e_2\right)\Theta\left(\frac{\zb_{12}}{\zb_{23}}\e_1\e_3\right)\tilde{X}\left(\zb_{ij}, J_i, \D_i\right) \mathcal{S}\left(\zb_{ij}, J_i\right) \, \mathcal{N}\left(\D_i, \e_i \right)\\
     &\nonumber \text{sin}\pi \D_1 \, \text{sin}\pi \D_2 \, \text{sin}\pi \D_3  \int_{0}^{+\infty} d\tu_1 \, d\tu_2 \, d\tu_3   \tu_1^{-\D_1}  \tu_2^{-\D_2}  \tu_3^{-\D_3} \frac{ \left(i \, \e_1 \, \text{sign}\left(\zb_{23}\right)\right)^{J_1+J_2+J_3-1}\Gamma\left(J_1+J_2+J_3-1\right)}{\left(\tu_1 - \tu_2+\tu_3 + i \e_1\, \text{sign} (\zb_{23}) \varepsilon\right)^{J_1+J_2+J_3-1}}. 
\end{align}
We have defined $\tilde{X}\left(\zb_{ij}, \D_i, J_i \right) \equiv  \left|\zb_{12}\right|^{\D_1+J_1+J_2-2}\left|\zb_{23}\right|^{\D_2+J_2+J_3-2} \left|\zb_{31}\right|^{\D_3+J_3+J_1-2}$. The three integrals can now be performed successively to get
\begin{align}
    \mathcal{M}_3  =& -\frac{i}{4}\,\kappa_{J_1,J_2,J_3} \delta\left(z_{12}\right)\delta\left(z_{23}\right) 
     \Theta\left(-\frac{\zb_{13}}{\zb_{23}}\e_1\e_2\right)\Theta\left(\frac{\zb_{12}}{\zb_{23}}\e_1\e_3\right)\tilde{X}\left(\zb_{ij}, J_i,\D_i\right) \mathcal{S}\left(\zb_{ij}, J_i\right)\\
     &\nonumber  (-i\e_1)^{\D_1}(-i\e_2)^{\D_2}(-i\e_3)^{\D_3} \left(i\e_1 \text{sign} \zb_{23} \right)^{\nu}\lim_{\varepsilon \to 0} \Gamma \left(\nu\right) \varepsilon^{-\nu},
\end{align}
where $\nu = \D_1+ \D_2+\D_3+J_1+J_2+J_3-4$. Using the limit formula
\begin{align}
    \lim_{\varepsilon \to 0} \Gamma \left(\nu\right) \varepsilon^{-\nu} = 2\pi \, \delta (\nu),
\end{align}
we get an expression for the 3 point celestial amplitude
\begin{align}
    \mathcal{M}_3  &= \frac{(-i)^{J_1+J_2+J_3} \, \pi }{2}\,\kappa_{J_1,J_2,J_3} \delta\left(z_{12}\right)\delta\left(z_{23}\right) 
     \Theta\left(-\frac{\zb_{13}}{\zb_{23}}\e_1\e_2\right)\Theta\left(\frac{\zb_{12}}{\zb_{23}}\e_1\e_3\right)\tilde{X}\left(\zb_{ij}, J_i, \D_i\right) \mathcal{S}\left(\zb_{ij}, J_i\right)\nonumber\\
     &\qquad \times  (\e_1)^{\D_1}(\e_2)^{\D_2}(\e_3)^{\D_3}  \delta \left(\D_1+ \D_2+\D_3+J_1+J_2+J_3-4 \right).
\end{align}
Particularizing the above formula to $J_1 = 1, J_2 = 1, J_3 = -1$, this reproduces the 3 gluon celestial amplitude of \cite{Pasterski:2017ylz} upto an overall phase. Furthermore, denoting $\bar{h}_k = \frac{\D_k-J_k}{2}$, we can replace
\begin{align}
    \zb_{12}^{\D_3+J_1+J_2-2}\zb_{31}^{\D_2+J_1+J_3-2}\zb_{23}^{\D_1+J_2+J_3-2} \rightarrow \frac{1}{\zb_{12}^{\hb_1+\hb_2-\hb_3}\zb_{23}^{\hb_2+\hb_3-\hb_1}\zb_{12}^{\hb_3+\hb_1-\hb_2}} ,
\end{align}
on the support of the  delta distribution of conformal dimensions. The right-hand side above is the standard form for the conformal three point function.

\section{Four-point amplitudes}
\label{sec:Four-point amplitude}
In this section, we compute four-point Carrollian amplitudes in Yang-Mills and gravity. These results can be recovered from (3.11) and (3.17) of \cite{Banerjee:2019prz} by setting  $\lambda_i$ to appropriate values. As in a 2D CFT, the two- and three-point correlation functions in a Carrollian CFT are completely fixed by symmetries while the four-point functions are not. They contain dynamical information and constitute constraints on the putative dual Carrollian CFT . At tree-level the only non-zero 4-point amplitudes are the MHV amplitudes, which are given by\footnote{For Yang-Mills theory, we are working with colour ordered amplitudes and as in the 3 particle case, we are suppressing the colour indices on all the particles. For more details, we refer the reader to the review \cite{Mangano:1990by} and the reference therein.}
\begin{align}
    &\mA_4 \left(1^{+1}, 2^{-1}, 3^{-1}, 4^{+1}\right) = \kappa_{1,1,-1}^2\frac{\an{23}^4}{\an{12}\an{23}\an{34}\an{41}} = \kappa_{1,1,-1}^2\frac{\om_2 \om_3}{\om_1 \om_4} \frac{z_{23}^3}{z_{12}z_{34}z_{41}} , \label{eq:4ptgluonamp}\\
    &\mA_4 \left(1^{+2}, 2^{-2}, 3^{-2}, 4^{+2}\right) = \kappa_{2,2,-2}^2\left(\an{23}\sq{14}\right)^4 \frac{1}{s t u} =  \kappa_{2,2,-2}^2\frac{\om_2\om_3\om_4}{\om_1} \frac{z_{23}^4\zb_{14}^4}{z_{12}\zb_{12}z_{13}\zb_{13}z_{14}\zb_{14}}\label{eq:4ptgravamp}
\end{align}
where $s= (p_1 + p_2)^2$, $t= (p_1 + p_3)^2$ and $u = (p_1 + p_4)^2$ are the Mandelstam variables. Again, we have left the momentum conserving $\delta$ function implicit in the above formulas. We will continue  using \eqref{eq:mompar22} to parameterize the momentum, implicitly working in Klein space. However, the results are also valid in Minkowski spacetime upon interpreting $\zb_i$ to be the complex conjugate of $z_i$. 

\subsection{Carrollian amplitude}

In order to evaluate the Fourier transforms of \eqref{eq:4ptgluonamp} and \eqref{eq:4ptgravamp}, we first write the delta distribution as
\begin{multline}
\label{delta function explicit}
      \delta^{(4)}\left(p_1+p_2+p_3+p_4\right) = \frac{1}{4\om_4\left|z_{24}\zb_{13}\right|^2} \delta\left(\om_1 + z \left|\frac{z_{24}}{z_{12}}\right|^2 \e_1 \e_4 \om_4 \right)\\
      \times \delta\left(\om_2 - \frac{1-z}{z} \left|\frac{z_{34}}{z_{23}}\right|^2 \e_2 \e_4 \om_4 \right) \delta\left(\om_3 + \frac{1}{1-z} \left|\frac{z_{14}}{z_{13}}\right|^2 \e_3 \e_4 \om_4 \right) \delta\left(z-\zb\right)
\end{multline}
where $z = \frac{z_{12}z_{34}}{z_{13}z_{24}}$ is the cross ratio and $\left|z_{ij}\right|^2 = z_{ij} \zb_{ij} $. Note that this is just a shorthand in Klein space where $z_{ij}$ and $\zb_{ij}$ are real and independent. As in the previous sections, we will be interested in evaluating both the Carrollian amplitude $\mc_4$ and the correlator of $u-$descendants $\tilde{\mc}_4$. Plugging \eqref{delta function explicit} into \eqref{eq:AtoC} and \eqref{eq:AtoCnews} for $n=4$, we get respectively
\begin{multline}
\label{eq:4pt}
    \mc_4 = \frac{1}{\left(2\pi\right)^4}\delta\left(z-\zb\right) \Theta\left(-z \left|\frac{z_{24}}{z_{12}}\right|^2 \e_1 \e_4\right) \Theta\left(\frac{1-z}{z} \left|\frac{z_{34}}{z_{23}}\right|^2 \e_2 \e_4 \right)\Theta\left( -\frac{1}{1-z} \left|\frac{z_{14}}{z_{13}}\right|^2 \e_3 \e_4 \right)\\
    \times \int_0^{+\infty} d\om_4 e^{i \e_4 \om_4 \left(-u_1 z \left|\frac{z_{24}}{z_{12}}\right|^2 +u_2 \frac{1-z}{z}\left|\frac{z_{34}}{z_{23}}\right|^2 - \frac{u_3}{1-z} \left|\frac{z_{14}}{z_{13}}\right|^2 + u_4\right) } \frac{1}{\om_4} \mA_4^*,
\end{multline} 
\begin{multline}
\label{eq:4ptnews}
    \tilde{\mc}_4 = \frac{\e_1 \e_2 \e_3 \e_4}{\left(2\pi\right)^4}\delta\left(z-\zb\right) \Theta\left(-z \left|\frac{z_{24}}{z_{12}}\right|^2 \e_1 \e_4\right) \Theta\left(\frac{1-z}{z} \left|\frac{z_{34}}{z_{23}}\right|^2 \e_2 \e_4 \right)\Theta\left( -\frac{1}{1-z} \left|\frac{z_{14}}{z_{13}}\right|^2 \e_3 \e_4 \right)\\
    \times \left|\frac{z_{24}z_{34}z_{14}}{z_{12}z_{23}z_{13}}\right|^2\int_0^{+\infty} d\om_4 e^{i \e_4 \om_4 \left(-u_1 z \left|\frac{z_{24}}{z_{12}}\right|^2 +u_2 \frac{1-z}{z}\left|\frac{z_{34}}{z_{23}}\right|^2 - \frac{u_3}{1-z} \left|\frac{z_{14}}{z_{13}}\right|^2 + u_4\right) } \om_4^3 \mA_4^*,
\end{multline}
where $\mA_4^*$ represents the 4-point amplitude evaluated on the support of \eqref{delta function explicit}. For Yang-Mills, $\mc_4$ is IR divergent but $\tilde{\mc}_4$ is not, while for gravity, both are IR finite. 

\paragraph{4 gluon amplitude:} 
We can now evaluate the four-point Carrollian amplitude for gluons by inserting \eqref{eq:4ptgluonamp} into \eqref{eq:4pt} to get 
\begin{multline}
\label{eq:4ptgluon}
\mc_4\left(1^{+1},2^{-1},3^{-1},4^{+1}\right) =\frac{ \kappa_{1,1,-1}^2 \, \e_1\e_2\e_3\e_4}{\left(2\pi\right)^4} \frac{(1-z)\zb_{14}^2}{z z_{14}^2 \zb_{13}^2 \zb_{24}^2} \delta\left(z-\zb\right)  \Theta\left(-z \left|\frac{z_{24}}{z_{12}}\right|^2 \e_1 \e_4\right) \\  \Theta\left(\frac{1-z}{z} \left|\frac{z_{34}}{z_{23}}\right|^2 \e_2 \e_4 \right) \Theta\left( -\frac{1}{1-z} \left|\frac{z_{14}}{z_{13}}\right|^2 \e_3 \e_4 \right) \\
   \times \mathcal{I}_0 \left(u_4 - u_1 z \left|\frac{z_{24}}{z_{12}}\right|^2+u_2 \frac{1-z}{z}\left|\frac{z_{34}}{z_{23}}\right|^2 - u_3\frac{1}{1-z}\left|\frac{z_{14}}{z_{13}}\right|^2\right),
\end{multline}
where 
$\mathcal{I}_0$ is the same integral as the one discussed in \eqref{divergent integral} for the two-point function. In particular, it can be regularized and written as in \eqref{curly I beta}. An identical computation for $\tilde{\mc}_4$ gives
\begin{multline}
\label{eq:4ptgluonnews}
\tilde{\mc}_4\left(1^{+1},2^{-1},3^{-1},4^{+1}\right) = \frac{\kappa_{1,1,-1}^2 }{\left(2\pi\right)^4} \frac{z_{34}^2 \zb_{14}^4 \zb_{34}^2}{z^3 (1-z) z_{13}^3 z_{24}\zb_{13}^5 \zb_{24}^3}  \delta\left(z-\zb\right)  \Theta\left(-z \left|\frac{z_{24}}{z_{12}}\right|^2 \e_1 \e_4\right)  \\
  \Theta\left(\frac{1-z}{z} \left|\frac{z_{34}}{z_{23}}\right|^2 \e_2 \e_4 \right)\Theta\left( -\frac{1}{1-z} \left|\frac{z_{14}}{z_{13}}\right|^2 \e_3 \e_4 \right)\\
    \times \frac{3!}{\left(u_4 - u_1 z \left|\frac{z_{24}}{z_{12}}\right|^2+u_2 \frac{1-z}{z}\left|\frac{z_{34}}{z_{23}}\right|^2 - u_3\frac{1}{1-z}\left|\frac{z_{14}}{z_{13}}\right|^2\right)^4},
\end{multline}
which is IR finite as expected.

\paragraph{4 graviton amplitude: } Similarly, the Carrollian amplitude corresponding to the four-point graviton amplitude is obtained by inserting \eqref{eq:4ptgravamp} into \eqref{eq:4pt}, which yields
\begin{multline}
\label{eq:4ptgrav}
    \mc_4 \left(1^{+2},2^{-2},3^{-2},4^{+2}\right)= \kappa_{2,2,-2}^2 \frac{\e_1\e_2\e_3\e_4}{\left(2\pi\right)^4}\frac{1}{\left|z_{12}\right|^2}\frac{z_{14}^3}{\zb_{14}^4\left|1-z\right|^2 }\delta\left(z-\zb\right) \\
   \Theta\left(-z \left|\frac{z_{24}}{z_{12}}\right|^2 \e_1 \e_4\right)  \Theta\left(\frac{1-z}{z} \left|\frac{z_{34}}{z_{23}}\right|^2 \e_2 \e_4 \right) \Theta\left( -\frac{1}{1-z} \left|\frac{z_{14}}{z_{13}}\right|^2 \e_3 \e_4 \right) \\
   \times   \frac{1}{\left(u_4 - u_1 z \left|\frac{z_{24}}{z_{12}}\right|^2+u_2 \frac{1-z}{z}\left|\frac{z_{34}}{z_{23}}\right|^2 - u_3\frac{1}{1-z}\left|\frac{z_{14}}{z_{13}}\right|^2\right)^2},
\end{multline}
or into \eqref{eq:4ptnews}, leading to
\begin{multline}
\label{eq:4ptgravnews}
    \tilde{\mc}_4 \left(1^{+2},2^{-2},3^{-2},4^{+2}\right)= \kappa_{2,2,-2}^2 \frac{1}{\left(2\pi\right)^4}\frac{1}{\left|z_{12}\right|^2}\left|\frac{z_{24}z_{34}z_{14}}{z_{12}z_{23}z_{13}}\right|^2\frac{z_{14}^3}{\zb_{14}^4\left|1-z\right|^2 }\delta\left(z-\zb\right) \\
    \Theta\left(-z \left|\frac{z_{24}}{z_{12}}\right|^2 \e_1 \e_4\right) \Theta\left(\frac{1-z}{z} \left|\frac{z_{34}}{z_{23}}\right|^2 \e_2 \e_4 \right)
    \Theta\left( -\frac{1}{1-z} \left|\frac{z_{14}}{z_{13}}\right|^2 \e_3 \e_4 \right)\\
   \times \frac{5!}{\left(u_4 - u_1 z \left|\frac{z_{24}}{z_{12}}\right|^2+u_2 \frac{1-z}{z}\left|\frac{z_{34}}{z_{23}}\right|^2 - u_3\frac{1}{1-z}\left|\frac{z_{14}}{z_{13}}\right|^2\right)^6} .
\end{multline}
In contrast to the gluon case, both of these are IR finite. 
\subsection{Celestial from Carrollian}
\label{sec:crosscheck4pt}
We will conclude this section by showing that the Carrollian amplitude \eqref{eq:4ptgluonnews} reproduces the corresponding celestial amplitude upon application of \eqref{Btransfom2}. We are using $\tilde{\mc}_4$ rather than $\mc_4$ to avoid dealing with the IR-divergence. Starting from 
\begin{equation}
    \mathcal{M}_4 = \prod_{i=1}^4 \left(  \int_{-\infty}^{+\infty}  du_i \, (- i \epsilon_i)^{\Delta_i} \Gamma(\D_i-1)\, u_i^{1-\D_i} \right) \tilde{\mc}_4,
\end{equation}
we change integration variables to 
\begin{align}
    \label{eq:varchange}
    \tilde{u}_1 = u_1 z \left|\frac{z_{24}}{z_{12}}\right|^2, \,\, \tilde{u}_2 = u_2 \frac{1-z}{z} \left|\frac{z_{34}}{z_{23}}\right|^2, \,\, \tilde{u}_3 = u_3 \frac{1}{1-z} \left|\frac{z_{14}}{z_{23}}\right|^2, \,\, \tu_4 = u_4
\end{align}
resulting in the simpler form
\begin{multline}
\label{step comp}
    \mathcal{M}_4 = \frac{\kappa_{1,1,-1}^2}{\left(2\pi\right)^4}\frac{z_{34}^2 \zb_{14}^4 \zb_{34}^2}{z^3 (1-z) z_{13}^3 z_{24}\zb_{13}^5 \zb_{24}^3}  \left(z \left|\frac{z_{24}}{z_{12}}\right|^2\right)^{\D_1-2}\left(\frac{1-z}{z}\left|\frac{z_{34}}{z_{23}}\right|^2 \right)^{\D_2-2}\left(\frac{1}{1-z}\left|\frac{z_{14}}{z_{13}}\right|^2\right)^{\D_3-2}\\
    \delta\left(z-\zb\right) \Theta\left(-z \left|\frac{z_{24}}{z_{12}}\right|^2 \e_1 \e_4\right) \Theta\left(\frac{1-z}{z} \left|\frac{z_{34}}{z_{23}}\right|^2 \e_2 \e_4 \right)\Theta\left( -\frac{1}{1-z} \left|\frac{z_{14}}{z_{13}}\right|^2 \e_3 \e_4 \right) M_4,
\end{multline}
where 
\begin{equation}
   M_4 = \prod_{i=1}^4 \left( \int^{+\infty}_{-\infty}  d\tilde{u}_i \, \Gamma(\D_i-1)\,(-i\e_i)^{\D_i} \tilde{u}^{1-\D_i}_i \right) \frac{3!}{\left(\tu_4 - \tilde{u}_1 +\tilde{u}_2 - \tilde{u}_3 + i \e_4\varepsilon\right)^4 }.
\end{equation}
We now focus on the evaluation of $M_4$. The contours of all these integrals can be deformed to pick up the discontinuity across the branch cuts along the negative $u_i$ axes resulting in 
\begin{equation}
    M_4 =2^4\prod_{k=1}^4 (-i \e_k)^{\D_k} \text{sin} \pi \D_k\int_{0}^{+\infty} d\tilde{u}_k \, \tu_k^{1-\D_k} \frac{3!}{\left(-\tu_4 + \tilde{u}_1 -\tilde{u}_2 + \tilde{u}_3+i \e_4\varepsilon\right)^4}.
\end{equation}
All these integrals can be performed in a straightforward manner upon retaining the $i \e_4 \varepsilon$ till the end. The result is 
\begin{equation}
    M_4 = \frac{\prod_{i=1}^4(-i\e_i)^{\D_i}(-1)^{\D_1+\D_3}(2\pi)^4}{6} \lim_{\varepsilon \to 0}\Gamma \left(\beta \right) \varepsilon^{-\beta} = \frac{\prod_{i=1}^4(-i\e_i)^{\D_i}(-1)^{\D_1+\D_3}(2\pi)^4}{6}  2\pi \delta \left(\beta \right)
\end{equation}
where $\beta = \D_1 + \D_2+ \D_3 + \D_4 - 4$. Finally, re-injecting this expression into \eqref{step comp} (with $h_i = \frac{\D_i+J_i}{2}, \hb_i =  \frac{\D_i-J_i}{2}$) leads to
\begin{multline}
     \mathcal{M}_4  =\prod_{i=1}^4(-i\e_i)^{\D_i} z^{-\frac{1}{3}}\left(1-z\right)^{\frac{5}{3}} \prod_{i<j} z_{ij}^{\frac{h}{3}-h_i-h_j}\zb_{ij}^{\frac{\hb}{3}-\hb_i-\hb_j}  (-1)^{\D_2+\D_4+1}2\pi \delta\left(\D_1+\D_2+\D_3+\D_4-4\right)\\
    \delta\left(z-\zb\right) \Theta\left(-z \left|\frac{z_{24}}{z_{12}}\right|^2 \e_1 \e_4\right) \Theta\left(\frac{1-z}{z} \left|\frac{z_{34}}{z_{23}}\right|^2 \e_2 \e_4 \right)\Theta\left( -\frac{1}{1-z} \left|\frac{z_{14}}{z_{13}}\right|^2 \e_3 \e_4 \right) .
\end{multline}
This matches the result of \cite{Pasterski:2017ylz} adapted to the helicity configuration considered here up to a phase. An identical procedure can be used to reproduce the 4 graviton celestial amplitude starting from \eqref{eq:4ptgrav} or \eqref{eq:4ptgravnews}.

\section{MHV \texorpdfstring{$n$}{n}-point amplitudes}
\label{sec:MHV Carrollian amplitudes}

\subsection{Carrollian amplitude}

In this section, we evaluate MHV Carrollian amplitude for arbitrary multiplicities. We will first demonstrate that the computation of the Carrollian amplitude in both Yang-Mills and Einstein gravity reduces to the evaluation of the same integral. Following \cite{Schreiber:2017jsr}, we can rewrite the momentum conserving delta function as 
\begin{align}
\label{eq:momsolve}
    \delta^{(4)} \left(\sum_{i=1}^n p_i \right) = \frac{1}{\left|\mU_{1234}\right|}\prod_{I=1}^4 \delta \left(\om_I - \om_I^*\right) , \quad \text{with} \quad  \om_I^* = -\frac{1}{\mU_{1234}}\sum_{i=5}^n \om_i \mU_{Ii}  
\end{align}
where
\begin{align}
    \label{eq:Udefs}
    \mU_{1234} = \text{det}\left(q_1^{\mu}, \dots , q_4^{\mu} \right), \qquad \mU_{Ii} = \mU_{1234}|_{I \to i}, \quad I = 1,2,3,4; i = 5, \dots , n
\end{align}
and $q_i^{\mu}$ are defined in \eqref{eq:mompar31} and \eqref{eq:mompar22} for (3,1) and (2,2) signatures, respectively. Note that $\mU_{Ii}$ are independent of all the $\om_i$ and it was shown in \cite{Mizera:2022sln} that they evaluate to 
\begin{equation}
    \label{eq:Uformula}
    \mU_{ijkl} = 8 \e_i \e_j \e_k \e_l \left|z_{ik}z_{jl}\right|^2 \text{Im}\frac{z_{ij}z_{kl}} {z_{ik}z_{jl}} .
\end{equation}
The equation above is also valid in Klein space if we interpret Im$\frac{z_{ij}z_{kl}} {z_{ik}z_{jl}}  = \frac{z_{ij}z_{kl}} {z_{ik}z_{jl}}  - \frac{\zb_{ij}\zb_{kl}} {\zb_{ik}\zb_{jl}}$ with $z_i, \zb_i$ being real and independent. 
\paragraph{{\bf Yang-Mills}} The colour ordered MHV gluon amplitude (with $n+1$ identified with 1) is 
\begin{equation}
    \label{eq:nptgluonamp}
    \mA_n\left(1^-,2^-,3^+, \dots, n^+\right) = \kappa_{1,1,-1}^{n-2}\frac{\an{12}^4}{\prod_{j=1}^n \an{jj+1}} = \kappa_{1,1,-1}^{n-2}\frac{\om_1 \om_2}{\prod_{j=3}^n \om_j} \, \frac{z_{12}^3}{\prod_{j=2}^n z_{jj+1}}
\end{equation}
where we kept the momentum conserving $\delta$ function implicit. We are now ready to compute the position space amplitude using \eqref{eq:momsolve}. The Fourier transform of \eqref{eq:nptgluonamp} is divergent as in the 4 point case and we will instead compute the correlator of $u-$descendants. Applying the integral transform \eqref{eq:AtoCnews} on \eqref{eq:nptgluonamp} and performing the integrals over $\om_1, \dots ,\om_4$, we end up with 
\begin{align}
\label{eq:nptgluonpos1}
   &\tilde{\mc}_n\left(1^-,2^-,3^+,\dots ,n^+\right) \nonumber \\
   &= \frac{\kappa_{1,1,-1}^{n-2}}{(2\pi)^n\left|\mU_{1234}\right|}\frac{z_{12}^3}{\prod_{j=2}^n z_{jj+1}}\int \prod_{j=5}^n d\om_j \om_j e^{i \e_j \om_j u_j}  \left[\prod_{I=1}^4 e^{i\e_I \om^*_I u_I} \Theta\left(\om_I^*\right) \om_I^*\right] \frac{\om_1^*\om_2^*}{\om_3^* \om_4^*\prod_{j=5}^n \om_j} \\
    & =  \frac{\kappa_{1,1,-1}^{n-2}}{(2\pi)^n\left|\mU_{1234}\right|}\frac{z_{12}^3}{\prod_{j=2}^n z_{jj+1}}\frac{\partial^4}{\partial u_1^2\, \partial u_2^2}\int \prod_{j=5}^n d\om_j  e^{i \frac{\om_j}{\mU_{1234}} \left(\e_j u_j \mU_{1234}- \sum_{J=1}^4 \e_J u_{J} \mU_{Jj}\right)}  \prod_{I=1}^4 \Theta\left(\om_I^*\right) \nonumber \\
    & \equiv \frac{\kappa_{1,1,-1}^{n-2}}{(2\pi)^n\left|\mU_{1234}\right|}\frac{z_{12}^3}{\prod_{j=2}^n z_{jj+1}}\frac{\partial^4}{\partial u_1^2\, \partial u_2^2}I_n. \nonumber
\end{align}
In the last line, we have introduced the integral
\begin{equation}
    \label{eq:Indef}
    I_n = \int \prod_{j=5}^n d\om_j  e^{i \om_j L_j}  \prod_{I=1}^4 \Theta\left(\om_I^*\right),
\end{equation}
with $L_j = \left(\e_j u_j - \sum_{J=1}^4 \e_J u_J\frac{\mU_{Jj}}{\mU_{1234}} \right) $. This integral will also show up in the computation of gravitational MHV amplitudes.

\paragraph{{\bf Gravity}} The MHV amplitude is
\begin{equation}
    \label{eq:MHVgrav}
    \mA_n \left(1^{--}, 2^{--}, 3^{++}, \dots , n^{++} \right) =\kappa_{2,2,-2}^{n-2} \left| \Phi_{abc}^{def} \right| \frac{\an{12}^8}{\an{ab}\an{bc}\an{ca}\an{de}\an{ef}\an{fd}},
\end{equation} where we kept the momentum conserving $\delta$ function implicit. Here $\left|\Phi^{def}_{abc}\right|$ is the determinant of the matrix $\Phi$ with rows $a,b,c$ and columns $d,e,f$ removed. The amplitude is independent of the choice of $a, \dots, f$. $\Phi$ has matrix elements
\begin{align}
    \Phi_{ij} = \frac{\sq{ij}}{\an{ij}} = \frac{\zb_{ij}}{z_{ij}} \quad (i\neq j), \qquad \Phi_{ii} = -\sum_{j\neq i}\frac{\sq{ij} \an{jr}\an{js}}{\an{ij}\an{ir}\an{is}} = -\sum_{j\neq i}\frac{\zb_{ij} z_{jr}z_{js}}{z_{ij}z_{ir}z_{is}}\frac{\om_j}{\om_i} .
\end{align}
Here $r, s$ are arbitrary reference spinors. $\Phi_{ii}$ is independent of the choice of $r, s$. Note that this is also the leading soft factor (see $S^{(0)}$ in \eqref{soft theorem} in Section \ref{sec:Soft limits and memories} below). The integral we want to compute is given by 
\begin{align}
    \label{eq:MHVCtilde1}
    \tilde{\mathcal C}_n = \frac{1}{(2\pi)^n}\int \prod_{j=1}^n \, d\om_j \, i \e_j \om_j e^{i \e_j \om_j u_j } \mA_n .
\end{align}
In order to proceed, we first observe that $\om_1 \dots \om_n \mA_n$ has no negative powers of $\om_i$. To see this, note that negative powers of $\om_i$ occur only on the diagonal of $\Phi$. Eliminating rows and columns $a, \dots , f$ eliminates negative powers of $\om_a, \dots , \om_f$ from the determinant. Multiplying the determinant by the remaining  $\om_i$  eliminates all negative powers. This suggests to define $\left|\tilde{\Phi}^{def}_{abc}\right|_{\om_i} = \prod_{k=1, k\neq {a, \dots f}}^n \om_k \,\left| \Phi_{abc}^{def} \right| $. This has no negative powers of $\om_i$ by the above reasoning and we can now write 
\begin{align}
    \label{eq:MHVCtilde2}
    \tilde{\mathcal C}_n = \frac{\kappa_{2,2,-2}^{n-2}}{(2\pi)^n}\frac{i^n \, z_{12}^8}{z_{ab}z_{bc}z_{ca}z_{de}z_{ef}z_{fd}}\prod_{k=1}^n \e_k \left|\tilde{\Phi}^{def}_{abc}\right|_{\omega_i \to -i \e_i \frac{\partial}{\partial u_i}}  \frac{\partial^4}{\partial u_1^4}\frac{\partial^4}{\partial u_2^4}\int \prod_{j=1}^n \, d\om_j \, e^{i \e_j \om_j u_j } \delta^{(4)} \left(\sum_k p_k \right) .
\end{align}
Plugging in the solution for $\om_1, \dots ,\om_4$ from \eqref{eq:momsolve}, we get
\begin{align}
    \label{eq:MHVCtilde3}
    \tilde{\mathcal C}_n = \frac{\kappa_{2,2,-2}^{n-2}}{\left|\mU_{1234}\right|(2\pi)^n}\frac{i^n \, z_{12}^8}{z_{ab}z_{bc}z_{ca}z_{de}z_{ef}z_{fd}}&\prod_{k=1}^n \e_k \left|\tilde{\Phi}^{def}_{abc}\right| \left(-i \e_i \frac{\partial}{\partial u_i}\right) \frac{\partial^4}{\partial u_2^4}\frac{\partial^4}{\partial u_2^4} I_n,
\end{align}
with $I_n$ defined in \eqref{eq:Indef}.

\paragraph{{\bf Evaluating $I_n$}} In order to proceed, we must solve the inequalities arising from the  $\Theta$ functions -- an incredibly difficult task whose complexity increases rapidly with $n$. These inequalities can be solved only for specific configurations of $z_i, \zb_i$ and $ \e_i$, thus implying that the corresponding correlators are not supported everywhere on the celestial sphere \cite{Mizera:2022sln}. Each valid configuration specifies domains of integration for the $\om_i$. As an example, consider the case $n=6$ where the constraints are  
\begin{align}
\label{eq:6ptineqs}
    \begin{pmatrix}
        \frac{\mU_{5234}}{\mU_{1234}} & \frac{\mU_{6234}}{\mU_{1234}} \\
        \frac{\mU_{1254}}{\mU_{1234}} & \frac{\mU_{1634}}{\mU_{1234}} \\
        \frac{\mU_{1254}}{\mU_{1234}} & \frac{\mU_{1264}}{\mU_{1234}} \\
        \frac{\mU_{1235}}{\mU_{1234}} & \frac{\mU_{1236}}{\mU_{1234}} \\
    \end{pmatrix} \begin{pmatrix}
        \om_5 \\
        \om_6
    \end{pmatrix} \leq 0 .
\end{align}
We will not list all valid kinematic regions but merely present two examples. 
\paragraph{Example 1:} One solution to the inequalities \eqref{eq:6ptineqs} is the non-trivial region in kinematic space carved out by 
\begin{align}
\label{eq:exampler1}
    &\nonumber \mU_{1534}\, \mU_{1236} - \mU_{1634}\, \mU_{1236}  \leq 0,\qquad \mU_{1534} \, \mU_{1264} - \mU_{1634}\, \mU_{1254} \leq 0,\qquad \mU_{1234}\, \mU_{5234} \geq 0, \qquad \mU_{1234} \,\mU_{1235} \geq 0, \\
    &\nonumber \hspace{2cm} \mU_{1234} \, \mU_{6234} \geq 0, \qquad \mU_{1234} \, \mU_{1534} \leq 0, \qquad \mU_{1234} \, \mU_{1235} \leq 0, \qquad  \mU_{1234} \, \mU_{1264} \leq 0.
\end{align}
In this region, the domain of integration is
\begin{align}
\om_5 \geq 0, \qquad \ - \frac{\mU_{1534}}{\mU_{1634}} \om_5 \leq \, \om_6 \, \leq -\frac{\mU_{1254}}{\mU_{1264}}\om_5.
\end{align}
\paragraph{Example 2:} Another solution to ($\ref{eq:6ptineqs}$) is the simpler region
\begin{align}
\label{eq:exampler2}
    \frac{\mU_{Ii}}{\mU_{1234}}<0, \qquad I=1, 2, 3, 4; i = 5, 6.
\end{align}
In this case, the domains of integration are simply $\om_5 \geq 0, \, \om_6 \geq 0$. Without loss of generality, we can take the generic situation to be
\begin{align}
    \label{eq:region}
    B_n \leq \om_n \leq B'_n, \qquad \sum_{a=k+1}^n B_{k a}\, \om_a \leq \om_k \leq   \sum_{a=k+1}^n B'_{k a}\, \om_a .
\end{align}
We have assumed that the inequalities have been solved in such a way that the bounds on the integration domain of $\om_k$ depend only on $\om_{k+1}, \dots \om_n$. It is convenient to define the vectors
\begin{align}
    \label{eq:domainvecs}
    B_k = \left(0, \dots , B_{kk+1}, \dots , B_{kn}\right), \qquad  B'_k = \left(0, \dots , B'_{kk+1}, \dots ,B'_{kn}\right), \qquad \om = \left(\om_5, \dots, \om_n \right).
\end{align}
Note that $B_k \cdot \om$ reduces to the combination appearing in \eqref{eq:region}. With this, the integral \eqref{eq:Indef} can be written as 
\begin{align}
     I_n = \int_0^{\infty} \prod_{a=5}^n d\om_a \, e^{i L_a \, \om_a  }  \Theta \left(\om_n - B_n\right)\Theta \left(B'_n - \om_n\right)\prod_{k=5}^{n-1} \Theta \left(\om_k - B_k \cdot \om\right)\Theta \left(B'_k \cdot \om - \om_k\right) .
\end{align}
We relegate the details of the computation of this integral to Appendix \ref{sec:Incomp} merely presenting the final result,
\begin{align}
     I_n =\left(\frac{1}{2\pi i}\right)^{2(n-4)}\int \prod_{k=5}^n\frac{d\tau_k d\tau_k'}{(\tau_k+i \e)(\tau_k' + i \e)} \frac{1}{X_k\left(\tau\right) - i \e}\text{exp}\left[-i B_n \tau_n + i B'_n \tau'_n \right]
\end{align}
where $X_k \left(\tau\right)= \tau_k - \tau'_k - \sum_{a=5}^{k-1} \tau_a B_{ak} + \sum_{a=5}^{k-1} \tau'_a B'_{ak} - L_k$. These integrals can be evaluated in a straightforward manner by computing the appropriate residues. Writing down a general formula is cumbersome and we present results for 5 and 6 point amplitudes:
\begin{align}
    &I_5 = \frac{e^{-i B_5 L_5} - e^{-i B'_5 L_5}}{L_5} ,\\
    &\nonumber I_6 = \frac{1}{L_5 \left(B_{56}L_5 + L_6\right)}\left[e^{i B'_6(B_{56}L_5+L_6)} - e^{i B_6(B_{56}L_5+L_6)} \right] \\
    &\nonumber \qquad\qquad\qquad + \frac{1}{L_5 \left(B'_{56}L_5 + L_6\right)}\left[e^{i B_6(B'_{56}L_5+L_6)} - e^{i B'_6(B'_{56}L_5+L_6)} \right] .
\end{align}
These expressions demonstrate that the Carrollian amplitude takes on a very simple functional form in generic kinematic regions of the type shown in example 1. We can provide a compact form for $I_n$ in the region corresponding to example 2, where all the $\om_i$ are integrated between $0$ and $\infty$. As we explain in Appendix \ref{sec:Incomp}, this is done simply by computing the residues at $\tau_k = \tau'_k = 0$ yielding
\begin{align}
    \label{eq:Inspecial}
    I_n = (-1)^{n-4} \prod_{j=5}^n \frac{1}{L_j} .
\end{align}
As we already pointed out at the beginning of this section, these functions are much simpler than their celestial counterparts, which for gluons have been evaluated in \cite{Schreiber:2017jsr}.

\subsection{From Carrollian to Celestial}
\label{sec:nptcrosscheck}
The gluon MHV celestial amplitudes have only been evaluated in the region \eqref{eq:exampler2}. In this section we will work in this restricted kinematic regime where the MHV Carrollian amplitude of $u$-descendants is
\begin{align}
     \tilde{\mc}_n\left(1^-,2^-,3^+,\dots ,n^+\right) = (-1)^{n-4} \frac{\kappa_{1,1,-1}^{n-2}}{(2\pi)^n\left|\mU_{1234}\right|}\frac{z_{12}^3}{\prod_{j=2}^n z_{jj+1}}\frac{\partial^4}{\partial u_1^2\, \partial u_2^2} \prod_{j=5}^n \frac{1}{L_j}.
\end{align}
Applying \eqref{Btransfom2} on this expression, we can immediately perform the integrals over $u_5, \dots , u_n$ by methods analogous to those used in Sections \ref{sec:crosscheck3pt} and \ref{sec:crosscheck4pt}, and by noting that these variables only occur in $L_5, \dots , L_n$, respectively. This gives
\begin{align}
    \mathcal{M}_n\left(1^-,2^-,3^+,\dots ,n^+\right) =&\frac{ (-1)^{n-4} \kappa_{1,1,-1}^{n-2}}{(2\pi)^n\left|\mU_{1234}\right|}\frac{z_{12}^3}{\prod_{j=2}^n z_{jj+1}} \mathcal{N} \times \mathcal{I}_{n-4}
\end{align}
where 
\begin{align}
    \mathcal{N} = 16 (2\pi i)^{n-4}\Gamma\left(\D_1+1\right)\Gamma\left(\D_2+1\right)\Gamma\left(\D_3-1\right)\Gamma\left(\D_4-1\right) \prod_{k=1}^4(-i \e_k)^{\D_k} \,\text{sin} \pi \D_k \prod_{j=5}^n i^{\D_j+1} \Gamma\left(\D_j-1\right),
\end{align}
and 
\begin{align}
    \mathcal{I}_{n-4} = \int du_1 \, du_2 \, du_3 \, du_4 \, u_1^{-\D_1-1}\,u_2^{-\D_2-1}\,u_3^{1-\D_3}\,u_4^{1-\D_4}  \prod_{j=5}^n \left(\sum_{J=1}^4 \e_J u_J \frac{\mU_{jJ}}{\mU_{1234}}\right)^{1-\D_j}.
\end{align}
This can be identified as an Aomoto-Gelfand hypergeometric function \cite{book:596104}. It is an integral over 4 variables which naively appears to be different from the $n-4$ fold integral obtained in \cite{Schreiber:2017jsr}. However, this corresponds to the dual representation of the same hypergeometric function. For more details, see the Appendix of \cite{Schreiber:2017jsr} and also \cite{book:596104}. This completes the cross-check of the MHV Carrollian amplitudes. 

\section{UV and IR behaviours of Carrollian amplitudes}
\label{sec:Soft limits and memories}

In this section, we will analyze how Carrollian amplitudes encode the deep UV and IR behaviours of scattering amplitudes and show that they are reflected in the singularity structure of the Carrollian amplitudes in $u$. This should be compared with the celestial amplitudes whose singularity structure in the space of conformal dimensions encodes the UV and IR behaviours of scattering amplitudes \cite{Arkani-Hamed:2020gyp}.

Consider first the behaviour of Carrollian amplitudes as $u \to 0$, where $u$ is represents a placeholder for some translation invariant combination of the $u_i$ involved in the amplitude. Examples of these are those appearing in the denominators of \eqref{eq:3ptnewsneg} and \eqref{eq:4ptgluonnews}. In this limit, we expect that the Fourier transform is dominated by the UV behaviour of the corresponding momentum space amplitudes. If the amplitude falls off as $\mA_n \xrightarrow[]{\om \to \infty} \om^{-k}$ for large $\om$, we have 
\begin{align}
\label{eq:FTHE}
    \frac{1}{2\pi}\int_{\Lambda}^{\infty} d\om \, e^{i u \om - \varepsilon \om} \,\om^{-k} \xrightarrow[\Lambda \sim \frac{1}{u}]{u \to 0} \alpha \, u^{k-1}
\end{align}
for some numerical constant $\alpha$ implying that the Carrollian amplitude {\it vanishes} as $u^{k-1}$ as $u \to 0$.  We can also consider amplitudes which are exponentially suppressed in the UV with string amplitudes being the prototypical example. Thus the behaviour $\mA_n \xrightarrow[]{\om \to \infty} e^{-\alpha \om^2}$ yields a Fourier transform
\begin{align}
    \label{eq:FTHEstring}
    \frac{1}{2\pi}\int_{\Lambda}^{\infty} d\om \, e^{i u \om - \alpha \om^2}  \xrightarrow[\Lambda \sim \frac{1}{u}]{u \to 0} \frac{u}{2\alpha} e^{-\frac{\alpha}{u^2}}
\end{align}
which also vanishes exponentially.

In the large $u$ limit, we expect the Fourier transform to be dominated by limit of amplitudes as $\om \to 0$. In momentum space, this limit is controlled by the soft theorems which relate $(n+1)$-point scattering amplitudes with one of the particles being soft, and $n$-point scattering amplitudes without the soft particle. At tree-level, they take the form 
\begin{equation}
    \mathcal{A}_{n+1} \xrightarrow[]{\om_1 \to 0} \left( \frac{1}{\omega_1} S^{(0)} + \omega_1^0 S^{(1)} + \mathcal{O}(\omega_1)  \right) \mathcal{A}_n.
    \label{soft theorem}
\end{equation} 
The above equation has been written for the case of particle 1 becoming soft. For Yang-Mills and QED, the leading and subleading soft theorems are universal \cite{Weinberg:1965nx, Casali:2014xpa, Low:1958sn}, while in gravity, the leading, subleading and sub-subleading soft graviton theorems are universal \cite{Weinberg:1965nx, Cachazo:2014fwa}. More generally, recursion relations show that scattering amplitudes in momentum space admit a decomposition of the form 
\begin{align}
    \mA_{n+1} \left(1, \dots , n\right) = \mA_{n+1}^{c}  \left(1, \dots , n\right)+ \mA_{n+1}^{nc}  \left(1, \dots , n\right)+ \mA_{n+1}^{\infty} \left(1, \dots , n  \right),  
\end{align}
with $\mA_{n+1}^c$ corresponding to the part of the amplitude containing two particle factorization channels, $\mA_{n+1}^{nc}$ to non-collinear or multi-particle channels and $\mA_{n+1}^{\infty}$ to terms at $\infty$. More explicitly,
\begin{align}
    &\mA_{n+1}^c \left(1, \dots , n\right) \sim \mA_3  \left(1,2, -P \right) \frac{1}{(p_1+p_2)^2} \mA_{n-1} \left(P, \dots n\right),\\
    &\nonumber\mA_{n+1}^{nc} \left(1, \dots , n\right) \sim \mA_k  \left(1, \dots , k-1, -P \right) \frac{1}{(p_1+\dots p_{k-1})^2} \mA_{n-1} \left(P, k, \dots , n\right).
\end{align}
We will not elaborate on $\mA_{n+1}^{\infty}$ here and refer the reader to \cite{Cachazo:2014fwa} for more details. Furthermore, the collinear part of the amplitude $\mA_{n+1}^c$ is completely controlled by the soft theorems and exhibits universal soft factorization to all order. For MHV amplitudes, $\mA_{n+1}^{nc} = \mA_{n+1}^{\infty} = 0$ and this factorization extends to the complete amplitude. Thus, we can write  
\begin{equation}
    \mathcal{A}^c_{n+1} = \left( \sum_{k=-\infty}^{1} \omega_1^{-k} S^{(-k)}\right) \mathcal{A}^c_n,
     \label{soft theorem MHV}
\end{equation} where the $S^{(k)}$ do not depend on $\om_1, \dots , \om_n$ but do depend on all the $z_i, \zb_i$. For the rest of this paper, we will focus on MHV amplitudes and drop the superscript $c$ with the understanding that all the statements made here can be extended to the collinear parts of any amplitude beyond MHV. In order to avoid dealing with divergent integrals, we consider correlator of $u-$descendants $\tilde{\mc}$ instead of the Carrollian amplitude $\mc$. We have 
\begin{align}
\label{fourier transform on soft theorem}
   \tilde{\mc}_n \left({u_1, z_1, \zb_1}, \dots , {u_n, z_n, \zb_n}\right) \xrightarrow[]{u_1 \to \infty} &\underset{\om \approx 0}{\int}  \frac{d\om_1}{2\pi}  \left( \, e^{i\e_1 \om_1 u_1}  \sum_{k=-\infty}^{1} \omega_1^{1-k} S^{(-k)}\right)\tilde{\mc}_{n-1} \left({u_2, z_2, \zb_2}, \dots , {u_n, z_n, \zb_n}\right)  \nonumber\\
    =& \frac{1}{2\pi}\, \sum_{k=-\infty}^{1} (-i\epsilon_1)^{k-2} (1-k)! u_1^{k-2} S^{(-k)}\tilde{\mc}_{n-1} \left({u_2, z_2, \zb_2}, \dots , {u_n, z_n, \zb_n}\right) .
\end{align} 
Hence we see that the Carrollian amplitude vanishes as $\frac{1}{u^{2-k}}$ as $u \to \infty$. Note that the exponent is positive as have $k=1, 0, -1, -2, \dots$. A corollary of this is that the soft factor $S^{(-k)}$ can be extracted via the residue integral
\begin{equation}
\label{eq:softresidue}
   S^{(-k)}\left(z_1, \zb_1\right) \tilde{\mc}_{n-1} =  -\frac{(-i\epsilon)^{2-k}}{(1- k)!}\underset{u_1 = \infty}{\oint} du_1 u_1^{-k+1}  \tilde{\mathcal{C}}_{n+1}
\end{equation} where the residue is taken around the pole at infinity. Finally, as discussed in \cite{Donnay:2022sdg}, the soft behaviors of the position space amplitude \eqref{fourier transform on soft theorem} can be directly related to the memory effects in gravity and gauge theories \cite{Strominger:2014pwa,Pasterski:2015zua,1502.06120PSZ}.

\section{Collinear limits and Carrollian OPEs}
\label{sec:Collinear limits and Carrollian OPEs}
In this section, we analyze the collinear limit of amplitudes in position space. The analogous property is well studied in momentum space \cite{Mangano:1990by,Berends:1988zn} and forms the basis for the derivation of celestial OPEs \cite{Fan:2019emx, Himwich:2021dau}. Using the Carrollian holography dictionary, we show that this limit naturally yields a notion of OPE in the putative dual Carrollian CFT. 

\subsection{Derivation of Carrollian OPEs}

We will restrict ourselves to the collinear limit of two outgoing particles, i.e. $\e_1 = \e_2 = +1$. The other cases can be handled similarly. At tree-level, scattering amplitudes have collinear poles and the corresponding residues factorize. This can be written as\footnote{The notion of collinear factorization is well defined in both Minkowski and Klein spacetime. In Minkowski spacetime, where the three point amplitudes involving massless particles vanish, this is defined by considering amplitudes with one leg slightly off shell. We will work in split signature for convenience.}
\begin{align}
    \mA_n \left(1^{J_1}, 2^{J_2}, 3^{J_3}, \dots , n^{J_n} \right) \xrightarrow[]{1 || 2} \sum_{J}\mA_3 \left(1^{J_1}, 2^{J_2}, -P^{-J} \right) \frac{1}{\an{12}\sq{21}} \mA_{n-1} \left(P^{J}, 3^{J_3}, \dots ,n^{J_n}\right) .
\end{align}
Here $J$ corresponds to the helicity of the exchanged particle. In general every massless particle that can be exchanged leads to a collinear pole.
 Since $P$ is a massless momentum, we can write
\begin{equation}
\label{eq:Ppar}
    P =  \om \left(1+z \zb, z + \zb, z - \zb , 1 - z \zb \right).
\end{equation}
A convenient change of variables is $\om_1 = t \om, \om_2 = \left(1-t\right) \om$. We can now  solve $p_1 + p_2 - P = 0$ for $z, \zb$ in terms of $z_1, z_2, \zb_1, \zb_2, t$, giving
\begin{align}
     z = z_1, \qquad\qquad \zb = t \zb_1 + (1-t)\zb_2 .
\end{align}
In this limit, the position space amplitude behaves as
\begin{multline}
    \mc_n  \xrightarrow[]{1 || 2} \frac{1}{4\pi^2} \int_0^{+\infty} d\om \, \int_0^1 dt \,\om \,e^{i\left(t u_1+ (1-t)u_2\right)\om} \sum_{J}\mA_3 \left(1^{J_1}, 2^{J_2}, -P^{-J}\right) \\ \times \frac{1}{\an{12}\sq{21}}\mA_{n-1} \left(P^{J}, 3^{J_3}, \dots , n^{J_n}\right)  e^{i \e_3 u_3 \om_3+ \dots i \e_n u_n \om_n} .
\end{multline}
For the holomorphic collinear limit, $z_{12} \to 0$, the only non-zero three particle amplitudes are the anti-holomorphic ones. Thus, we must have $J_1+J_2-J >0$ and plugging in the appropriate three point amplitude from \eqref{eq:threepoints}, we get
\begin{multline}
\label{eq:carrollcollinear}
    \mc_n  \xrightarrow[]{1 || 2} -\frac{\kappa_{J_1, J_2, -J}}{4\pi^2}\frac{\zb_{12}^{p}}{z_{12}}\int d\om \, \om^{p}  \left[\int dt\,  e^{it\om\left(u_1-u_2\right)} t^{J_2-J-1} (1-t)^{J_1-J-1}\right]\\
    \times e^{i u_2 \om + \dots i \e_n u_n \om_n}  \mA_{n-1} \left(\left\lbrace \om, z_2, \zb_2 + t \zb_{12}\right\rbrace, 3^{J_3}, \dots , n^{J_n}\right) 
\end{multline}
where $p = J_1+J_2-J-1 \geq 0$. In writing the above equations, we have left implicit a sum over all allowed values of $p$. These are deduced by solving the inequalities  
\begin{align}
    \label{eq:prestrictions}
    p\geq 0, \qquad \left|J_1+J_2-p-1\right| \leq  2 \quad \text{and} \quad \left|J_1\right| \leq 2, \qquad \left|J_2\right| \leq 2 . 
\end{align}
The first of these is just taking into account the fact that only anti-holomorphic three point amplitudes  contribute to the holomorphic collinear limit, while the remaining ones remind us to neglect massless higher spins. The right-hand side of \eqref{eq:carrollcollinear} depends non trivially on $\zb_1, \zb_2, u_1$ and $u_2$. This reflects the fact that we have considered the holomorphic collinear limit and left the remaining variables at generic values. In this case, following the Carrollian holography dictionary \eqref{carrollian identification}, we can write
\begin{align}
    \label{eq:carrollOPEblock}
    &\Phi_{J_1} \left(u_1, z_1, \zb_1 \right) \Phi_{J_2} \left(u_2, z_2, \zb_2 \right) \\
    &\nonumber \qquad\qquad \sim  -\frac{\kappa_{J_1, J_2, -J}}{4\pi^2}\frac{\zb_{12}^{p}}{z_{12}} \int d\om \, \om^{p}  \left[\int dt\,  e^{it\om\left(u_1-u_2\right)} t^{J_2-J-1} (1-t)^{J_1-J-1}\right]  \\
    &\nonumber \hspace{9cm} \times e^{i u_2 \om }\,  \Phi_{J}\left(\om, z_2, \zb_2 + t \zb_{12}\right) .
\end{align}
Here we have denoted the boundary operator corresponding to a helicity  $J$ particle in the bulk by $\Phi_{(k,\bar k)}(u,z, \zb) \equiv \Phi_J(u,z, \zb)$ with the understanding that they have Carrollian weights $(k,\bar k) = ( \tfrac{1+J}{2}, \tfrac{1-J}{2})$ as in \eqref{fiexed Carrollian weights}. This expression constitutes a \textit{Carrollian OPE block}, which is the analogue of a conformal OPE block in CFT \cite{Ferrara:1971vh}. From here, we can obtain three seemingly different but equivalent formulas for the OPE. Each of them has the benefit of making certain features of the OPE manifest and we will present all three of them.
\paragraph{Formula 1.} This is the most compact expression and includes all the $u-$ and $\zb-$ descendants in one integral formula. To obtain this, we perform the $\om$ integral by making the identification 
\begin{equation}
    \frac{1}{2\pi}\int d\om \, \om^{p} e^{i \left(u_2 + t u_{12} \right) \,\om }\,  \Phi_{J}\left(\om, z_2, \zb_2+t \zb_{12}\right)  = \partial_u^{p}\Phi_{J}\left(u, z_2, \zb_2+t \zb_{12}\right)\rvert_{u = u_2 + t u_{12}}
\end{equation}
with $u_{12} \equiv u_1 - u_2$ and land on 
\begin{align}
    \label{eq:carrollOPEformula1}
    &\Phi_{J_1} \left(u_1, z_1, \zb_1 \right) \Phi_{J_2} \left(u_2, z_2, \zb_2 \right) \\
    &\nonumber \qquad\qquad \sim  -\frac{\kappa_{J_1, J_2, -J}}{2\pi}\frac{\zb_{12}^{p}}{z_{12}} \int_0^1 dt\,  t^{J_2-J-1} (1-t)^{J_1-J-1} \,  \left(\frac{\partial}{\partial u}\right)^{p}\Phi_{J}\left(u, z_2,\zb_2 +t \zb_{12}\right)\rvert_{u = u_2 + t u_{12}}.
\end{align}
The integral compactly encodes the contributions from all descendants and represents a $u, \zb$ Carrollian OPE block. This also closely resembles the celestial OPE block \cite{Guevara:2021abz}. The explicit relation between Carrollian and celestial OPE blocks will be established in Section \ref{sec:OPEFrom Carrollian to celestial}.

\paragraph{Formula 2.} This formula makes the $u$ OPE block explicit. To derive this, we first Taylor expand
\begin{equation}
    \label{eq:zbarexp}
   \Phi_{J}\left(\om, z_2, \zb_2 + t \zb_{12}\right)= \sum_{m=0}^{\infty} \frac{t^m \zb_{12}^m}{m!} \left(\frac{\partial}{\partial \zb_2}\right)^m \Phi_{J}\left(\om, z_2, \zb_2 \right)
\end{equation}
in \eqref{eq:carrollOPEblock}, and then perform the integral over $t$:
\begin{multline}
\label{eq:collinearlimitint}
    \int dt\,  e^{it\left(u_1-u_2\right)\om} t^{J_2-J+m-1} (1-t)^{J_1-J-1} = B(J_2-J+m, J_1-J) \\
    \hypfo \left(J_2-J+m, J_1+J_2-2J+m, i \left(u_1 - u_2\right)\om \right).
\end{multline}
Here $B(x,y)$ is the Euler Beta function and $\hypfo (a,b, z)$ is the confluent hypergeometric function which admits the power series expansion 
\begin{multline}
\label{eq:hypfexp}
    B(J_2-J+m, J_1-J) \hypfo \left(J_2-J+m, J_1+J_2-2J+m, i \left(u_1 - u_2\right)\om \right) \xrightarrow[]{u_1 \to u_2}  \\ \sum_{n=0}^{\infty} 
    \frac{\left(iu_{12}\right)^n \om^n}{n!} B(J_1-J,J_2-J+m+n) .
\end{multline}
Noting that the hypergeometric function only has positive integer valued powers of $\om$, we deduce the formal expression for the OPE
\begin{align}
\label{eq:carrollOPEformula2}
   &\Phi_{J_1} \left(u_1, z_1, \zb_1 \right) \Phi_{J_2} \left(u_2, z_2, \zb_2 \right) \sim   -\frac{\kappa_{J_1, J_2, -J}}{2\pi z_{12}}\sum_{m=0}^{\infty} \frac{\zb_{12}^{p+m}}{m!} B(J_2-J+m, J_1-J)  \frac{\partial^m \, \mathcal{F} \left[\Phi_J \left(u_2, z_2, \zb_2 \right)\right] }{\partial \zb_2^m} \,  
\end{align}
where
\begin{align}
   & \mathcal{F} \left[\Phi_J \left(u_2, z_2, \zb_2 \right)\right]  = \hypfo \left(J_2-J+m, J_1+J_2-2J+m, \left(u_1 - u_2\right)\frac{\partial}{\partial u_2}\right) \nonumber \left(\frac{\partial}{\partial u_2}\right)^{p}\Phi_{J}\left(u_2, z_2, \zb_2 \right).
\end{align}
This function represents the $u$ Carrollian OPE block. The formula also includes an explicit infinite sum over $\zb$ descendants.

\paragraph{Formula 3.} Finally, we can make the presence of both the $u$- and $\zb$-descendants explicit simply by further expanding the hypergeometric function in \eqref{eq:carrollOPEformula2} using \eqref{eq:hypfexp} to arrive at:
\begin{align}
\label{eq:carrollOPEformula3}
   &\Phi_{J_1} \left(u_1, z_1, \zb_1 \right) \Phi_{J_2} \left(u_2, z_2, \zb_2 \right) \\
   &\qquad\sim -\frac{\kappa_{J_1, J_2, -J}}{2\pi z_{12}} \sum_{m,n=0}^{\infty} B(J_2-J+m+n, J_1-J) \frac{\zb_{12}^{p+m}u_{12}^n}{m!n!}\left(\frac{\partial}{\partial \zb_2}\right)^m\left(\frac{\partial}{\partial u_2}\right)^{p+n}\Phi_{J}\left(u_2, z_2, \zb_2 \right). \nonumber
\end{align}
The following formula for the OPE of Carrollian descendants is readily obtained by making the replacements $J_1 \to J_1 + 1, J_2 \to J_2 + 1, J \to J+1$ in the OPE coefficient and the exponent of the $u_2$ derivative: 
\begin{multline}
    \label{eq:OPEnews}
    \partial_{u_1} \Phi_{J_1} (u_1, z_1, \zb_1)  \partial_{u_2} \Phi_{J_2} (u_2 , z_2, \zb_2) \sim -\frac{\kappa_{J_1, J_2, -J}}{2\pi z_{12}}\sum_{m,n=0}^{\infty} B(J_2-J+m+n+1, J_1-J+1) \frac{\zb_{12}^{p+m}u_{12}^n}{m!n!}\\
    \left(\frac{\partial}{\partial \zb_2}\right)^m\left(\frac{\partial}{\partial u_2}\right)^{p+n+1}  \partial_{u_2} \Phi_J (u_2 , z_2, \zb_2).
\end{multline}

\paragraph{Collinear limit of MHV gluon amplitude} As a check of our formulae, we will compute the collinear limit of the $7$ point gluon MHV amplitude \eqref{eq:nptgluonpos1} in the special kinematic configuration corresponding to \eqref{eq:Inspecial} and thereby directly verify the OPE \eqref{eq:OPEnews}.\footnote{The Beta function appearing in the OPE coefficient in \eqref{eq:carrollOPEformula3} is divergent for gluons and we consider instead the OPE of the Carrollian descendants \eqref{eq:OPEnews}.} The OPE of two positive helicity gluons in Yang-Mills has $p = 0$, which yields
\begin{multline}
    \label{eq:OPEgluon}
    \partial_{u_1} \Phi_{+1} (u_1,z_1, \zb_1)  \partial_{u_2} \Phi_{+1} (u_2, z_2, \zb_2) \sim -\frac{\kappa_{1, 1,-1}}{2\pi z_{12}}\sum_{m,n=0}^{\infty} \frac{1}{m+n+1} \frac{\zb_{12}^{m}u_{12}^n}{m!n!}\\
    \left(\frac{\partial}{\partial \zb_2}\right)^m\left(\frac{\partial}{\partial u_2}\right)^{n+1} \partial_{u_2}\Phi_{+1}\left(z_2, \zb_2, u_2\right).
\end{multline}
We will verify this directly from the 7 point MHV gluon amplitude (in the special kinematic configuration \eqref{eq:exampler2} for which we had an explicit expression for generic $n$):
\begin{align}
\label{eq:MHVgluon7pt}
    \tilde{\mc}_7\left(1^{-}, 2^{-}, 3^{+}, \dots , 7^{+} \right) = -\frac{\kappa_{1,1,-1}^{5}}{(2\pi)^7\left|\mU_{1234}\right|}\frac{z_{12}^3}{\prod_{j=2}^7 z_{jj+1}}\frac{\partial^4}{\partial u_1^2\, \partial u_2^2}\left[\frac{1}{L_5 L_6 L_7}\right]
\end{align}
where $L_j = \left(\e_j u_j - \sum_{J=1}^4 \e_J u_J\frac{\mU_{Jj}}{\mU_{1234}} \right) $. For convenience, consider the true collinear limit $z_6 \to z_7, \zb_6 \to \zb_7$ of \eqref{eq:MHVgluon7pt} in which case the OPE \eqref{eq:OPEgluon} implies
\begin{align}
\label{eq:colllimit7pt}
    \tilde{\mc}_7 \xrightarrow[z_{67} \to 0]{\zb_{67} \to 0} -\frac{\kappa_{1, 1,-1}}{2\pi z_{67}} \frac{1}{(n+1)!} \left(\frac{\partial}{\partial u_2}\right)^{n+1} \tilde{\mc}_6.
\end{align}
To verify this, we start by noting that $L_6$ is the only relevant quantity that has a non-trivial behaviour in this limit. Setting $\e_6 = \e_7 = 1$, we find
\begin{align}
    L_6 &= \,  u_6  - \sum_{J=1}^4 u_J \e_J \, \frac{\mU_{J6}}{\mU_{1234}}   \xrightarrow[z_{67} \to 0]{\zb_{67} \to 0}  u_6 - \sum_{J=1}^4 u_J \e_J \, \frac{\mU_{J7}}{\mU_{1234}} = L_7 + u_{67}.
\end{align}
The fact that $\mU_{J6}   \xrightarrow[z_{67} \to 0]{\zb_{67} \to 0}  \mU_{J7}$ follows from the definition \eqref{eq:Uformula}. The final step in verifying \eqref{eq:colllimit7pt} is to note that we can write
\begin{equation}
    \frac{1}{L_6 L_7} \xrightarrow[z_{67} \to 0]{\zb_{67} \to 0}  \sum_{n=0}^{\infty} \frac{\left(-1\right)^n}{L_7^2} \left(\frac{u_{67}}{L_7}\right)^n = - \sum_{n=0}^{\infty} \frac{1}{(n+1)!}\frac{\partial^{n+1}}{\partial_{u_7}^{n+1}} \frac{1}{L_7}.
\end{equation}
This completes the check of \eqref{eq:colllimit7pt}.

\subsection{Consistency of Carrollian OPEs with symmetries}

Defining OPEs in a Carrollian CFT is a delicate issue due to the ultra-local nature of the theory. We show here that the Carrollian OPEs derived above are compatible with the global conformal Carrollian symmetries generated by \eqref{BMS vectors} with parameters in \eqref{Poincaré}, and hence are in agreement with ultra-locality. For convenience, we will check the invariance on the formula 1 given in \eqref{eq:carrollOPEformula1} under infinitesimal transformations generated by $\mathcal{T}, \mathcal{Y}$ and $ \bar{\mathcal{Y}}$ separately. 
\paragraph{Invariance under $\mathcal{T}$} The action on the left-hand side (LHS) of \eqref{eq:carrollOPEformula1} is
\begin{equation}
\begin{split}
    &\delta_{\mathcal{T}} (\text{LHS}) \\
    &= \mathcal{T}(z_1,\bar z_{1}) \partial_{u_1} \Phi_{J_1} \left(u_1, z_1, \zb_1 \right) \Phi_{J_2} \left(u_2, z_2, \zb_2 \right)  + \Phi_{J_1} \left(u_1, z_1, \zb_1 \right) 
 \mathcal{T}(z_2,\bar z_{2}) \partial_{u_2} \Phi_{J_2} \left(u_2, z_2, \zb_2 \right) \\
    &\sim  -\frac{\kappa_{J_1, J_2, -J}}{2\pi}\frac{\zb_{12}^{p}}{z_{12}} \int_0^1 dt\,  t^{J_2-J-1} (1-t)^{J_1-J-1} [\mathcal{T}(z_2,\bar z_{1}) t + \mathcal{T}(z_2,\bar z_{2}) (1-t) ] \\
    &\qquad\qquad\qquad\qquad\qquad\qquad\qquad\qquad\qquad  \times\partial_u^{p+1}\Phi_{J}\left(u, z_2,\zb_2 +t \zb_{12}\right)\rvert_{u = u_2 + t u_{12}}
    \label{computation step1}
\end{split}
\end{equation} where we used successively \eqref{carrollian primary}, \eqref{eq:carrollOPEformula1} and the identification of $z_1$ with $z_2$ at leading order. Now acting directly on the right-hand side (RHS) of \eqref{eq:carrollOPEformula1} leads to
\begin{align}
        \delta_{\mathcal{T}} (\text{RHS}) &= -\frac{\kappa_{J_1, J_2, -J}}{2\pi}\frac{\zb_{12}^{p}}{z_{12}} \int_0^1 dt\,  t^{J_2-J-1} (1-t)^{J_1-J-1} \, \partial_u^{p} \delta_{\mathcal{T}}\Phi_{J}\left(u, z_2,\zb_2 +t \zb_{12}\right)\rvert_{u = u_2 + t u_{12}} \label{computation step2} \\
        &= -\frac{\kappa_{J_1, J_2, -J}}{2\pi}\frac{\zb_{12}^{p}}{z_{12}} \int_0^1 dt\,  t^{J_2-J-1} (1-t)^{J_1-J-1} \,  \mathcal{T} (z_2, \bar z_2 + t \bar{z}_{12})\partial_u^{p+1} \Phi_{J}\left(u, z_2,\zb_2 +t \zb_{12}\right)\rvert_{u = u_2 + t u_{12}} . \nonumber
\end{align} For each $\mathcal{T}(z,\bar z)=1,z,\bar z, z \bar z$, one can check explicitly that \eqref{computation step1} coincides with \eqref{computation step2}, which shows the invariance of the Carrollian OPEs under translation. 
\paragraph{Invariance under $\mathcal{Y}$} This check is very similar to the previous one and here we merely present the result, drawing attention only to the subtler points. We have
\begin{align}
        &\delta_{\mathcal{Y}} (\text{LHS}) =  -\frac{\kappa_{J_1, J_2, -J}}{2\pi}\frac{\zb_{12}^{p}}{z_{12}} \int_0^1 dt\,  t^{J_2-J-1} (1-t)^{J_1-J-1}  \mathcal{D} \left[\partial_{u}^P \Phi_J \left(u, z_2, \zb\right) \right],
\end{align}
to leading order in $z_2$, with
\begin{align}
   \mathcal{D} &= \frac{u}{2} \partial_{z_2}\mathcal{Y}(z_2)\partial_{u} + (k_1+k_2) \partial_{z_2}\mathcal{Y}(z_2) + \frac{1}{z_{12}}\left(\mathcal{Y}(z_2)-\mathcal{Y}(z_1) \right)+ \mathcal{Y}(z_2) \partial_{z_2}, \nonumber \\ 
   &= \frac{u}{2} \partial_{z_2}\mathcal{Y}(z_2)\partial_{u} + (k_1+k_2-1) \partial_{z_2}\mathcal{Y}(z_2)+ \mathcal{Y}(z_2) \partial_{z_2}.
\end{align}
In arriving at this result, we have made use of the following relations:
\begin{align}
    \partial_{u_1}\Phi = t \partial_{u} \Phi, \qquad   \partial_{u_2}\Phi = (1-t) \partial_{u} \Phi, \qquad \text{where} \qquad u = t u_1 + (1-t) u_2 .
\end{align}
It is straightforward to check that this is identical to the action of $\delta_{\mathcal{Y}}$ on the RHS since the weight of $\partial_u^p \Phi_J$ is $k = \frac{p}{2}+ \frac{1+J_1+J_2-p-1}{2} = k_1+k_2-1$.
\paragraph{Invariance under $\bar{\mathcal{Y}}$} This check is more involved since there is no implied limit on the $\zb_i$. We will also need to apply integration by parts identities in order to see the invariance. With this in mind, we compute
\begin{align}
\label{eq:ybarcheck1}
        &\delta_{\my} (\text{LHS}) - \delta_{\my} (\text{RHS}) =  -\frac{\kappa_{J_1, J_2, -J}}{2\pi}\frac{\zb_{12}^{p}}{z_{12}} \int_0^1 dt\,  t^{J_2-J-1} (1-t)^{J_1-J-1}  \bar{\mathcal{D}} \left[\partial_{u}^P \Phi_J \left(u, z_2, \zb\right) \right],
\end{align}
where
\begin{align}
  \bar{\mathcal{D}} &= \left[t\frac{u_1}{2} \left(\partial_{\zb_1}\my(\zb_1) - \partial_{\zb}\my(\zb)\right)+(1-t)\frac{u_2}{2} \left(\partial_{\zb_2}\my(\zb_2) - \partial_{\zb}\my(\zb) \right)\right] \partial_{u}+ \left[t \my \left(\zb_1 \right) + (1-t) \my  \left(\zb_2 \right) - \my  \left(\zb \right) \right] \partial_{\zb} \nonumber\\
  &\quad + \frac{1}{\zb_{12}}\left[\zb_{12} \bar{k}_1 \left(\partial_{\zb_1}\my(\zb_1) - \partial_{\zb}\my(\zb)\right) + \bar{k}_2  \left(\partial_{\zb_2}\my(\zb_2) - \partial_{\zb}\my(\zb) \right)+ p \left(\my(\zb_1)-\my(\zb_2) - \zb_{12} \partial_{\zb} \my \right)\right].
\end{align}
This manifestly vanishes for $\my = 1, \zb$ while for $\my = \zb^2$, we must verify that it integrates to zero. To this end, we note the following relations:
\begin{equation}
    \partial_u \Phi_J \left(u, z_2, \zb \right) = \frac{1}{u_{12}}  \partial_t \Phi_J \left(u, z_2, \zb \right),  \qquad \partial_{\zb} \Phi_J \left(u, z_2, \zb \right) = \frac{1}{\zb_{12}}  \partial_t \Phi_J \left(u, z_2, \zb \right).
\end{equation}
Using these, integrating by parts and simplifying the resulting expression yields
\begin{align}
    \delta_{\my} &(\text{LHS}) - \delta_{\my} (\text{RHS}) \\
    &=  \frac{\kappa_{J_1, J_2, -J}}{2\pi}\frac{\zb_{12}^{p+1}}{z_{12}} \int_0^1 dt\,  t^{J_2-J-1} (1-t)^{J_1-J-1} \left(t (J_1-J)-(1-t)(J_2-J)\right) = 0. \nonumber
\end{align}
Note that this result depends crucially on the precise value of the exponents. This suggests that the OPE coefficient can be fixed by symmetries in a manner analogous to \cite{Pate:2019lpp}. 
\subsection{From Carrollian to celestial}
\label{sec:OPEFrom Carrollian to celestial}

Finally, as a consistency check, we compare Carrollian and celestial OPEs. It is easier to start from \eqref{eq:carrollOPEformula1} and apply \eqref{Btransform1} to get
\begin{align}
    &\mo_{\D_1, J_1} \left(z_1, \zb_1 \right)\mo_{\D_2, J_2} \left(z_2, \zb_2 \right) \sim  -\frac{\kappa_{J_1, J_2, -J}}{2\pi}\frac{\zb_{12}^{p}}{z_{12}} (-i)^{\D_1+\D_2} \Gamma (\D_1) \Gamma(\D_2) \int_{-\infty}^{\infty} du_1 \, du_2 \, u_1^{-\D_1} \, u_2^{-\D_2} \\
    &\nonumber \hspace{5.5cm} \times\int_0^1 dt\,  t^{J_2-J-1} (1-t)^{J_1-J-1} \,  \left(\frac{\partial}{\partial u_2}\right)^{p}\Phi_{J}\left(u_2 + t u_{12}, z_2, \zb_2 + t \zb_{12}\right) . 
\end{align}
As before, we deform the $u_i$, pick up the discontinuity across the branch cut on the negative $u_i$ axes and reduce them to integrals over the positive real line. We further define $\tu_1 = t u_1, \tu_2 = (1-t)u_2$ and are led to
\begin{align}
    &\mo_{\D_1, J_1} \left(z_1, \zb_1 \right)\mo_{\D_2, J_2} \left(z_2, \zb_2 \right) \sim  \frac{2\pi(-i)^{\D_1+\D_2}\kappa_{J_1, J_2, -J}}{\Gamma (1-\D_1) \Gamma(1-\D_2) }\frac{\zb_{12}^{p}}{z_{12}} \int_0^1 dt\,  t^{2\hb_1+p-1} (1-t)^{2\hb_2-1} \,  \\
    &\nonumber \hspace{5.5cm} (-1)^p\times\int_{0}^{\infty} d\tu_1 \, d\tu_2 \, \tu_1^{-\D_1} \, \tu_2^{-\D_2} \left(\frac{\partial}{\partial \tu_2}\right)^{p}\Phi_{J}\left(\tu_1+\tu_2, z_2, \zb_2 + t \zb_{12}\right) 
\end{align}
where $(h, \bar{h}) = \left(\frac{\D+J}{2},\frac{\D-J}{2}\right)$ are the conformal weights. Integrating the $u_2$ derivative by parts and then making a final change of variables $\tu_1 = x u, \tu_2 = (1-x) u$ allows us to factorize the integrals. The integral over $x$ is then readily performed and the expression simplifies considerably to 
\begin{align}
    &\mo_{\D_1, J_1} \left(z_1, \zb_1 \right)\mo_{\D_2, J_2} \left(z_2, \zb_2 \right) \sim  -\kappa_{J_1, J_2, -J}(-1)^p\frac{\text{sin} \pi(\D_2+p)}{\text{sin} \pi\D_2 }\frac{\zb_{12}^{p}}{z_{12}}  \int_0^1 dt\,  t^{2\hb_1+p-1} (1-t)^{2\hb_2+p-1}\\
    & \hspace{5.5cm}\frac{2\pi(-i)^{\D_1+\D_2+1}}{\Gamma(2-\D_1-\D_2-p)} \, \int_0^{\infty} du u^{1-\D_1-\D_2-p} \Phi_{J}\left(\tu_1 + \tu_2, z_2, \zb_2 + t \zb_{12}\right) .\nonumber 
\end{align}
Identifying the expression on the final line with the Celestial operator (after a contour deformation) and noting that $\frac{\text{sin} \pi(\D_2+p)}{\text{sin} \pi\D_2 } = (-1)^p$, we arrive at the celestial OPE block \cite{Fan:2019emx, Himwich:2021dau}
\begin{align}
    &\mo_{\D_1, J_1} \left(z_1, \zb_1 \right)\mo_{\D_2, J_2} \left(z_2, \zb_2 \right) \sim  -\kappa_{J_1, J_2, -J}\frac{\zb_{12}^{p}}{z_{12}}  \int_0^1 dt\,  t^{2\hb_1+p-1} (1-t)^{2\hb_2+p-1}\mo_{\D_1+\D_2+p-1, J}.
\end{align}

\section{Soft symmetries of Carrollian amplitudes}
\label{sec:Celestial symmetries}

As emphasized in the introduction, one of the merits of celestial amplitudes is that they allow to render the soft symmetries manifest thanks to the OPE structure \cite{Guevara:2021abz,Strominger:2021mtt, Mago:2021wje, Ren:2022sws, Bhardwaj:2022anh, Jiang:2021ovh, Drozdov:2023qoy, Ahn:2021erj, Ahn:2022oor}. Understanding the action of these soft or celestial symmetries from the Carrollian point of view at null infinity is a legitimate and well-posed question, which we address in this section. In order to determine how the soft symmetries act on Carrollian correlators, we adopt the following definition for the (outgoing) soft operators:
\begin{align}
    \label{eq:softoperatordef}
    H^{k}_J \equiv \lim_{\D \to k} (\D-k) \Gamma (\D-1) (-i)^{\D} \int_{-\infty}^{\infty} du \, u^{1-\D} \partial_{u} \Phi_J \left(u, z, \zb \right), \qquad k = 1,0,-1, -2, \ldots
\end{align}
We will use this definition to first compute the OPE of one celestial and one Carrollian operator and then take the relevant limit.  Starting from the integral representation of the Carrollian OPEs \eqref{eq:carrollOPEblock}, deriving with respect to $u_1$ and $u_2$, and applying \eqref{Btransfom2} on the first operator, we get:
\begin{align}
\label{eq:cel-carrOPE}
    &\mo_{\D_1, J_1} \left(z_1, \zb_1\right)\partial_{u_2} \Phi_{J_2} \left(u_2,z_2, \zb_2\right)\\
    \nonumber &\qquad \sim -\frac{\kappa_{J_1, J_2, -J}}{2\pi}\frac{\zb_{12}^p}{z_{12}}\int_0^{\infty} d\om \, \om^{p+\D_1}
\int_0^1 dt \, e^{i \om (1-t)u_2} t^{J_2-J+\D_1-2} (1-t)^{J_1-J}  \mo_J \left(\om, z_2, \zb_2 + t \zb_{12} \right)\\ 
& \qquad \nonumber = -\frac{\kappa_{J_1, J_2, -J}}{2\pi\, z_{12}}\sum_0^{\infty} \frac{\zb_{12}^{p+m}}{m!}B(\D_1+J_2-J-1+m,J_1-J+1) \int_0^{\infty} d\om \, \om^{p+\D_1} \\
    \nonumber &\hspace{2cm} \times \hypfo \left(\D_1+J_2-J-1+m,\D_1+p-J+1+m, -i u_2 \om \right) \, \partial_{\zb_2}^m \mo_J\left(\om, z_2, \zb_2\right)e^{i u_2 \om} .
\end{align}
This master formula allows us to compute the action of soft currents on various Carrollian operators. While it might be possible to work out a general formula for arbitrary $J_1, J_2$ and $J$, in this paper, we will restrict ourselves to three cases of interest -- the action of soft graviton currents on Carrollian gluon and graviton operators and the action of soft gluon currents on Carrollian gluon operators.  The $w_{1+\infty}$ and $S$ algebra currents are related to the soft graviton and soft gluon currents via light-ray transforms \cite{Strominger:2021lvk, Himwich:2021dau}. We analyze each of these cases below. 

\paragraph{Action of $w_{1+\infty}$ currents on gravitons} Let us now consider the first case of interest -- $w_{1+\infty}$ currents acting on Carrollian graviton operators. To this end, we first derive the action of soft graviton currents. Setting $J_1 = J_2 = J = 2$ in \eqref{eq:cel-carrOPE} yields
\begin{multline}
\label{eq:celgravcarrgravOPE}
    \mo_{\D_1, 2} \left(z_1, \zb_1\right)\partial_{u_2} \Phi_{2} \left(u_2,z_2, \zb_2\right) \sim  -\frac{\kappa_{2,2,-2}}{2\pi}\frac{\zb_{12}}{z_{12}} \sum_0^{\infty} \frac{\zb_{12}^m}{m!} B(\D_1-1+m,1)\\
    \int_0^{\infty} d\om \, \om^{\D_1+1} \hypfo \left(\D_1+m-1,\D_1+m, -i u_2 \om \right) \, \partial_{\zb_2}^m \mo_2\left(\om, z_2, \zb_2\right)e^{i u_2 \om} .
\end{multline}
Taking the limit \eqref{eq:softoperatordef} on the first operator gives the following action of the soft graviton operator on the Carrollian graviton operator:
\begin{equation}
    \label{eq:softgrav-gravOPE}
    H^k_2 \left(z_1, \zb_1\right) \partial_{u_2}\Phi_{2} \left(u_2,z_2, \zb_2\right) \sim  -\frac{\kappa_{2,2,-2}}{z_{12}}\sum_{m=0}^{1-k} \frac{\zb_{12}^{m+1}}{m!} \frac{(-i u_2)^{1-k-m}}{(1-k-m)!} \left(-i \partial_{u_2}\right)^{2-m}\partial_{\zb_2}^m \Phi_{2} \left(u_2, z_2, \zb_2\right).
\end{equation}
Note that this action is local only for $k \geq -1$, corresponding to the universal soft theorems. In arriving at the above formula, we have made use of the limit
\begin{equation}
    \label{eq:limitformula1}
    \lim_{\D_1 \to k} \left(\D_1-k\right)\hypfo \left(\D_1+m-1,\D_1+m, -i u_2 \om \right)B(\D_1-1+m,1) = \begin{cases}
        &\frac{\left(-i u_2 \om\right)^{1-k-m}}{(1-k-m)!} \qquad m \leq 1-k \\
        &0  \hspace{2.6cm} m > 1-k .
    \end{cases}
\end{equation}
We can define the $w_{1+\infty}$ currents corresponding to soft currents by slightly modifying the definition in \cite{Himwich:2021dau}\footnote{Alternatively, we could apply the unmodified definition to \eqref{eq:celgravcarrgravOPE}.} to 
\begin{align}
\label{eq:wcurrentdef}
    w^q \left(z, \zb \right) = \lim_{\varepsilon \to 0} \frac{(-1)^{2q}\Gamma(2q)}{2\pi i\, \varepsilon} \int_{-\infty}^{\infty} \frac{d \bar{w}}{(\zb - \bar{w})^{2q}} \mo_{4-2q+\varepsilon, 2} \left(z, \bar{w} \right),  
\end{align}
where $q = \frac{4-k}{2}$. Note that $\mo_{4-2q+\varepsilon, 2} \left(z, \bar{w} \right) \xrightarrow[]{\varepsilon \to 0} \frac{1}{\varepsilon} H^{k}_2 $. Expanding the current in modes as\footnote{The truncated mode expansion can be motivated from various perspectives for currents that enter the collinear part of the amplitudes, see \cite{Banerjee:2019aoy, Banerjee:2019tam, Guevara:2021abz, Pasterski:2021dqe}.} 
\begin{align}
\label{eq:wmodeexp}
      w^q \left(z, \zb \right) = \sum_{m=q-1}^{1-q} \frac{1}{\zb^{m+1-q}} w^q_m, \qquad q = \frac{3}{2}, 2, \frac{5}{2}, \dots 
\end{align}
it has been shown that these modes satisfy the $w_{1+\infty}$ algebra \cite{Strominger:2021lvk},
\begin{align}
    \label{eq:wlagebra}
    \left[w^p_m, w^q_n \right] = \left(m(q-1)-n(p-1) \right)w^{p+q-2}_{m+n}.
\end{align}
Implementing this in \eqref{eq:softgrav-gravOPE} gives the $w_{1+\infty}$ action on Carrollian graviton operators:
\begin{multline}
    \label{eq:softgrav-gravOPEw}
    w^q \left(z_1, \zb_1\right) \partial_{u_2}\Phi_{2} \left(u_2,z_2, \zb_2\right) \\
    \sim  -\frac{\kappa_{2,2,-2}}{z_{12}} \sum_{m=0}^{2q-3} (-1)^m (m+1)\zb_{12}^{m+2-2q} (-i u_2)^{2q-3-m}(-i \partial_{u_2})^{2-m}\partial_{\zb_2}^m \Phi_{2} \left(u_2, z_2, \zb_2\right).
\end{multline}
Finally, as a consistency check, we will obtain the action of the $w_{1+\infty}$ currents on celestial graviton operators. This is done by applying \eqref{Btransfom2} to the second operator in \eqref{eq:softgrav-gravOPEw} to get  
\begin{align}
\label{eq:softactioncheck1}
    w^q \left(z_1, \zb_1\right) \mo_{\D_2,2} \left(z_2, \zb_2\right) \sim  -\frac{\kappa_{2,2,-2}}{z_{12}}& \sum_{m=0}^{2q-3} (-1)^m (-i)^{\D_2+2q-1-2m} \Gamma(\D_2-1) (m+1)\zb_{12}^{m+2-2q}\\
    &\nonumber\qquad \times  \int_{-\infty}^{\infty} du_2 \, (u_2)^{2q-2-m-\D_2}(\partial_{u_2})^{2-m}\partial_{\zb_2}^m \Phi_{2} \left(u_2, z_2, \zb_2\right).
\end{align}
Integrating by parts and identifying the celestial graviton operator, we find
\begin{align}
    \label{eq:softgrav-gravOPEw-celestial}
    & w^q \left(z_1, \zb_1\right) \mo_{\D_2,2} \left(z_2, \zb_2\right) \sim  -\frac{\kappa_{2,2,-2}}{z_{12}}\sum_{m=0}^{2q-3}\frac{(m+1)\Gamma(\D_2-1)}{\Gamma \left(2+\D_2+m-2q \right)} \zb_{12}^{m+2-2q} \mo_{\D_2+4-2q,2}\left(z_2, \zb_2\right),
\end{align}
which is in agreement with the result of \cite{Himwich:2021dau}. 
\paragraph{Action of $w_{1+\infty}$ currents on gluons} A similar procedure can be used to obtain the action of $w_{1+\infty}$ currents on Carrollian gluon operators. Setting $J_1 = 2, J_2 = 1, J = 1$ in \eqref{eq:cel-carrOPE} and suppressing colour indices and the structure constants, we get
\begin{align}
\label{eq:cel-carrOPE2}
    \mo_{\D_1,2} \left(z_1, \zb_1\right)\partial_{u_2} \Phi_{1} \left(u_2,z_2, \zb_2\right) &\sim -\frac{\kappa_{2, 1, -1}}{2\pi}\frac{\zb_{12}}{z_{12}} \sum_0^{\infty} \frac{\zb_{12}^m}{m!}B(\D_1-1+m,2) \int_0^{\infty} d\om \, \om^{1+\D_1} \\
    \nonumber &\hspace{1cm}\hypfo \left(\D_1-1+m,\D_1+1+m, -i u_2 \om \right) \, \partial_{\zb_2}^m \mo_1\left(\om, z_2, \zb_2\right)e^{i u_2 \om} .
\end{align}
In order to proceed here, we need the formula
\begin{multline}
    \label{eq:limitformula2}
    \lim_{\D_1 \to k} \left(\D_1-k\right)\hypfo \left(\D_1+m,\D_1+1+m, -i u_2 \om \right)B(\D_1-1+m,2) \\
    = \begin{cases}
        &-\frac{\left(-i u_2 \om\right)^{-k-m}}{(-k-m)!} + \frac{\left(-i u_2 \om\right)^{1-k-m}}{(1-k-m)!} \hspace{1.65cm} m \leq 1-k \\
        &0  \hspace{6.35cm} m > 1-k 
    \end{cases}
\end{multline}
which implies the OPE
\begin{align}
    \label{eq:softgrav-gluonOPE}
    &H^k_2 \left(z_1 ,\zb_1\right) \partial_{u_2}\Phi_1 \left(u_2,z_2, \zb_2\right) \\
    &\nonumber \qquad \sim  -\frac{\kappa_{2,1,-1}}{z_{12}} \sum_{m=0}^{1-k} \frac{\zb_{12}^{m+1}}{m!} \left[-\frac{(-iu_2)^{-k-m}}{(-k-m)!} 
   (-i\partial_{u_2})^{1-m} + \frac{(-i u_2)^{1-k-m}}{(1-k-m)!} (-i\partial_{u_2})^{2-m}  \right] \partial_{\zb_2}^m \Phi_1 \left(u_2, z_2, \zb_2\right).
\end{align}
The action of $w_{1+\infty}$ on the Carrollian gluon operator, derived by following the same procedure as in \eqref{eq:wcurrentdef} and \eqref{eq:softgrav-gravOPEw} is 
\begin{align}
    \label{eq:softgrav-gluonOPEw}
    w^{q} \left(z_1 ,\zb_1\right) \partial_{u_2}\Phi_1 \left(u_2,z_2, \zb_2\right) \sim  -\frac{\kappa_{2,1,-1}}{z_{12}} \sum_{m=0}^{2q-3} &(-1)^m (m+1) \zb_{12}^{m+2-2q} \left[(2q+m-5)(-iu_2)^{2q-4-m} 
   (-i\partial_{u_2})^{1-m} \right. \nonumber\\
   & \left. + (-i u_2)^{2q-3-m} (-i\partial_{u_2})^{2-m}  \right] \partial_{\zb_2}^m \Phi_1 \left(u_2, z_2, \zb_2\right).
\end{align}
\paragraph{Action of $S$ algebra currents on gluons} The action of the $S$ algebra can be obtained by following the same procedure as above. Setting $J_1 = J_2 = J = 1$ in \eqref{eq:cel-carrOPE} and taking the limit in \eqref{eq:softoperatordef}, we get
\begin{align}
    \label{eq:softgluon-gluonOPE}
    &H^k_1 \left(z_1 ,\zb_1\right) \partial_{u_2}\Phi_1 \left(u_2,z_2, \zb_2\right)  \sim  -\frac{\kappa_{1,1,-1}}{z_{12}} \sum_{m=0}^{1-k} \frac{\zb_{12}^{m}}{m!} \frac{(-i u_2)^{1-k-m}}{(1-k-m)!} (-i\partial_{u_2})^{1-m} \partial_{\zb_2}^m \Phi_1 \left(u_2, z_2, \zb_2\right).
\end{align}
The actions of the soft operators with $k=1,0$ are local while the rest are non-local. This is similar to the gravitational case. In arriving at this result, we have made use of the formula
\begin{equation}
    \lim_{\D_1 \to k} \frac{(\D_1-k)}{(\D_1-1+m)} \hypfo \left(\D_1-1+m, \D_1+m, -i u_2 \om \right) = \frac{(-i u_2 \om)^{1-k-m}}{(1-k-m)!}.
\end{equation}
The $S$ algebra currents are defined as
\begin{equation}
    \label{eq:Scurrents}
    S^{q,a} \left(z, \zb \right) = \lim_{\varepsilon \to 0} \frac{(-1)^{2q} \Gamma (2q)}{2\pi i \varepsilon} \int_{\infty}^{\infty} \frac{d\bar{w}}{(\zb - \bar{w})^{2q}} \mo^a_{3-2q+\varepsilon, 1}, \qquad q = 1, \frac{3}{2}, 2, \dots 
\end{equation}
where $q = \frac{3-k}{2}$ and $\mo_{3-2q+\varepsilon, 1} \left(z, \bar{w} \right) \xrightarrow[]{\varepsilon \to 0} \frac{1}{\varepsilon} H^{k}_1$. Note that in the above equation, we have reintroduced the colour index on the gluon operator. These currents admit a mode expansion similar to the $w^q$ in \eqref{eq:wmodeexp} and are known to satisfy the algebra \cite{Strominger:2021lvk},
\begin{align}
    \label{eq:salgebra}
    \left[S_n^{q,a}, S_m^{p,b}\right] = -if^{ab}_c S_{m+n}^{p+q-1,c},
\end{align}
where $f^{abc}$ are the structure constants of the gauge group. Implementing the redefinition \eqref{eq:Scurrents} in \eqref{eq:softgluon-gluonOPE}, we get the action of the $S$ algebra on Carrollian current 
\begin{align}
    \label{eq:softgluon-gluonOPEw}
    &S^{q,a} \left(z_1 ,\zb_1\right) \partial_{u_2}\Phi^b_1 \left(u_2,z_2, \zb_2\right)  \\
    \nonumber &\qquad\sim  -f^{abc}\frac{\kappa_{1,1,-1}}{z_{12}} \sum_{m=0}^{2q-2} (-1)^m (m+1)\frac{\zb_{12}^{m+1-2q}}{m!}(-i u_2)^{2q-2-m} (-i\partial_{u_2})^{1-m} \partial_{\zb_2}^m \Phi^c_1 \left(u_2, z_2, \zb_2\right).
\end{align}
In arriving at the above result, it was necessary to reintroduce colour indices and the structure constants into \eqref{eq:softgluon-gluonOPE}. We can derive the action of  $S$ algebra currents on celestial gluon operators by a computation identical to \eqref{eq:softactioncheck1} and \eqref{eq:softgrav-gravOPEw-celestial}, upon which we find a result in agreement with \cite{Himwich:2021dau}.  

\section{Momentum, Carrollian and twistorial transforms}
\label{sec:twistors}

In this section we give a proof of the Fourier transform between on-shell momentum space wave functions and asymptotic data for massless fields.  Although  arguments for this transform have been given elsewhere, \cite{Ashtekar:1981hw, Ashtekar:1981sf,Donnay:2022wvx}, the version presented here makes the conformal invariance of the transform manifest by introducing homogeneous coordinates that explicitly encode  the action of conformal transformations of $\mathscr{S}$ and hence Lorentz invariance.  It will also expedite the connections with twistor space and the twistor encoding of the data.

The twistor encoding of zero rest mass fields can happen in a variety of ways owing to the cohomological nature of the twistor representatives arising from the Penrose transform.  Two choices are relevant to our discussions here depending on the choice of signature.  In Lorentz signature, as described in \cite{Sparling:1990, Eastwood:1982,Mason:1986}, asymptotic data at $\scri$ directly give rise to certain  preferred Dolbeault twistor cohomology classes adapted to null infinity.  This does not make a significant difference to the discussion above and and we postpone discussion of these to Section \ref{sec:Discussion}.  In Klein space or split signature there is a version more analogous to Cech cohomology but where the cohomological freedom is fixed and the Penrose transform coincides with the so-called X-ray transform discussed below.  There is further a nontrivial transform from asymptotic data to twistor data discussed in   \cite{Mason:2022hly} and proved below. This will allow us to establish a direct connection between twistor space and Carrollian CFT. 

\subsection{Homogeneous coordinates on $\scri$ and on-shell momentum space} 
This homogeneous framework for $\scri$ was originally introduced by Sparling \cite{Sparling:1990,Eastwood:1982} and our conventions in this section are adapted from these references and will slightly differ from those of the previous sections. This choice allows us to connect more easily with the twistor literature later.
The homogeneous coordinates will allow us to work with manifest Lorentz and conformal invariance and write more compact formulae that keep track of weights under transformations.  The following formulae are all valid in both Lorentz and in split signature although we will pivot to split signature in the twistor subsection. 

To connect with the notation in \eqref{eq:shvars} we introduce the momentum spinors
\begin{equation}
k_{\alpha\dot\alpha}=\kappa_\alpha\tilde\kappa_{\dot\alpha}\, , \quad \mbox{ where } \quad\kappa_\alpha=\sqrt{\omega}z_\alpha\, ,\qquad \tilde \kappa_{\dot \alpha}=\sqrt{\omega}\tilde z_{\dot\alpha}\, ,
\label{k-gauge}
\end{equation}
where 
\begin{equation}
z_\alpha=(1,z)\,  , \qquad \tilde z_{\dot\alpha}=(1,\tilde z)\, .
\end{equation}
For helicity\footnote{Here we will take the sign of the helicity as if the momentum were future pointing so that positive helicity corresponds to self-duality. } $n/2$  an  on-shell momentum space wave function $\hat \phi (\kappa_\alpha,\tilde\kappa_{\dot\alpha})$  is required to have weight $n$ 
\begin{equation}
\hat \phi (\alpha\kappa_\alpha, \alpha^{-1}\tilde\kappa_{\dot\alpha})= \alpha^n \hat\phi\, .
\end{equation} 
Note that  here $\omega\in\R$ not $\R^+$ but with this rescaling freedom, it only needs to appear in integral powers in the transform formulae so no signs or factors of $i$ need to be introduced.

Similarly, at $\scri$, the Bondi coordinates $(u_B,z,\bar z)$ introduced in Section \ref{sec:Elements of Carrollian CFT} can be wrapped up into  homogeneous spinorial quantities  
\begin{equation}
(u,\lambda_\alpha,\lambda_{\dot \alpha}) =\left(\lambda_0\tilde \lambda_{\dot 0} u_B,\lambda_0z_\alpha\, ,\tilde\lambda_{\dot 0}\bar z_{\dot\alpha}\right)\, ,
\end{equation}
where a function $f(u,\lambda,\tilde{\lambda})$ is now said to have weights $(p,q)$ if
\begin{equation}
f(\alpha\tilde \alpha
u,\alpha\lambda_\alpha,\tilde \alpha\tilde\lambda_{\dot\alpha})=\alpha^p\tilde \alpha^q f(u,\lambda_\alpha,\tilde\lambda_{\dot\alpha})\, . \label{Bondi-gauge}
\end{equation} 
Homogeneity weights $(p,q)$ correspond to Carrollian weights $(-2 k,-2\bar k)$.  This follows because the 1-forms 
\begin{equation}
D\lambda :=\langle \lambda d\lambda\rangle, \qquad  D\tilde \lambda:=    [\tilde \lambda d\tilde \lambda]
\end{equation}
have respective homogeneity weights $(2,0)$ and $(0,2)$ so that $f (D\lambda)^{-\frac p 2} (D\tilde \lambda)^{-\frac q 2}$ defines a section of $(\Omega^{(1,0)}_{\mathscr S})^{-\frac p 2} \otimes (\Omega^{(0,1)}_{\mathscr S})^{-\frac q 2}$ with homogeneity weights zero: thus it descends to a tensor (or a conformal Carrollian primary field) on $\scri$. Thus, homogeneity weights $(p,q)$ are equivalent to Carrollian weights $(k,\bar k)=\left(-\frac{p}{2},-\frac{q}{2}\right)$ according to the definition \eqref{carrollian primary} and encode the conformal invariance properties via standard Lie derivatives of tensors. We see that in the homogeneous framework, a fixing of the rescaling freedom is equivalent to the choice of a spin-frame on the sphere, and that the construction is essentially local depending only on the local conformal structure.  It also allows the global $SL(2,\mathbb{C})$ symmetry to be manifested when one is working globally on the Riemann sphere $\mathbb{CP}^1$ in terms of its global homogeneous coordinates.

Thus our basic Carrollian fields $\Phi_J^\epsilon$ have homogeneity weights $(-1-\epsilon J, -1+\epsilon J)$.
At $\scri$, according to the Newman-Penrose spin-coefficient formalism of \cite{Newman:1962cia, Penrose:1986uia}, $(p,q)$ are encoded into spin weights  $s=(p-q)/2$ and boost weights $w=(p+q)/2$. One finds that
the radiation field $\widetilde \Psi^0_n$ for a massless field with $n=2\epsilon J \geq 0$ has  weights $(p,q)=(-1,-1-n)$  whereas $\Psi^0_n$ with $n= -2\epsilon J\geq 0$  has $(p,q)=(-1-n,-1)$.   Thus, on the basis of weights, for integral spin, we can identify
\begin{equation}
     \widetilde\Psi^0_n =\p_u^{\frac{n}2}\Phi_J^\epsilon     \, , \quad n= 2\epsilon J \, ,\qquad     \Psi^0_n =\p_u^{\frac{n}2}\Phi_J^\epsilon\, , \quad n=-2\epsilon J\,.
\end{equation}
In particular for spin two we have $\Psi^0_n=\p_u N=\p^2_u \bar \sigma$ where $N$ is the Bondi news, and $\bar\sigma$ the asymptotic shear. Thus $\Phi_2$ can indeed be identified with the shear.

As an application of the homogeneous formalism, we express the Carrollian OPE \eqref{eq:carrollOPEformula1} conformally invariantly as
\begin{align}
    \label{eq:carrollOPEformula1-inv}
    &\Phi_{J_1} \left(u_1, \lambda_1, \tilde \lambda_1 \right) \Phi_{J_2} \left(u_2, \lambda_2, \tilde\lambda_2 \right) \\
    &\nonumber \qquad\qquad \sim  -\frac{\kappa_{J_1, J_2, -J}}{2\pi}\frac{[\tilde \lambda_1\tilde \lambda_2]^{p}}{\langle \lambda_1\lambda_2\rangle} \int_{t_1=0}^{t_2=0} Dt\,  t_1^{J_2-J-1} t_2^{J_1-J-1} \,  \left(\frac{\partial}{\partial u}\right)^{p}\Phi_{J}\left(t_1 u_1+t_2u_2, \lambda_2,t_1\tilde\lambda_1+ t_2\tilde\lambda_2\right) \, .
\end{align}
Here $Dt:=t_2dt_1 -t_1dt_2$ where $t_i$ have weights $(0,-1)$ in $(u_i,\lambda_i,\tilde\lambda_i)$ respectively and $J=J_1+J_2-p-1$. As we are taking the limit $\langle\lambda_1\lambda_2\rangle\rightarrow 0$, we can identify the holomorphic weights in the OPE using the ratio $\lambda_1:\lambda_2$ in the limit. In this framework, the fact that the weights in this formula balance  proves its conformal invariance.

\subsection{From momentum space to asymptotic data and back}
\label{Fourier-scri}

Massless fields  of spin $n/2$ are  given respectively  by  momentum space wave functions $\hat \phi(\kappa,\tilde \kappa)$ of weight $n$,  or characteristic data at $\scri $, $\Psi^0_n(u,\lambda,\tilde \lambda)$ of weight $(-n-1,-1)$ for negative helicity and  of weight $(-1, -n-1)$ for positive helicity.  Thus for negative $\epsilon J=-n/2$ the space-time fields are given by
the formulae 
\begin{align}
\Psi_{\alpha_1\ldots \alpha_n}(x)&=\int_{\R^4} \kappa_{\alpha_1}\ldots\kappa_{\alpha_n} \hat \phi(\kappa,\tilde\kappa) \re^{ik\cdot x} \delta(k^2)\, d^4k \,, \label{Fourier}\\ 
&=\frac{1}{2\pi}\int_{\mathscr{S}}  \lambda_{\alpha_1}\ldots \lambda_{\alpha_n} \p_u\Psi^0_n( x^{\alpha\dot\alpha}\lambda_\alpha \tilde \lambda_{\dot\alpha},\lambda,\tilde \lambda) D\lambda \, D\tilde \lambda \, ,\label{dAdhemar}
\end{align}
Here $D\lambda:=\langle \lambda d \lambda\rangle$ and $D\tilde \lambda:= [\tilde \lambda d\tilde \lambda]$.
In both cases it is quickly checked that the integrands are invariant under their respective rescalings. The rescaling invariance for \eqref{dAdhemar} means that the $\lambda_\alpha$ and $\tilde\lambda_{\dot\alpha}$ integrals reduce, in Lorentz signature where $\tilde \lambda$ is the complex conjugate of $\lambda$, to the double integral over the sphere.
In split signature the $(\lambda,\tilde\lambda)$ integrals are over  projective lines $\RP^1$ with topology $S^1$ or simply $\R$ if a scaling is chosen so that $\lambda_0=1$.  

These  formulae are essentially ambidextrous, and the positive  helicity versions are given by the conjugate formulae.
The first of these in \eqref{Fourier} is the standard Fourier transform,  and the second in \eqref{dAdhemar} is an adaptation of the d'Adhémar integral formula of, for example, \cite{Penrose:1985bww}, and taken out to $\scri$ in either signature.

The main tool used in this paper is what might be known as the \emph{third-Fourier transform}: 
\begin{lemma}\label{Fourier-scri-lemma}
For spin $n/2$, we have the formulae
\begin{align}
\Psi_n^0(u,\lambda,\tilde{\lambda})&=i\int \hat\phi(\lambda, s\tilde\lambda) \re^{-isu} ds=i\int s^{\frac{n}2}\hat\phi(s^{\frac12}\lambda, s^{\frac12}\tilde\lambda) \re^{-isu} ds\, , \qquad \epsilon J\leq 0\, ,\label{third-}\\
\widetilde \Psi_n^0(u,\lambda,\tilde{\lambda})&=i\int \hat\phi(s\lambda, \tilde\lambda) \re^{-isu} ds=i\int s^{\frac{n}2}\hat\phi(s^{\frac12}\lambda, s^{\frac12}\tilde\lambda) \re^{-isu} ds\, ,\, , \qquad \epsilon J\geq 0\, .\label{third+}
\end{align}
with  the second equality following from the homogeneity.
\end{lemma}
In particular, for gravity we see  that with $\Psi^0_4=\ddot{\bar\sigma}$ we can write 
\begin{equation}
    \bar\sigma(u,\lambda,\bar\lambda)=i\int \hat\phi(s^{\frac12}\lambda, s^{\frac12}\tilde\lambda) \re^{-isu} ds\, ,
\end{equation}
agreeing with the identification following \eqref{eq:AtoC}. 
A well-known corollary is that if $\hat \phi $ is positive or negative frequency, then $\phi^0_n$ extends to the upper-half $u$-plane.

\noindent
{\bf Proof:}
To see this for the negative helicity case, we rewrite \eqref{dAdhemar} as an integral over $\scri$ using the Fourier representation of the delta function
\begin{equation}
2\pi \delta(u-x^{\alpha\dot\alpha}\lambda_\alpha \tilde \lambda_{\dot\alpha})= \int \re^{i
  s(u-x\cdot\lambda \tilde\lambda)}\rd s\, .
\end{equation}
Putting this together with 
Fourier inversion of \eqref{Fourier} 
\begin{equation}
\kappa_{\alpha_1}\ldots\kappa_{\alpha_n} \,\hat\phi(\kappa,\tilde\kappa) \, \delta(k^2)=\frac{1}{(2\pi)^4}\int_{\R^4} \re^{-ik\cdot x} \Psi_{\alpha_1\ldots\alpha_n} \rd^4x\, , \label{invFourier}
\end{equation}
we obtain
\begin{align}
\kappa_{\alpha_1}\ldots \kappa_{\alpha_n}\hat\phi \, \delta(k^2)&= \frac1{(2\pi)^5} \int d^4x \, ds \int_\scri \re^{i(k\cdot x +s(u-x\cdot\lambda\tilde\lambda))}\lambda_{\alpha_1}\ldots \lambda_{\alpha_n} \dot \Psi^0_n \, du D\lambda D\tilde \lambda \\
&=\frac1{2\pi}\int ds \int_\scri \delta^{(4)}(k_{\alpha\dot\alpha} -s \lambda_\alpha\tilde \lambda_{\dot\alpha})\, \e^{isu }\lambda_{\alpha_1}\ldots \lambda_{\alpha_n} \dot\Psi^0_n \,du D\lambda D\tilde \lambda\, ,
\end{align}
with the $x$-integrations directly yielding the delta functions. These imply the $\delta(k^2)$ delta function which can be factored out, and  the $(s,\lambda,\tilde\lambda)$ integrals can then be evaluated against the delta function. To do this it is useful to revert to the gauge fixing of \eqref{k-gauge} and \eqref{Bondi-gauge} with $\lambda_0=1=\tilde\lambda_{\dot 0}$, so that $s$-integral against the delta function  identifies $s=\omega$ to give
\begin{align}
\omega^{n/2}\hat\phi(\kappa,\tilde\kappa)&=\frac1{2\pi}\int_\R du_B \, \frac{\re^{i\omega u_B}}{\omega}\,\dot \phi^0_n(u_B, z, \tilde z)=\frac{-i}{2\pi}\int du_B \, \re^{i\omega u_B}\, \phi^0_n(u_B, z,\tilde z)
\end{align}
where in the second equality we integrated by parts.
Multiplying through by $\lambda_0^{-n}$, introducing $s=\omega/\lambda_0\tilde\lambda_{\dot 0}$ and using homogeneity we obtain
\begin{equation}
    \hat\phi(\lambda, s\tilde\lambda) =\frac{-i}{2\pi} \int du \,\re^{isu} \phi^0_n(u,\lambda,\tilde\lambda)\, .\label{invthird-}
\end{equation}
Inverting the transform we therefore obtain
\eqref{third-}
\begin{equation}
\phi^0_n\left(u,\lambda_\alpha,\tilde \lambda_{\dot\alpha}\right)=i\int d s \,\re^{-is u}  \hat\phi( \lambda_\alpha,s\tilde \lambda_{\dot \alpha})\, .
\end{equation}
The details are much the same for obtaining the positive helicity version \eqref{third+}. $\Box$


\subsection{Transforms to twistor space and back}

In order to have well-defined twistor functions, we work in Klein space with split signature as the Penrose transform there reduces to the X-ray transform expressing massless fields on space-time in terms of smooth functions on real twistor space $\RP^3$ with homogeneous coordinates $(\lambda_\alpha,\mu^{\dot\alpha})\sim \alpha (\lambda_\alpha,\mu^{\dot\alpha})$. In split signature  all coordinates  will be taken to be real.

We introduce homogeneous coordinates $(\lambda_\alpha,\mu^{\dot\alpha})$ on twistor space $\RP^3$ with twistor function $f(\lambda_\alpha,\mu^{\dot\alpha})$ said to be of homogeneity $n$ if 
\begin{equation}
f(\alpha\lambda_\alpha,\alpha\mu^{\dot\alpha})= \alpha^nf(\lambda_\alpha,\mu^{\dot\alpha})\,  .
\end{equation}
The twistor integral formulae are starting with twistor functions $f(\lambda_\alpha,\mu^{\dot\alpha})$ of weight $n-2$,
 for spin $n/2$.  These are chiral transforms given differently for positive and negative helicity massless fields
\begin{align}
\Psi_{\dot\alpha_1\ldots \dot\alpha_n}(x)&=\int_{S^1}
\left.\frac{\p^n  f}{\p\mu^{\dot\alpha_1}\ldots\p\mu^{\dot\alpha_n}}\right|_{\mu^{\dot\alpha}=x^{\alpha\dot\alpha}\lambda_\alpha} D\lambda \, , \qquad \epsilon J> 0\label{X-ray+}\\
\Psi_{\alpha_1\ldots \alpha_n}(x)&=\int_{S^1}\lambda_{\alpha_1}\ldots \lambda_{\alpha_n} f(\lambda, x^{\alpha\dot\alpha}\lambda_\alpha)\, D\lambda \,, \qquad \epsilon J\leq 0. \label{X-ray}
\end{align}
It is quickly checked that the integrands are invariant under their respective rescalings so that, in split signature, the $\lambda_\alpha$ integrals in \eqref{X-ray+}-\eqref{X-ray} reduce to integrals over the projective line $\RP^1$ with topology $S^1$ or simply $\R$ if a scaling is chosen so that $\lambda_0=1$.

 These integrals are higher spin versions of the so-called  \emph{X-ray transform}, because they integrate $f$ over lines in $\RP^3$, providing solutions to higher spin versions of  the ultrahyperbolic scalar wave equation, \cite{John:1938}.\footnote{Usually, the Penrose transform reformulates massless fields as cohomology classes on regions in twistor space subject to constraints and gauge freedom.  The X-ray transform has the advantage of being a correspondence with functions on $\RP^3$.} These are  manifestly conformally invariant as written.

Concatenating \eqref{invFourier}  with the twistor integral formulae leads to Witten's \cite{Witten:2003nn} \emph{half-Fourier} transform between momentum space and twistor space
\begin{equation}
f(\lambda, \mu)=\int \hat \phi(\lambda,\tilde \lambda) \re^{i [\mu\, \tilde \lambda]} d^2\tilde \lambda\, , \qquad \hat\phi(\lambda,\tilde \lambda)=\frac{1}{(2\pi)^2} \int f (\lambda,\mu) \re^{-i[\mu\, \tilde{\lambda}]} d^2\mu\,, \label{half-fourier}
\end{equation}
see appendix B of \cite{Mason:2009sa} for  full  details of the derivation.  
To obtain the map between asymptotic data and twistor space, we can concatenate this last formula with the half Fourier transform to obtain for negative helicity
\begin{align}
\Psi^0_n\left(u,\lambda_\alpha,\tilde \lambda_{\dot\alpha}\right)&=\frac{i}{4\pi^2}\int d s d^2\mu\,\re^{-is( u -[\mu\tilde\lambda])}  f( \lambda_\alpha,\mu^{\dot \alpha})\, ,\\
&=\frac{i}{2\pi}\int d^2\mu\,\delta( u -[\mu\tilde\lambda])  f( \lambda_\alpha,\mu^{\dot \alpha})\, .
\end{align}
However, for positive helicity we obtain 
\begin{align}
\phi^0_n\left(u,\lambda_\alpha,\tilde \lambda_{\dot\alpha}\right)&=\frac{i}{4\pi^2}\int s^n d s d^2\mu\,\re^{-is( u -[\mu\tilde\lambda])}  f( \lambda_\alpha,\mu^{\dot \alpha})\, ,\\
&=\frac{i}{2\pi}(i\p_u)^n\int d^2\mu\,\delta( u -[\mu\tilde\lambda])  f( \lambda_\alpha,\mu^{\dot \alpha})\, .
\end{align}
For positive helicity gravitons, $n=4$ and we write $\phi^0_4=\Psi^0_4=\ddot \sigma$. It is easy to see that, after integrating twice with respect to $u$ we obtain for the asymptotic shear with  $f$ of weight 2 
\begin{equation}
    \sigma(u,\lambda,\tilde \lambda)= \frac{i}{2\pi} \p_u^2 \int d^2\mu\,\delta( u -[\mu\tilde\lambda])  f( \lambda_\alpha,\mu^{\dot \alpha})\, .
\end{equation}
Integrating out the delta function,  this formula reduces to  (A.10) in \cite{Mason:2022hly} thereby giving a proof of that formula. 

For Maxwell/Yang Mills we have $n=2$ and $\phi^0_2=\p_u a$ where $a$ is the $D\tilde \lambda$ component of the gauge potential in a gauge in which the $du$ component is zero.  In this case we can similarly write for $f$ now of weight zero
\begin{equation}
    a(u,\lambda,\tilde \lambda)= \frac{i}{2\pi} \p_u \int d^2\mu\,\delta( u -[\mu\tilde\lambda])  f( \lambda_\alpha,\mu^{\dot \alpha})\, .
\end{equation}

 The above integral transforms provide a direct relation between operators $f$ on twistor space and what we holographically identify with Carrollian operators in the putative dual theory at $\mathscr{I}$. This can be exploited to study properties of Carrollian amplitudes using twistor theory, without resorting to momentum space. We defer this analysis for future works.

\section{Discussion}
\label{sec:Discussion}

Scattering amplitudes are usually expressed in terms of on-shell momentum space wave functions. 
There has nevertheless been a long tradition of studying classical scattering from null infinity, already implicit in the work of Bondi, Sachs et al.\ \cite{Bondi:1962px,Sachs:1962wk,Sachs:1962zza}  made explicit\footnote{The scattering theory of  of Lax and Philips \cite{Lax:1967} was similarly re-interpreted  as scattering from $\scri$ by Friedlander \cite{Friedlander:1980}. } in the work of Newman and Penrose \cite{Newman:1961qr,Newman:1962cia,Penrose:1962ij,Penrose:1964ge,Newman:1966ub} and later by Ashtekar and others \cite{Ashtekar:1981hw,Mason:2004lqj}. Quantum scattering has also been studied from null infinity in \cite{Ashtekar:1981sf,Ashtekar:1987tt} and more recently, although perhaps more implicitly in the  conformal collider discussion of \cite{Hofman:2008ar}. 
Only relatively recently has there been much interaction with the modern study of scattering amplitudes.  For example, the ambitwistor strings of \cite{Mason:2013sva,Geyer:2014fka} were reformulated in terms of strings at $\scri$ in \cite{Adamo:2014yya,Geyer:2014lca,Adamo:2019ipt,Adamo:2021lrv} with amplitudes constructed from vertex operators built from radiation fields at null infinity. Nevertheless, in these articles, the focus was on  momentum eigenstates or Mellin states even if expressed at $\scri$, in order to identify soft theorems.

In this article we have instead systematically reformulated massless amplitudes in terms of  their radiation data at $\scri$ via a Fourier relationship between energy and the Bondi retarded time.  As examples we have been able to fully transform two-point, generic Lorentz-invariant three-point amplitudes, and gauge and gravity MHV amplitudes with many points.  This has enabled us to establish a bottom-up approach to a dictionary for Carrollian holography by proposing that such Carrollian amplitudes arise as correlators of some system of operators defining a Carrollian CFT at $\scri$.  The OPEs of these operators have been obtained from the transform of the three-point functions. Although the Fourier transform between momentum space and radiative data in its most explicit form and hence the Carrollian OPE do not make Lorentz invariance manifest, in Section \ref{Fourier-scri} we saw that they can be formulated with manifest conformal invariance.

The interest of position space amplitudes at $\mathscr I$ for flat space holography \cite{Donnay:2022aba,Bagchi:2022emh,Donnay:2022wvx} is based on the observation that the position space amplitudes for a massless scattering in flat space can be encoded in Carrollian correlation functions at null infinity. The $n$-point Carrollian amplitudes and OPEs derived in this work offer strong constraints on the potential dual theory at null infinity. This opens the road to a bootstrap program for Carrollian CFT.

The Carrollian holography dictionary discussed in this work can be seen as an extrapolate-type of dictionary: the Carrollian correlators at the boundary can be obtained as boundary values of the bulk correlators (see the discussion below \eqref{carrollian identification}). This is very much in the spirit of AdS/CFT, and in that case, the bulk path integral can be interpreted as the generating functional for boundary correlators \cite{Witten:1998qj}. The first steps towards such an identification in flat space have been taken in \cite{Gonzo:2022tjm,Kim:2023qbl,Jain:2023fxc}, where the authors use the bulk path integral to explicitly realize the $\mathcal{S}$-matrix (in specific cases) as boundary correlators. It would be interesting to compare these with partition functions of Carrollian theories.

Furthermore, we believe that Carrollian amplitudes are the natural objects obtained from the flat limit of AdS amplitudes with suitable boundary conditions. Indeed, from the classical Einstein equations, we know that the flat limit in the bulk ($\Lambda \to 0$) implies a Carrollian limit ($c\to 0$) at the boundary \cite{Barnich:2012aw,Ciambelli:2018wre,Compere:2019bua,Compere:2020lrt,Ruzziconi:2020cjt,Ciambelli:2020ftk,Ciambelli:2020eba,Fiorucci:2020xto,Ruzziconi:2020wrb,Campoleoni:2022wmf,Geiller:2022vto,Campoleoni:2023fug,Adami:2023fbm}. We expect that a similar procedure applies for scattering amplitudes (see e.g. \cite{Hijano:2019qmi,deGioia:2022fcn,deGioia:2023cbd,Bagchi:2023fbj,Bagchi:2023cen} for recent works in that direction).  

In the last part of this work, we initiated a connection between Carrollian holography and twistor theory by expressing the asymptotic data at null infinity, corresponding to Carrollian operators, and operators defined on twistor space. The latter is the natural framework to provide a geometric interpretation for the celestial symmetries \cite{Adamo:2021lrv,Mason:2022hly,Bu:2022iak,Bittleston:2023bzp}. We plan to exploit this interplay in future works. 

Finally, although here we have focused on bottom-up methods for Carrollian holography, but there are by now a number of top-down approaches in the literature.  In particular, worldsheet theories with target $\scri$ have been constructed  in \cite{Adamo:2014yya,Geyer:2014lca,Adamo:2019ipt,Adamo:2021bej}.  These  are all essentially ambitwistor string theories except for \cite{Adamo:2021bej} which is a classical sigma model.   They encode the radiative sector into worldsheet vertex operators whose correlation functions yield amplitudes at tree level and beyond.  Such vertex operators also   implement both BMS  and higher celestial symmetries.   Other top-down methods, also have a twistorial focus.  The twisted holography work of Costello, Sharma and Paquette \cite{Costello:2022jpg,Costello:2022upu} 
should also have a Carrollian realization.  The twistor actions of \cite{Mason:2005zm,Mason:2007ct,Mason:2009afn, Adamo:2011pv,Adamo:2013tja} already   have an interesting realization in this Carrollian framework, because a natural  gauge fixing for the twistor cohomology classes on which the twistor actions are based arises from asymptotic data at $\scri$ following \cite{Sparling:1990,Eastwood:1982}; this has already been exploited in the radiative approach to amplitudes in \cite{Adamo:2020yzi,Adamo:2022mev}. 
On the other hand, split signature approaches to the transform to twistor space \cite{Witten:2003nn,Mason:2009sa} give different insights into the structures studied here, in particular because actions of the enhanced symmetry algebras, $Lw_{1+\infty}$ and the $S$-algebra, are locally realized on twistor space in this signature as seen in \cite{Mason:2022hly};  the Lorentz signature approach of \cite{Adamo:2021lrv} required a \v Cech-Dolbeault correspondence to be implemented before a local geometric realization could be obtained, and so a cleaner picture should be available in this framework.

\paragraph{Acknowlegements.}
RR is supported by the Titchmarsh Research Fellowship at the Mathematical Institute and by the Walker Early Career Fellowship in Mathematical Physics at Balliol College. AYS and LJM are supported by the STFC grant ST/X000761/1. This work was supported by the Simons Collaboration on Celestial Holography. LJM would like to thank Juan Maldacena and David Skinner for related correspondence and discussion in 2009.

\appendix

\section{Computation of \texorpdfstring{$I_n$}{In}}
\label{sec:Incomp}
We will perform the integral over one such generic region which without loss of generality we will take to be
\begin{align}
    \label{eq:region2}
    B_n \leq \om_n \leq B'_n, \qquad \sum_{a=k+1}^n B_{k a}\, \om_a \leq \om_k \leq   \sum_{a=k+1}^n B'_{k a}\, \om_a .
\end{align}
We have assumed that the inequalities have been solved in such a way that the bounds on the integration domain of $\om_k$ depend only on $\om_{k+1}, \dots , \om_n$. It is convenient to define the vectors
\begin{align}
    \label{eq:domainvecs2}
    B_k = \left(0, \dots , B_{kk+1}, \dots , B_{kn}\right), \qquad  B'_k = \left(0, \dots, B'_{kk+1}, \dots ,B'_{kn}\right), \qquad \om = \left(\om_5, \dots ,\om_n \right).
\end{align}
Note that $B_k \cdot \om$ reduces to the combination appearing in \eqref{eq:region2}. With this, the integral \eqref{eq:Indef} can be written as 
\begin{align}
     I_n = \int_0^{\infty} \prod_{a=5}^n d\om_a \, e^{i L_a \, \om_a  }  \Theta \left(\om_n - B_n\right)\Theta \left(B'_n - \om_n\right)\prod_{k=5}^{n-1} \Theta \left(\om_k - B_k \cdot \om\right)\Theta \left(B'_k \cdot \om - \om_k\right)
\end{align}
We can now replace all the $\Theta$ functions by their integral representations
\begin{equation}
\label{eq:theta}
    \Theta \left(x\right) = \frac{1}{2\pi i} \int_{-\infty}^{\infty} \frac{d\tau}{\tau + i \varepsilon} e^{-i x \tau}
\end{equation}
with the understanding that $\varepsilon$ is an infinitesimal parameter. We introduce a variable $\tau_i$ for every lower bound and $\tau'_i$ for every upper bound and get
\begin{align}
     I_n &= \nonumber \left(\frac{1}{2\pi i}\right)^{2(n-4)}\int_0^{\infty} \prod_{a=5}^n d\om_a \, e^{i L_a \, \om_a  } \int \prod_{k=5}^{n-1}\frac{d\tau_k d\tau_k'}{(\tau_k+i \varepsilon)(\tau_k' + i \varepsilon)} \text{exp}\left[i \tau_k \left(\om_k - B_k \cdot \om \right) - i \tau'_k \left(\om_k - B'_k \cdot \om \right) \right]\\
     &\nonumber \hspace{6cm} \times\int \frac{d\tau_n d\tau'_n}{(\tau_n+i \varepsilon)(\tau_n' + i \varepsilon)}  \text{exp}\left[-i B_n \tau_n + i B'_n \tau'_n \right]\\
     &=\left(\frac{1}{2\pi i}\right)^{2(n-4)}\int_0^{\infty} \prod_{a=5}^n d\om_a \, \int \prod_{k=5}^n\frac{d\tau_k d\tau_k'}{(\tau_k+i \varepsilon)(\tau_k' + i \varepsilon)} \text{exp}\left[-i \om_k X_k \left(\tau \right) \right]\text{exp}\left[-i B_n \tau_n + i B'_n \tau'_n \right]\,
\end{align}
where $X_k \left(\tau\right)= \tau_k - \tau'_k - \sum_{a=5}^{k-1} \tau_a B_{ak} + \sum_{a=5}^{k-1} \tau'_a B'_{ak} - L_k$. We can now perform all the $\om_k$ integrals to get
\begin{align}
\label{eq:Inresidue}
     I_n =\left(\frac{1}{2\pi i}\right)^{2(n-4)}\int \prod_{k=5}^n\frac{d\tau_k d\tau_k'}{(\tau_k+i \varepsilon)(\tau_k' + i \varepsilon)} \frac{1}{X_k\left(\tau\right) - i \varepsilon}\text{exp}\left[-i B_n \tau_n + i B'_n \tau'_n \right]\,
\end{align}
Finally, for certain kinematic regions, we can have $B_k = 0$ and $B'_k = \infty$. All the $\Theta$ functions reduce to 
 $1$ in this limit. In the integral representation, we see that the $x \to \infty$ limit is dominated by the $\tau = 0$. Thus, if we use \eqref{eq:Inresidue} to compute $I_n$ when all $B_k = 0$ and $B'_k = \infty$, we must compute the residue around $\tau_k = \tau'_k = 0$.

\subsection*{References}
\bibliographystyle{style}
\renewcommand\refname{\vskip -1.3cm}
\bibliography{references}

\providecommand{\href}[2]{#2}\begingroup\raggedright\begin{thebibliography}{100}

\bibitem{Arcioni:2003xx}
G.~Arcioni and C.~Dappiaggi, \emph{{Exploring the holographic principle in asymptotically flat space-times via the BMS group}}, Nucl. Phys. B {\bf 674} (2003) 553--592,
\href{http://www.arXiv.org/abs/hep-th/0306142}{{\tt hep-th/0306142}}

\bibitem{Arcioni:2003td}
G.~Arcioni and C.~Dappiaggi, \emph{{Holography in asymptotically flat space-times and the BMS group}}, Class. Quant. Grav. {\bf 21} (2004) 5655,
\href{http://www.arXiv.org/abs/hep-th/0312186}{{\tt hep-th/0312186}}

\bibitem{Dappiaggi:2005ci}
C.~Dappiaggi, V.~Moretti  and N.~Pinamonti, \emph{{Rigorous steps towards holography in asymptotically flat spacetimes}}, Rev. Math. Phys. {\bf 18} (2006) 349--416,
\href{http://www.arXiv.org/abs/gr-qc/0506069}{{\tt gr-qc/0506069}}

\bibitem{Barnich:2006av}
G.~Barnich and G.~Compere, \emph{{Classical central extension for asymptotic symmetries at null infinity in three spacetime dimensions}}, Class. Quant. Grav. {\bf 24} (2007) F15--F23,
\href{http://www.arXiv.org/abs/gr-qc/0610130}{{\tt gr-qc/0610130}}

\bibitem{Barnich:2010eb}
G.~Barnich and C.~Troessaert, \emph{{Aspects of the BMS/CFT correspondence}}, JHEP {\bf 05} (2010) 062,
\href{http://www.arXiv.org/abs/1001.1541}{{\tt 1001.1541}}

\bibitem{Bagchi:2010zz}
A.~Bagchi, \emph{{Correspondence between Asymptotically Flat Spacetimes and Nonrelativistic Conformal Field Theories}}, Phys. Rev. Lett. {\bf 105} (2010) 171601,
\href{http://www.arXiv.org/abs/1006.3354}{{\tt 1006.3354}}

\bibitem{Barnich:2012xq}
G.~Barnich, \emph{{Entropy of three-dimensional asymptotically flat cosmological solutions}}, JHEP {\bf 10} (2012) 095,
\href{http://www.arXiv.org/abs/1208.4371}{{\tt 1208.4371}}

\bibitem{Barnich:2012rz}
G.~Barnich, A.~Gomberoff  and H.~A. Gonz\'alez, \emph{{Three-dimensional Bondi-Metzner-Sachs invariant two-dimensional field theories as the flat limit of Liouville theory}}, Phys. Rev. D {\bf 87} (2013), no.~12, 124032,
\href{http://www.arXiv.org/abs/1210.0731}{{\tt 1210.0731}}

\bibitem{Bagchi:2012xr}
A.~Bagchi, S.~Detournay, R.~Fareghbal  and J.~Sim\'on, \emph{{Holography of 3D Flat Cosmological Horizons}}, Phys. Rev. Lett. {\bf 110} (2013), no.~14, 141302,
\href{http://www.arXiv.org/abs/1208.4372}{{\tt 1208.4372}}

\bibitem{Bagchi:2014iea}
A.~Bagchi, R.~Basu, D.~Grumiller  and M.~Riegler, \emph{{Entanglement entropy in Galilean conformal field theories and flat holography}}, Phys. Rev. Lett. {\bf 114} (2015), no.~11, 111602,
\href{http://www.arXiv.org/abs/1410.4089}{{\tt 1410.4089}}

\bibitem{Bagchi:2015wna}
A.~Bagchi, D.~Grumiller  and W.~Merbis, \emph{{Stress tensor correlators in three-dimensional gravity}}, Phys. Rev. D {\bf 93} (2016), no.~6, 061502,
\href{http://www.arXiv.org/abs/1507.05620}{{\tt 1507.05620}}

\bibitem{Bagchi:2016bcd}
A.~Bagchi, R.~Basu, A.~Kakkar  and A.~Mehra, \emph{{Flat Holography: Aspects of the dual field theory}}, JHEP {\bf 12} (2016) 147,
\href{http://www.arXiv.org/abs/1609.06203}{{\tt 1609.06203}}

\bibitem{Ciambelli:2018wre}
L.~Ciambelli, C.~Marteau, A.~C. Petkou, P.~M. Petropoulos  and K.~Siampos, \emph{{Flat holography and Carrollian fluids}}, JHEP {\bf 07} (2018) 165,
\href{http://www.arXiv.org/abs/1802.06809}{{\tt 1802.06809}}

\bibitem{Donnay:2022aba}
L.~Donnay, A.~Fiorucci, Y.~Herfray  and R.~Ruzziconi, \emph{{Carrollian Perspective on Celestial Holography}}, Phys. Rev. Lett. {\bf 129} (2022), no.~7, 071602,
\href{http://www.arXiv.org/abs/2202.04702}{{\tt 2202.04702}}

\bibitem{Bagchi:2022emh}
A.~Bagchi, S.~Banerjee, R.~Basu  and S.~Dutta, \emph{{Scattering Amplitudes: Celestial and Carrollian}}, Phys. Rev. Lett. {\bf 128} (2022), no.~24, 241601,
\href{http://www.arXiv.org/abs/2202.08438}{{\tt 2202.08438}}

\bibitem{Donnay:2022wvx}
L.~Donnay, A.~Fiorucci, Y.~Herfray  and R.~Ruzziconi, \emph{{Bridging Carrollian and celestial holography}}, Phys. Rev. D {\bf 107} (2023), no.~12, 126027,
\href{http://www.arXiv.org/abs/2212.12553}{{\tt 2212.12553}}

\bibitem{Leblond}
J.-M. L\'evy-Leblond, \emph{{Une nouvelle limite non-relativiste du groupe de Poincar\'e}}, A. Inst. Henri Poincar\'e III 1
(1965)

\bibitem{deBoer:2003vf}
J.~de~Boer and S.~N. Solodukhin, \emph{{A Holographic reduction of Minkowski space-time}}, Nucl. Phys. {\bf B665} (2003) 545--593,
\href{http://www.arXiv.org/abs/hep-th/0303006}{{\tt hep-th/0303006}}

\bibitem{He:2015zea}
T.~He, P.~Mitra  and A.~Strominger, \emph{{2D Kac-Moody Symmetry of 4D Yang-Mills Theory}}, JHEP {\bf 10} (2016) 137,
\href{http://www.arXiv.org/abs/1503.02663}{{\tt 1503.02663}}

\bibitem{Pasterski:2016qvg}
S.~Pasterski, S.-H. Shao  and A.~Strominger, \emph{{Flat Space Amplitudes and Conformal Symmetry of the Celestial Sphere}}, Phys. Rev. {\bf D96} (2017), no.~6, 065026,
\href{http://www.arXiv.org/abs/1701.00049}{{\tt 1701.00049}}

\bibitem{Cheung:2016iub}
C.~Cheung, A.~de~la Fuente  and R.~Sundrum, \emph{{4D scattering amplitudes and asymptotic symmetries from 2D CFT}}, JHEP {\bf 01} (2017) 112,
\href{http://www.arXiv.org/abs/1609.00732}{{\tt 1609.00732}}

\bibitem{Pasterski:2017kqt}
S.~Pasterski and S.-H. Shao, \emph{{Conformal basis for flat space amplitudes}}, Phys. Rev. {\bf D96} (2017), no.~6, 065022,
\href{http://www.arXiv.org/abs/1705.01027}{{\tt 1705.01027}}

\bibitem{Strominger:2017zoo}
A.~Strominger, {\em {Lectures on the Infrared Structure of Gravity and Gauge Theory}}.
\newblock {Princeton University Press},
2018

\bibitem{Pasterski:2017ylz}
S.~Pasterski, S.-H. Shao  and A.~Strominger, \emph{{Gluon Amplitudes as 2d Conformal Correlators}}, Phys. Rev. {\bf D96} (2017), no.~8, 085006,
\href{http://www.arXiv.org/abs/1706.03917}{{\tt 1706.03917}}

\bibitem{Pasterski:2021rjz}
S.~Pasterski, \emph{{Lectures on celestial amplitudes}}, Eur. Phys. J. C {\bf 81} (2021), no.~12, 1062,
\href{http://www.arXiv.org/abs/2108.04801}{{\tt 2108.04801}}

\bibitem{Raclariu:2021zjz}
A.-M. Raclariu, \emph{{Lectures on Celestial Holography}},
\href{http://www.arXiv.org/abs/2107.02075}{{\tt 2107.02075}}

\bibitem{McLoughlin:2022ljp}
T.~McLoughlin, A.~Puhm  and A.-M. Raclariu, \emph{{The SAGEX review on scattering amplitudes chapter 11: soft theorems and celestial amplitudes}}, J. Phys. A {\bf 55} (2022), no.~44, 443012,
\href{http://www.arXiv.org/abs/2203.13022}{{\tt 2203.13022}}

\bibitem{Donnay:2023mrd}
L.~Donnay, \emph{{Celestial holography: An asymptotic symmetry perspective}},
\href{http://www.arXiv.org/abs/2310.12922}{{\tt 2310.12922}}

\bibitem{Ko:1977gw}
M.~Ko, E.~T. Newman  and R.~Penrose, \emph{{The Kahler Structure of Asymptotic Twistor Space}}, J. Math. Phys. {\bf 18} (1977)
58--64

\bibitem{Hansen:1978jz}
R.~O. Hansen, E.~T. Newman, R.~Penrose  and K.~P. Tod, \emph{{The Metric and Curvature Properties of H Space}}, Proc. Roy. Soc. Lond. {\bf A363} (1978)
445--468

\bibitem{Eastwood:1982}
M.~Eastwood and P.~Tod, \emph{{Edth -- a differential operator on the sphere}}, Math. Proc. Camb. Phil. Soc. {\bf 92} (1982)
317--330

\bibitem{Mason:1986}
L.~J. Mason, \emph{{Dolbeault representative from characteristic initial data at null infinity}}, in {\em {Further Advances in Twistor Theory, Vol 1}}, L.~J. Mason and L.~P. Hughston, eds., vol.~231, ch.~1.2.16.
\newblock Pitman Research Notes in Mathematics,
1990.
\newblock

\bibitem{Adamo:2014yya}
T.~Adamo, E.~Casali  and D.~Skinner, \emph{{Perturbative gravity at null infinity}}, Class. Quant. Grav. {\bf 31} (2014), no.~22, 225008,
\href{http://www.arXiv.org/abs/1405.5122}{{\tt 1405.5122}}

\bibitem{Geyer:2014lca}
Y.~Geyer, A.~E. Lipstein  and L.~Mason, \emph{{Ambitwistor strings at null infinity and (subleading) soft limits}}, Class. Quant. Grav. {\bf 32} (2015), no.~5, 055003,
\href{http://www.arXiv.org/abs/1406.1462}{{\tt 1406.1462}}

\bibitem{Adamo:2020yzi}
T.~Adamo, L.~Mason  and A.~Sharma, \emph{{Gluon Scattering on Self-Dual Radiative Gauge Fields}}, Commun. Math. Phys. {\bf 399} (2023) 1731--1771,
\href{http://www.arXiv.org/abs/2010.14996}{{\tt 2010.14996}}

\bibitem{Adamo:2021bej}
T.~Adamo, L.~Mason  and A.~Sharma, \emph{{Twistor sigma models for quaternionic geometry and graviton scattering}},
\href{http://www.arXiv.org/abs/2103.16984}{{\tt 2103.16984}}

\bibitem{Adamo:2021lrv}
T.~Adamo, L.~Mason  and A.~Sharma, \emph{{Celestial $w_{1+\infty}$ Symmetries from Twistor Space}}, SIGMA {\bf 18} (2022) 016,
\href{http://www.arXiv.org/abs/2110.06066}{{\tt 2110.06066}}

\bibitem{Adamo:2022mev}
T.~Adamo, L.~Mason  and A.~Sharma, \emph{{Graviton scattering in self-dual radiative space-times}}, Class. Quant. Grav. {\bf 40} (2023), no.~9, 095002,
\href{http://www.arXiv.org/abs/2203.02238}{{\tt 2203.02238}}

\bibitem{Adamo:2019ipt}
T.~Adamo, L.~Mason  and A.~Sharma, \emph{{Celestial amplitudes and conformal soft theorems}}, Class. Quant. Grav. {\bf 36} (2019), no.~20, 205018,
\href{http://www.arXiv.org/abs/1905.09224}{{\tt 1905.09224}}

\bibitem{Costello:2022jpg}
K.~Costello, N.~M. Paquette  and A.~Sharma, \emph{{Top-down holography in an asymptotically flat spacetime}},
\href{http://www.arXiv.org/abs/2208.14233}{{\tt 2208.14233}}

\bibitem{Costello:2022upu}
K.~Costello and N.~M. Paquette, \emph{{Associativity of One-Loop Corrections to the Celestial Operator Product Expansion}}, Phys. Rev. Lett. {\bf 129} (2022), no.~23, 231604,
\href{http://www.arXiv.org/abs/2204.05301}{{\tt 2204.05301}}

\bibitem{Mason:2022hly}
L.~Mason, \emph{{Gravity from holomorphic discs and celestial $Lw_{1+\infty}$ symmetries}},
\href{http://www.arXiv.org/abs/2212.10895}{{\tt 2212.10895}}

\bibitem{Cachazo:2014fwa}
F.~Cachazo and A.~Strominger, \emph{{Evidence for a New Soft Graviton Theorem}},
\href{http://www.arXiv.org/abs/1404.4091}{{\tt 1404.4091}}

\bibitem{Kapec:2014opa}
D.~Kapec, V.~Lysov, S.~Pasterski  and A.~Strominger, \emph{{Semiclassical Virasoro symmetry of the quantum gravity $ \mathcal{S}$-matrix}}, JHEP {\bf 08} (2014) 058,
\href{http://www.arXiv.org/abs/1406.3312}{{\tt 1406.3312}}

\bibitem{Kapec:2016jld}
D.~Kapec, P.~Mitra, A.-M. Raclariu  and A.~Strominger, \emph{{2D Stress Tensor for 4D Gravity}}, Phys. Rev. Lett. {\bf 119} (2017), no.~12, 121601,
\href{http://www.arXiv.org/abs/1609.00282}{{\tt 1609.00282}}

\bibitem{Arkani-Hamed:2020gyp}
N.~Arkani-Hamed, M.~Pate, A.-M. Raclariu  and A.~Strominger, \emph{{Celestial amplitudes from UV to IR}}, JHEP {\bf 08} (2021) 062,
\href{http://www.arXiv.org/abs/2012.04208}{{\tt 2012.04208}}

\bibitem{Puhm:2019zbl}
A.~Puhm, \emph{{Conformally Soft Theorem in Gravity}}, JHEP {\bf 09} (2020) 130,
\href{http://www.arXiv.org/abs/1905.09799}{{\tt 1905.09799}}

\bibitem{Schreiber:2017jsr}
A.~Schreiber, A.~Volovich  and M.~Zlotnikov, \emph{{Tree-level gluon amplitudes on the celestial sphere}}, Phys. Lett. {\bf B781} (2018) 349--357,
\href{http://www.arXiv.org/abs/1711.08435}{{\tt 1711.08435}}

\bibitem{Banerjee:2017jeg}
N.~Banerjee, S.~Banerjee, S.~Atul~Bhatkar  and S.~Jain, \emph{{Conformal Structure of Massless Scalar Amplitudes Beyond Tree level}}, JHEP {\bf 04} (2018) 039,
\href{http://www.arXiv.org/abs/1711.06690}{{\tt 1711.06690}}

\bibitem{Gonzalez:2020tpi}
H.~A. Gonz\'alez, A.~Puhm  and F.~Rojas, \emph{{Loop corrections to celestial amplitudes}}, Phys. Rev. D {\bf 102} (2020), no.~12, 126027,
\href{http://www.arXiv.org/abs/2009.07290}{{\tt 2009.07290}}

\bibitem{Albayrak:2020saa}
S.~Albayrak, C.~Chowdhury  and S.~Kharel, \emph{{On loop celestial amplitudes for gauge theory and gravity}}, Phys. Rev. D {\bf 102} (2020) 126020,
\href{http://www.arXiv.org/abs/2007.09338}{{\tt 2007.09338}}

\bibitem{Donnay:2018neh}
L.~Donnay, A.~Puhm  and A.~Strominger, \emph{{Conformally Soft Photons and Gravitons}}, JHEP {\bf 01} (2019) 184,
\href{http://www.arXiv.org/abs/1810.05219}{{\tt 1810.05219}}

\bibitem{Guevara:2019ypd}
A.~Guevara, \emph{{Notes on Conformal Soft Theorems and Recursion Relations in Gravity}},
\href{http://www.arXiv.org/abs/1906.07810}{{\tt 1906.07810}}

\bibitem{Fan:2019emx}
W.~Fan, A.~Fotopoulos  and T.~R. Taylor, \emph{{Soft Limits of Yang-Mills Amplitudes and Conformal Correlators}}, JHEP {\bf 05} (2019) 121,
\href{http://www.arXiv.org/abs/1903.01676}{{\tt 1903.01676}}

\bibitem{Pate:2019lpp}
M.~Pate, A.-M. Raclariu, A.~Strominger  and E.~Y. Yuan, \emph{{Celestial Operator Products of Gluons and Gravitons}},
\href{http://www.arXiv.org/abs/1910.07424}{{\tt 1910.07424}}

\bibitem{Guevara:2021abz}
A.~Guevara, E.~Himwich, M.~Pate  and A.~Strominger, \emph{{Holographic symmetry algebras for gauge theory and gravity}}, JHEP {\bf 11} (2021) 152,
\href{http://www.arXiv.org/abs/2103.03961}{{\tt 2103.03961}}

\bibitem{Strominger:2021mtt}
A.~Strominger, \emph{{$w_{1+\infty}$ Algebra and the Celestial Sphere: Infinite Towers of Soft Graviton, Photon, and Gluon Symmetries}}, Phys. Rev. Lett. {\bf 127} (2021), no.~22,
221601

\bibitem{Mago:2021wje}
J.~Mago, L.~Ren, A.~Y. Srikant  and A.~Volovich, \emph{{Deformed $w_{1+\infty}$ Algebras in the Celestial CFT}},
\href{http://www.arXiv.org/abs/2111.11356}{{\tt 2111.11356}}

\bibitem{Ren:2022sws}
L.~Ren, M.~Spradlin, A.~Yelleshpur~Srikant  and A.~Volovich, \emph{{On effective field theories with celestial duals}}, JHEP {\bf 08} (2022) 251,
\href{http://www.arXiv.org/abs/2206.08322}{{\tt 2206.08322}}

\bibitem{Bhardwaj:2022anh}
R.~Bhardwaj, L.~Lippstreu, L.~Ren, M.~Spradlin, A.~Yelleshpur~Srikant  and A.~Volovich, \emph{{Loop-level gluon OPEs in celestial holography}}, JHEP {\bf 11} (2022) 171,
\href{http://www.arXiv.org/abs/2208.14416}{{\tt 2208.14416}}

\bibitem{Jiang:2021ovh}
H.~Jiang, \emph{{Holographic chiral algebra: supersymmetry, infinite Ward identities, and EFTs}}, JHEP {\bf 01} (2022) 113,
\href{http://www.arXiv.org/abs/2108.08799}{{\tt 2108.08799}}

\bibitem{Drozdov:2023qoy}
P.~Drozdov and T.~Kimura, \emph{{Structure of deformed w1+\ensuremath{\infty} symmetry and topological generalization in Celestial CFT}}, Phys. Lett. B {\bf 847} (2023) 138272,
\href{http://www.arXiv.org/abs/2306.11693}{{\tt 2306.11693}}

\bibitem{Ahn:2021erj}
C.~Ahn, \emph{{Towards a supersymmetric w1+\ensuremath{\infty} symmetry in the celestial conformal field theory}}, Phys. Rev. D {\bf 105} (2022), no.~8, 086028,
\href{http://www.arXiv.org/abs/2111.04268}{{\tt 2111.04268}}

\bibitem{Ahn:2022oor}
C.~Ahn, \emph{{A deformed supersymmetric $w_{1+\infty }$ symmetry in the celestial conformal field theory}}, Eur. Phys. J. C {\bf 82} (2022), no.~7, 630,
\href{http://www.arXiv.org/abs/2202.02949}{{\tt 2202.02949}}

\bibitem{Freidel:2021ytz}
L.~Freidel, D.~Pranzetti  and A.-M. Raclariu, \emph{{Higher spin dynamics in gravity and $w_{1 + \infty}$ celestial symmetries}},
\href{http://www.arXiv.org/abs/2112.15573}{{\tt 2112.15573}}

\bibitem{Freidel:2023gue}
L.~Freidel, D.~Pranzetti  and A.-M. Raclariu, \emph{{On infinite symmetry algebras in Yang-Mills theory}},
\href{http://www.arXiv.org/abs/2306.02373}{{\tt 2306.02373}}

\bibitem{Bu:2022iak}
W.~Bu, S.~Heuveline  and D.~Skinner, \emph{{Moyal deformations, W$_{1+\infty}$ and celestial holography}}, JHEP {\bf 12} (2022) 011,
\href{http://www.arXiv.org/abs/2208.13750}{{\tt 2208.13750}}

\bibitem{Bittleston:2023bzp}
R.~Bittleston, S.~Heuveline  and D.~Skinner, \emph{{The celestial chiral algebra of self-dual gravity on Eguchi-Hanson space}}, JHEP {\bf 09} (2023) 008,
\href{http://www.arXiv.org/abs/2305.09451}{{\tt 2305.09451}}

\bibitem{Liu:2022mne}
W.-B. Liu and J.~Long, \emph{{Symmetry group at future null infinity I: scalar theory}},
\href{http://www.arXiv.org/abs/2210.00516}{{\tt 2210.00516}}

\bibitem{Salzer:2023jqv}
J.~Salzer, \emph{{An Embedding Space Approach to Carrollian CFT Correlators for Flat Space Holography}},
\href{http://www.arXiv.org/abs/2304.08292}{{\tt 2304.08292}}

\bibitem{Nguyen:2023miw}
K.~Nguyen, \emph{{Carrollian conformal correlators and massless scattering amplitudes}},
\href{http://www.arXiv.org/abs/2311.09869}{{\tt 2311.09869}}

\bibitem{Banerjee:2019prz}
S.~Banerjee, S.~Ghosh, P.~Pandey  and A.~P. Saha, \emph{{Modified celestial amplitude in Einstein gravity}}, JHEP {\bf 03} (2020) 125,
\href{http://www.arXiv.org/abs/1909.03075}{{\tt 1909.03075}}

\bibitem{Penrose:1962ij}
R.~Penrose, \emph{{Asymptotic properties of fields and space-times}}, Phys. Rev. Lett. {\bf 10} (1963)
66--68

\bibitem{Penrose:1964ge}
R.~Penrose,
\emph{{Conformal treatment of infinity}},

\bibitem{Newman:1966ub}
E.~T. Newman and R.~Penrose, \emph{{Note on the Bondi-Metzner-Sachs group}}, J. Math. Phys. {\bf 7} (1966)
863--870

\bibitem{Penrose:1986uia}
R.~Penrose and W.~Rindler, {\em {Spinors and Space-Time}}, vol.~2 of {\em Cambridge Monographs on Mathematical Physics}.
\newblock Cambridge Univ. Press, Cambridge, UK,
1986

\bibitem{Duval:2014uva}
C.~Duval, G.~W. Gibbons  and P.~A. Horvathy, \emph{{Conformal Carroll groups and BMS symmetry}}, Class. Quant. Grav. {\bf 31} (2014) 092001,
\href{http://www.arXiv.org/abs/1402.5894}{{\tt 1402.5894}}

\bibitem{Duval:2014lpa}
C.~Duval, G.~W. Gibbons  and P.~A. Horvathy, \emph{{Conformal Carroll groups}}, J. Phys. A {\bf 47} (2014), no.~33, 335204,
\href{http://www.arXiv.org/abs/1403.4213}{{\tt 1403.4213}}

\bibitem{1977asst.conf....1G}
R.~{Geroch}, \emph{{Asymptotic Structure of Space-Time}}, in {\em Asymptotic Structure of Space-Time}, F.~P. {Esposito} and L.~{Witten}, eds., p.~1.
\newblock
Jan., 1977.
\newblock

\bibitem{Ashtekar:2014zsa}
A.~Ashtekar, \emph{{Geometry and Physics of Null Infinity}},
\href{http://www.arXiv.org/abs/1409.1800}{{\tt 1409.1800}}

\bibitem{Bondi:1962px}
H.~Bondi, M.~G.~J. van~der Burg  and A.~W.~K. Metzner, \emph{{Gravitational waves in general relativity. 7. Waves from axisymmetric isolated systems}}, Proc. Roy. Soc. Lond. {\bf A269} (1962)
21

\bibitem{Sachs:1962wk}
R.~K. Sachs, \emph{{Gravitational waves in general relativity. 8. Waves in asymptotically flat space-times}}, Proc. Roy. Soc. Lond. {\bf A270} (1962)
103--126

\bibitem{Sachs:1962zza}
R.~Sachs, \emph{{Asymptotic symmetries in gravitational theory}}, Phys. Rev. {\bf 128} (1962)
2851--2864

\bibitem{Barnich:2009se}
G.~Barnich and C.~Troessaert, \emph{{Symmetries of asymptotically flat 4 dimensional spacetimes at null infinity revisited}}, Phys. Rev. Lett. {\bf 105} (2010) 111103,
\href{http://www.arXiv.org/abs/0909.2617}{{\tt 0909.2617}}

\bibitem{Bagchi:2022eav}
A.~Bagchi, A.~Banerjee, S.~Dutta, K.~S. Kolekar  and P.~Sharma, \emph{{Carroll covariant scalar fields in two dimensions}}, JHEP {\bf 01} (2023) 072,
\href{http://www.arXiv.org/abs/2203.13197}{{\tt 2203.13197}}

\bibitem{Gupta:2020dtl}
N.~Gupta and N.~V. Suryanarayana, \emph{{Constructing Carrollian CFTs}}, JHEP {\bf 03} (2021) 194,
\href{http://www.arXiv.org/abs/2001.03056}{{\tt 2001.03056}}

\bibitem{Rivera-Betancour:2022lkc}
D.~Rivera-Betancour and M.~Vilatte, \emph{{Revisiting the Carrollian scalar field}}, Phys. Rev. D {\bf 106} (2022), no.~8, 085004,
\href{http://www.arXiv.org/abs/2207.01647}{{\tt 2207.01647}}

\bibitem{Baiguera:2022lsw}
S.~Baiguera, G.~Oling, W.~Sybesma  and B.~T. S\o{}gaard, \emph{{Conformal Carroll Scalars with Boosts}},
\href{http://www.arXiv.org/abs/2207.03468}{{\tt 2207.03468}}

\bibitem{Duval:2014uoa}
C.~Duval, G.~W. Gibbons, P.~A. Horvathy  and P.~M. Zhang, \emph{{Carroll versus Newton and Galilei: two dual non-Einsteinian concepts of time}}, Class. Quant. Grav. {\bf 31} (2014) 085016,
\href{http://www.arXiv.org/abs/1402.0657}{{\tt 1402.0657}}

\bibitem{Bagchi:2019xfx}
A.~Bagchi, A.~Mehra  and P.~Nandi, \emph{{Field Theories with Conformal Carrollian Symmetry}}, JHEP {\bf 05} (2019) 108,
\href{http://www.arXiv.org/abs/1901.10147}{{\tt 1901.10147}}

\bibitem{Henneaux:2021yzg}
M.~Henneaux and P.~Salgado-Rebolledo, \emph{{Carroll contractions of Lorentz-invariant theories}}, JHEP {\bf 11} (2021) 180,
\href{http://www.arXiv.org/abs/2109.06708}{{\tt 2109.06708}}

\bibitem{Chen:2021xkw}
B.~Chen, R.~Liu  and Y.-f. Zheng, \emph{{On Higher-dimensional Carrollian and Galilean Conformal Field Theories}},
\href{http://www.arXiv.org/abs/2112.10514}{{\tt 2112.10514}}

\bibitem{deBoer:2021jej}
J.~de~Boer, J.~Hartong, N.~A. Obers, W.~Sybesma  and S.~Vandoren, \emph{{Carroll Symmetry, Dark Energy and Inflation}}, Front. in Phys. {\bf 10} (2022) 810405,
\href{http://www.arXiv.org/abs/2110.02319}{{\tt 2110.02319}}

\bibitem{Chen:2023pqf}
B.~Chen, R.~Liu, H.~Sun  and Y.-f. Zheng, \emph{{Constructing Carrollian Field Theories from Null Reduction}},
\href{http://www.arXiv.org/abs/2301.06011}{{\tt 2301.06011}}

\bibitem{Nguyen:2023vfz}
K.~Nguyen and P.~West, \emph{{Carrollian conformal fields and flat holography}},
\href{http://www.arXiv.org/abs/2305.02884}{{\tt 2305.02884}}

\bibitem{Grumiller:2020elf}
D.~Grumiller, J.~Hartong, S.~Prohazka  and J.~Salzer, \emph{{Limits of JT gravity}}, JHEP {\bf 02} (2021) 134,
\href{http://www.arXiv.org/abs/2011.13870}{{\tt 2011.13870}}

\bibitem{Gomis:2020wxp}
J.~Gomis, D.~Hidalgo  and P.~Salgado-Rebolledo, \emph{{Non-relativistic and Carrollian limits of Jackiw-Teitelboim gravity}}, JHEP {\bf 05} (2021) 162,
\href{http://www.arXiv.org/abs/2011.15053}{{\tt 2011.15053}}

\bibitem{Hansen:2021fxi}
D.~Hansen, N.~A. Obers, G.~Oling  and B.~T. S\o{}gaard, \emph{{Carroll Expansion of General Relativity}}, SciPost Phys. {\bf 13} (2022), no.~3, 055,
\href{http://www.arXiv.org/abs/2112.12684}{{\tt 2112.12684}}

\bibitem{Perez:2021abf}
A.~P\'erez, \emph{{Asymptotic symmetries in Carrollian theories of gravity}}, JHEP {\bf 12} (2021) 173,
\href{http://www.arXiv.org/abs/2110.15834}{{\tt 2110.15834}}

\bibitem{Fuentealba:2022gdx}
O.~Fuentealba, M.~Henneaux, P.~Salgado-Rebolledo  and J.~Salzer, \emph{{Asymptotic structure of Carrollian limits of Einstein-Yang-Mills theory in four spacetime dimensions}}, Phys. Rev. D {\bf 106} (2022), no.~10, 104047,
\href{http://www.arXiv.org/abs/2207.11359}{{\tt 2207.11359}}

\bibitem{Campoleoni:2022ebj}
A.~Campoleoni, M.~Henneaux, S.~Pekar, A.~P\'erez  and P.~Salgado-Rebolledo, \emph{{Magnetic Carrollian gravity from the Carroll algebra}}, JHEP {\bf 09} (2022) 127,
\href{http://www.arXiv.org/abs/2207.14167}{{\tt 2207.14167}}

\bibitem{deBoer:2023fnj}
J.~de~Boer, J.~Hartong, N.~A. Obers, W.~Sybesma  and S.~Vandoren, \emph{{Carroll stories}}, JHEP {\bf 09} (2023) 148,
\href{http://www.arXiv.org/abs/2307.06827}{{\tt 2307.06827}}

\bibitem{Hao:2022xhq}
P.-X. Hao, W.~Song, Z.~Xiao  and X.~Xie, \emph{{A BMS-invariant free fermion model}},
\href{http://www.arXiv.org/abs/2211.06927}{{\tt 2211.06927}}

\bibitem{Bagchi:2022eui}
A.~Bagchi, A.~Banerjee, R.~Basu, M.~Islam  and S.~Mondal, \emph{{Magic fermions: Carroll and flat bands}}, JHEP {\bf 03} (2023) 227,
\href{http://www.arXiv.org/abs/2211.11640}{{\tt 2211.11640}}

\bibitem{Banerjee:2022ocj}
A.~Banerjee, S.~Dutta  and S.~Mondal, \emph{{Carroll fermions in two dimensions}}, Phys. Rev. D {\bf 107} (2023), no.~12, 125020,
\href{http://www.arXiv.org/abs/2211.11639}{{\tt 2211.11639}}

\bibitem{Yu:2022bcp}
Z.-f. Yu and B.~Chen, \emph{{Free field realization of the BMS Ising model}}, JHEP {\bf 08} (2023) 116,
\href{http://www.arXiv.org/abs/2211.06926}{{\tt 2211.06926}}

\bibitem{Barducci:2018thr}
A.~Barducci, R.~Casalbuoni  and J.~Gomis, \emph{{Vector SUSY models with Carroll or Galilei invariance}}, Phys. Rev. D {\bf 99} (2019), no.~4, 045016,
\href{http://www.arXiv.org/abs/1811.12672}{{\tt 1811.12672}}

\bibitem{Bagchi:2022owq}
A.~Bagchi, D.~Grumiller  and P.~Nandi, \emph{{Carrollian superconformal theories and super BMS}},
\href{http://www.arXiv.org/abs/2202.01172}{{\tt 2202.01172}}

\bibitem{Bidussi:2021nmp}
L.~Bidussi, J.~Hartong, E.~Have, J.~Musaeus  and S.~Prohazka, \emph{{Fractons, dipole symmetries and curved spacetime}}, SciPost Phys. {\bf 12} (2022), no.~6, 205,
\href{http://www.arXiv.org/abs/2111.03668}{{\tt 2111.03668}}

\bibitem{Figueroa-OFarrill:2023vbj}
J.~Figueroa-O'Farrill, A.~P\'erez  and S.~Prohazka, \emph{{Carroll/fracton particles and their correspondence}}, JHEP {\bf 06} (2023) 207,
\href{http://www.arXiv.org/abs/2305.06730}{{\tt 2305.06730}}

\bibitem{Perez:2023uwt}
A.~P\'erez, S.~Prohazka  and A.~Seraj, \emph{{Fracton infrared triangle}},
\href{http://www.arXiv.org/abs/2310.16683}{{\tt 2310.16683}}

\bibitem{Barnich:2017jgw}
G.~Barnich, H.~A. Gonzalez  and P.~Salgado-Rebolledo, \emph{{Geometric actions for three-dimensional gravity}}, Class. Quant. Grav. {\bf 35} (2018), no.~1, 014003,
\href{http://www.arXiv.org/abs/1707.08887}{{\tt 1707.08887}}

\bibitem{Barnich:2022bni}
G.~Barnich, K.~Nguyen  and R.~Ruzziconi, \emph{{Geometric action for extended Bondi-Metzner-Sachs group in four dimensions}}, JHEP {\bf 12} (2022) 154,
\href{http://www.arXiv.org/abs/2211.07592}{{\tt 2211.07592}}

\bibitem{Merbis:2019wgk}
W.~Merbis and M.~Riegler, \emph{{Geometric actions and flat space holography}}, JHEP {\bf 02} (2020) 125,
\href{http://www.arXiv.org/abs/1912.08207}{{\tt 1912.08207}}

\bibitem{Barnich:2021dta}
G.~Barnich and R.~Ruzziconi, \emph{{Coadjoint representation of the BMS group on celestial Riemann surfaces}}, JHEP {\bf 06} (2021) 079,
\href{http://www.arXiv.org/abs/2103.11253}{{\tt 2103.11253}}

\bibitem{Ciambelli:2018xat}
L.~Ciambelli, C.~Marteau, A.~C. Petkou, P.~M. Petropoulos  and K.~Siampos, \emph{{Covariant Galilean versus Carrollian hydrodynamics from relativistic fluids}}, Class. Quant. Grav. {\bf 35} (2018), no.~16, 165001,
\href{http://www.arXiv.org/abs/1802.05286}{{\tt 1802.05286}}

\bibitem{Freidel:2021qpz}
L.~Freidel and D.~Pranzetti, \emph{{Gravity from symmetry: duality and impulsive waves}}, JHEP {\bf 04} (2022) 125,
\href{http://www.arXiv.org/abs/2109.06342}{{\tt 2109.06342}}

\bibitem{Ashtekar:1981hw}
A.~Ashtekar, \emph{{Radiative Degrees of Freedom of the Gravitational Field in Exact General Relativity}}, J. Math. Phys. {\bf 22} (1981)
2885--2895

\bibitem{Ashtekar:1981sf}
A.~Ashtekar, \emph{{Asymptotic Quantization of the Gravitational Field}}, Phys. Rev. Lett. {\bf 46} (1981)
573--576

\bibitem{Newman:1961qr}
E.~Newman and R.~Penrose, \emph{{An Approach to gravitational radiation by a method of spin coefficients}}, J. Math. Phys. {\bf 3} (1962)
566--578

\bibitem{Penrose1980GoldenON}
R.~Penrose, \emph{Null Hypersurface Initial Data for Classical Fields of Arbitrary Spin and for General Relativity}, General Relativity and Gravitation {\bf 12} (1980)
225--264

\bibitem{Penrose:1985bww}
R.~Penrose and W.~Rindler, {\em {Spinors and Space-Time}}.
\newblock Cambridge Monographs on Mathematical Physics. Cambridge Univ. Press, Cambridge, UK,
4, 2011

\bibitem{Banerjee:2018gce}
S.~Banerjee, \emph{{Null Infinity and Unitary Representation of The Poincare Group}}, JHEP {\bf 01} (2019) 205,
\href{http://www.arXiv.org/abs/1801.10171}{{\tt 1801.10171}}

\bibitem{Banerjee:2018fgd}
S.~Banerjee, \emph{{Symmetries of free massless particles and soft theorems}}, Gen. Rel. Grav. {\bf 51} (2019), no.~9, 128,
\href{http://www.arXiv.org/abs/1804.06646}{{\tt 1804.06646}}

\bibitem{Fiorucci:2023lpb}
A.~Fiorucci, D.~Grumiller  and R.~Ruzziconi, \emph{{Logarithmic Celestial Conformal Field Theory}},
\href{http://www.arXiv.org/abs/2305.08913}{{\tt 2305.08913}}

\bibitem{Pasterski:2021dqe}
S.~Pasterski, A.~Puhm  and E.~Trevisani, \emph{{Revisiting the conformally soft sector with celestial diamonds}}, JHEP {\bf 11} (2021) 143,
\href{http://www.arXiv.org/abs/2105.09792}{{\tt 2105.09792}}

\bibitem{Atanasov:2021oyu}
A.~Atanasov, A.~Ball, W.~Melton, A.-M. Raclariu  and A.~Strominger, \emph{{(2, 2) Scattering and the celestial torus}}, JHEP {\bf 07} (2021) 083,
\href{http://www.arXiv.org/abs/2101.09591}{{\tt 2101.09591}}

\bibitem{McGady:2013sga}
D.~A. McGady and L.~Rodina, \emph{{Higher-spin massless $S$-matrices in four-dimensions}}, Phys. Rev. D {\bf 90} (2014), no.~8, 084048,
\href{http://www.arXiv.org/abs/1311.2938}{{\tt 1311.2938}}

\bibitem{Mangano:1990by}
M.~L. Mangano and S.~J. Parke, \emph{{Multiparton amplitudes in gauge theories}}, Phys. Rept. {\bf 200} (1991) 301--367,
\href{http://www.arXiv.org/abs/hep-th/0509223}{{\tt hep-th/0509223}}

\bibitem{Mizera:2022sln}
S.~Mizera and S.~Pasterski, \emph{{Celestial geometry}}, JHEP {\bf 09} (2022) 045,
\href{http://www.arXiv.org/abs/2204.02505}{{\tt 2204.02505}}

\bibitem{book:596104}
M.~K.~a. Kazuhiko~Aomoto, {\em Theory of hypergeometric functions}.
\newblock Springer Monographs in Mathematics. Springer Tokyo, 1~ed.,
2011

\bibitem{Weinberg:1965nx}
S.~Weinberg, \emph{{Infrared photons and gravitons}}, Phys. Rev. {\bf 140} (1965)
B516--B524

\bibitem{Casali:2014xpa}
E.~Casali, \emph{{Soft sub-leading divergences in Yang-Mills amplitudes}}, JHEP {\bf 08} (2014) 077,
\href{http://www.arXiv.org/abs/1404.5551}{{\tt 1404.5551}}

\bibitem{Low:1958sn}
F.~Low, \emph{{Bremsstrahlung of very low-energy quanta in elementary particle collisions}}, Phys. Rev. {\bf 110} (1958)
974--977

\bibitem{Donnay:2022sdg}
L.~Donnay, S.~Pasterski  and A.~Puhm, \emph{{Goldilocks modes and the three scattering bases}}, JHEP {\bf 06} (2022) 124,
\href{http://www.arXiv.org/abs/2202.11127}{{\tt 2202.11127}}

\bibitem{Strominger:2014pwa}
A.~Strominger and A.~Zhiboedov, \emph{{Gravitational Memory, BMS Supertranslations and Soft Theorems}}, JHEP {\bf 01} (2016) 086,
\href{http://www.arXiv.org/abs/1411.5745}{{\tt 1411.5745}}

\bibitem{Pasterski:2015zua}
S.~Pasterski, \emph{{Asymptotic Symmetries and Electromagnetic Memory}}, JHEP {\bf 09} (2017) 154,
\href{http://www.arXiv.org/abs/1505.00716}{{\tt 1505.00716}}

\bibitem{1502.06120PSZ}
S.~Pasterski, A.~Strominger  and A.~Zhiboedov, \emph{{New Gravitational Memories}}, JHEP {\bf 12} (2016) 053,
\href{http://www.arXiv.org/abs/1502.06120}{{\tt 1502.06120}}

\bibitem{Berends:1988zn}
F.~A. Berends and W.~T. Giele, \emph{{Multiple Soft Gluon Radiation in Parton Processes}}, Nucl. Phys. B {\bf 313} (1989)
595--633

\bibitem{Himwich:2021dau}
E.~Himwich, M.~Pate  and K.~Singh, \emph{{Celestial operator product expansions and w$_{1+\infty}$ symmetry for all spins}}, JHEP {\bf 01} (2022) 080,
\href{http://www.arXiv.org/abs/2108.07763}{{\tt 2108.07763}}

\bibitem{Ferrara:1971vh}
S.~Ferrara, A.~F. Grillo  and R.~Gatto, \emph{{Manifestly conformal covariant operator-product expansion}}, Lett. Nuovo Cim. {\bf 2S2} (1971)
1363--1369

\bibitem{Strominger:2021lvk}
A.~Strominger, \emph{{w(1+infinity) and the Celestial Sphere}}, Phys. Rev. Lett. {\bf 127} (5, 2021) 221601,
\href{http://www.arXiv.org/abs/2105.14346}{{\tt 2105.14346}}

\bibitem{Banerjee:2019aoy}
S.~Banerjee, P.~Pandey  and P.~Paul, \emph{{Conformal properties of soft operators: Use of null states}}, Phys. Rev. D {\bf 101} (2020), no.~10, 106014,
\href{http://www.arXiv.org/abs/1902.02309}{{\tt 1902.02309}}

\bibitem{Banerjee:2019tam}
S.~Banerjee and P.~Pandey, \emph{{Conformal properties of soft-operators. Part II. Use of null-states}}, JHEP {\bf 02} (2020) 067,
\href{http://www.arXiv.org/abs/1906.01650}{{\tt 1906.01650}}

\bibitem{Sparling:1990}
G.~A.~J. Sparling, \emph{{Dynamically Broken Symmetry and Global Yang-Mills in Minkowski Space}}, in {\em {Further Advances in Twistor Theory}}, L.~J. Mason and L.~P. Hughston, eds., vol.~231, ch.~1.4.2.
\newblock Pitman Research Notes in Mathematics,
1990.
\newblock

\bibitem{Newman:1962cia}
E.~T. Newman and T.~W.~J. Unti, \emph{{Behavior of Asymptotically Flat Empty Spaces}}, J. Math. Phys. {\bf 3} (1962), no.~5,
891

\bibitem{John:1938}
F.~John, \emph{The ultrahyperbolic differential equation with four independent variables}, Duke Math. J. {\bf 4} (1938), no.~2,
300--322

\bibitem{Witten:2003nn}
E.~Witten, \emph{{Perturbative gauge theory as a string theory in twistor space}}, Commun. Math. Phys. {\bf 252} (2004) 189--258,
\href{http://www.arXiv.org/abs/hep-th/0312171}{{\tt hep-th/0312171}}

\bibitem{Mason:2009sa}
L.~J. Mason and D.~Skinner, \emph{{Scattering Amplitudes and BCFW Recursion in Twistor Space}}, JHEP {\bf 01} (2010) 064,
\href{http://www.arXiv.org/abs/0903.2083}{{\tt 0903.2083}}

\bibitem{Lax:1967}
P.~D. Lax and R.~S. Phillips, {\em Scattering theory}, vol.~Vol. 26 of {\em Pure and Applied Mathematics}.
\newblock Academic Press, New York-London,
1967

\bibitem{Friedlander:1980}
F.~G. Friedlander, \emph{Radiation fields and hyperbolic scattering theory}, Math. Proc. Cambridge Philos. Soc. {\bf 88} (1980), no.~3,
483--515

\bibitem{Mason:2004lqj}
L.~J. Mason and J.-P. Nicolas, \emph{{Conformal Scattering and the Goursat Problem}}, J. Hyperbol. Diff. Equat. {\bf 1} (2004), no.~02,
197--233

\bibitem{Ashtekar:1987tt}
A.~Ashtekar, \emph{{Asymptotic Quantization: Based on 1984 Naples Lectures}}, Naples, Italy: Bibliopolis (1987) 107 p. (Monographs And Textbooks In Physical Science, 2)
(1987)

\bibitem{Hofman:2008ar}
D.~M. Hofman and J.~Maldacena, \emph{{Conformal collider physics: Energy and charge correlations}}, JHEP {\bf 05} (2008) 012,
\href{http://www.arXiv.org/abs/0803.1467}{{\tt 0803.1467}}

\bibitem{Mason:2013sva}
L.~Mason and D.~Skinner, \emph{{Ambitwistor strings and the scattering equations}}, JHEP {\bf 07} (2014) 048,
\href{http://www.arXiv.org/abs/1311.2564}{{\tt 1311.2564}}

\bibitem{Geyer:2014fka}
Y.~Geyer, A.~E. Lipstein  and L.~J. Mason, \emph{{Ambitwistor Strings in Four Dimensions}}, Phys. Rev. Lett. {\bf 113} (2014), no.~8, 081602,
\href{http://www.arXiv.org/abs/1404.6219}{{\tt 1404.6219}}

\bibitem{Witten:1998qj}
E.~Witten, \emph{{Anti-de Sitter space and holography}}, Adv. Theor. Math. Phys. {\bf 2} (1998) 253--291,
\href{http://www.arXiv.org/abs/hep-th/9802150}{{\tt hep-th/9802150}}

\bibitem{Gonzo:2022tjm}
R.~Gonzo, T.~McLoughlin  and A.~Puhm, \emph{{Celestial holography on Kerr-Schild backgrounds}}, JHEP {\bf 10} (2022) 073,
\href{http://www.arXiv.org/abs/2207.13719}{{\tt 2207.13719}}

\bibitem{Kim:2023qbl}
S.~Kim, P.~Kraus, R.~Monten  and R.~M. Myers, \emph{{S-matrix path integral approach to symmetries and soft theorems}}, JHEP {\bf 10} (2023) 036,
\href{http://www.arXiv.org/abs/2307.12368}{{\tt 2307.12368}}

\bibitem{Jain:2023fxc}
D.~Jain, S.~Kundu, S.~Minwalla, O.~Parrikar, S.~G. Prabhu  and P.~Shrivastava, \emph{{The S-matrix and boundary correlators in flat space}},
\href{http://www.arXiv.org/abs/2311.03443}{{\tt 2311.03443}}

\bibitem{Barnich:2012aw}
G.~Barnich, A.~Gomberoff  and H.~A. Gonzalez, \emph{{The Flat limit of three dimensional asymptotically anti-de Sitter spacetimes}}, Phys. Rev. D {\bf 86} (2012) 024020,
\href{http://www.arXiv.org/abs/1204.3288}{{\tt 1204.3288}}

\bibitem{Compere:2019bua}
G.~Comp\`ere, A.~Fiorucci  and R.~Ruzziconi, \emph{{The $\Lambda$-BMS$_4$ group of dS$_4$ and new boundary conditions for AdS$_4$}}, Class. Quant. Grav. {\bf 36} (2019), no.~19, 195017, \href{http://www.arXiv.org/abs/1905.00971}{{\tt 1905.00971}},
[Erratum: Class.Quant.Grav. 38, 229501 (2021)]

\bibitem{Compere:2020lrt}
G.~Comp\`ere, A.~Fiorucci  and R.~Ruzziconi, \emph{{The $\Lambda$-BMS$_4$ charge algebra}}, JHEP {\bf 10} (2020) 205,
\href{http://www.arXiv.org/abs/2004.10769}{{\tt 2004.10769}}

\bibitem{Ruzziconi:2020cjt}
R.~Ruzziconi, {\em {On the Various Extensions of the BMS Group}}.
\newblock PhD thesis, U. Brussels, 2020.
\newblock
\href{http://www.arXiv.org/abs/2009.01926}{{\tt 2009.01926}}.
\newblock

\bibitem{Ciambelli:2020ftk}
L.~Ciambelli, C.~Marteau, P.~M. Petropoulos  and R.~Ruzziconi, \emph{{Fefferman-Graham and Bondi Gauges in the Fluid/Gravity Correspondence}}, PoS {\bf CORFU2019} (2020) 154,
\href{http://www.arXiv.org/abs/2006.10083}{{\tt 2006.10083}}

\bibitem{Ciambelli:2020eba}
L.~Ciambelli, C.~Marteau, P.~M. Petropoulos  and R.~Ruzziconi, \emph{{Gauges in Three-Dimensional Gravity and Holographic Fluids}}, JHEP {\bf 11} (2020) 092,
\href{http://www.arXiv.org/abs/2006.10082}{{\tt 2006.10082}}

\bibitem{Fiorucci:2020xto}
A.~Fiorucci and R.~Ruzziconi, \emph{{Charge algebra in Al(A)dS$_{n}$ spacetimes}}, JHEP {\bf 05} (2021) 210,
\href{http://www.arXiv.org/abs/2011.02002}{{\tt 2011.02002}}

\bibitem{Ruzziconi:2020wrb}
R.~Ruzziconi and C.~Zwikel, \emph{{Conservation and Integrability in Lower-Dimensional Gravity}}, JHEP {\bf 04} (2021) 034,
\href{http://www.arXiv.org/abs/2012.03961}{{\tt 2012.03961}}

\bibitem{Campoleoni:2022wmf}
A.~Campoleoni, L.~Ciambelli, A.~Delfante, C.~Marteau, P.~M. Petropoulos  and R.~Ruzziconi, \emph{{Holographic Lorentz and Carroll frames}}, JHEP {\bf 12} (2022) 007,
\href{http://www.arXiv.org/abs/2208.07575}{{\tt 2208.07575}}

\bibitem{Geiller:2022vto}
M.~Geiller and C.~Zwikel, \emph{{The partial Bondi gauge: Further enlarging the asymptotic structure of gravity}}, SciPost Phys. {\bf 13} (2022) 108,
\href{http://www.arXiv.org/abs/2205.11401}{{\tt 2205.11401}}

\bibitem{Campoleoni:2023fug}
A.~Campoleoni, A.~Delfante, S.~Pekar, P.~M. Petropoulos, D.~Rivera-Betancour  and M.~Vilatte, \emph{{Flat from anti-de Sitter}},
\href{http://www.arXiv.org/abs/2309.15182}{{\tt 2309.15182}}

\bibitem{Adami:2023fbm}
H.~Adami, A.~Parvizi, M.~M. Sheikh-Jabbari, V.~Taghiloo  and H.~Yavartanoo, \emph{{Hydro \& thermo dynamics at causal boundaries, examples in 3d gravity}}, JHEP {\bf 07} (2023) 038,
\href{http://www.arXiv.org/abs/2305.01009}{{\tt 2305.01009}}

\bibitem{Hijano:2019qmi}
E.~Hijano, \emph{{Flat space physics from AdS/CFT}}, JHEP {\bf 07} (2019) 132,
\href{http://www.arXiv.org/abs/1905.02729}{{\tt 1905.02729}}

\bibitem{deGioia:2022fcn}
L.~P. de~Gioia and A.-M. Raclariu, \emph{{Eikonal approximation in celestial CFT}}, JHEP {\bf 03} (2023) 030,
\href{http://www.arXiv.org/abs/2206.10547}{{\tt 2206.10547}}

\bibitem{deGioia:2023cbd}
L.~P. de~Gioia and A.-M. Raclariu, \emph{{Celestial Sector in CFT: Conformally Soft Symmetries}},
\href{http://www.arXiv.org/abs/2303.10037}{{\tt 2303.10037}}

\bibitem{Bagchi:2023fbj}
A.~Bagchi, P.~Dhivakar  and S.~Dutta, \emph{{AdS Witten diagrams to Carrollian correlators}}, JHEP {\bf 04} (2023) 135,
\href{http://www.arXiv.org/abs/2303.07388}{{\tt 2303.07388}}

\bibitem{Bagchi:2023cen}
A.~Bagchi, P.~Dhivakar  and S.~Dutta, \emph{{Holography in Flat Spacetimes: the case for Carroll}},
\href{http://www.arXiv.org/abs/2311.11246}{{\tt 2311.11246}}

\bibitem{Mason:2005zm}
L.~J. Mason, \emph{{Twistor actions for non-self-dual fields: A Derivation of twistor-string theory}}, JHEP {\bf 10} (2005) 009,
\href{http://www.arXiv.org/abs/hep-th/0507269}{{\tt hep-th/0507269}}

\bibitem{Mason:2007ct}
L.~J. Mason and M.~Wolf, \emph{{Twistor Actions for Self-Dual Supergravities}}, Commun. Math. Phys. {\bf 288} (2009) 97--123,
\href{http://www.arXiv.org/abs/0706.1941}{{\tt 0706.1941}}

\bibitem{Mason:2009afn}
L.~J. Mason and D.~Skinner, \emph{{Gravity, Twistors and the MHV Formalism}}, Commun. Math. Phys. {\bf 294} (2010) 827--862,
\href{http://www.arXiv.org/abs/0808.3907}{{\tt 0808.3907}}

\bibitem{Adamo:2011pv}
T.~Adamo, M.~Bullimore, L.~Mason  and D.~Skinner, \emph{{Scattering Amplitudes and Wilson Loops in Twistor Space}}, J. Phys. A {\bf 44} (2011) 454008,
\href{http://www.arXiv.org/abs/1104.2890}{{\tt 1104.2890}}

\bibitem{Adamo:2013tja}
T.~Adamo and L.~Mason, \emph{{Conformal and Einstein gravity from twistor actions}}, Class. Quant. Grav. {\bf 31} (2014), no.~4, 045014,
\href{http://www.arXiv.org/abs/1307.5043}{{\tt 1307.5043}}

\end{thebibliography}\endgroup

\end{document}